\documentclass[11pt]{article}
\usepackage{graphicx} 

\usepackage{epstopdf}

\usepackage[numbers]{natbib}

\usepackage{tikz}
\usetikzlibrary{positioning}
\usetikzlibrary{trees} 
\usetikzlibrary{calc,trees,positioning,arrows,chains,shapes.geometric,decorations.pathreplacing,decorations.pathmorphing,shapes,  matrix,shapes.symbols}

\usepackage{hyperref}
\usepackage{amsmath}
\usepackage{amssymb,amsthm,bm}
\usepackage{amsfonts}
\usepackage{graphics}
\usepackage{graphicx}         
\usepackage{color}
\usepackage{subfigure}
\usepackage{multirow}
\usepackage{multicol}
\usepackage{enumitem}
\usepackage{cancel}
\usepackage{booktabs}
\usepackage{tabulary}
\usepackage{geometry}
\usepackage{relsize}
\usepackage{csquotes}
\usepackage{rotating}
\usepackage{latexsym, amsfonts}

\usepackage{mathtools}

\newcommand{\bc}{\boldsymbol{c}}

\begin{document}


\title{Incorporating therapist capacity, availability and turnover in trials of therapist-delivered interventions: Non-orthogonal factorial designs}

\author{
  Arpan Singh\textsuperscript{1*}, Steven G Gilmour\textsuperscript{2} and Rebecca EA Walwyn\textsuperscript{1} \\
  \textsuperscript{1}Leeds Institute for Clinical Trials Research,\\ University of Leeds, Leeds, United Kingdom, LS2 9JT \\
\textsuperscript{2}Department of Mathematics,\\King’s College London, London, United Kingdom, WC2R 2LS \\
  \textsuperscript{*}Email: a.singh8@leeds.ac.uk
}
\date{}

\maketitle
\begin{abstract}

\noindent Disentangling therapist from intervention effects has long been considered a challenge in randomised trials of therapist-delivered interventions. When separating these effects is of interest, factorial designs have been proposed in which therapists and interventions are regarded as two potentially interacting components of a complex intervention, represented by two treatment factors (one random, one fixed). In these trial designs, every therapist delivers each intervention (therapists and interventions are crossed), and patients are randomised to the resulting therapist-intervention combinations. It is assumed that each therapist is allocated the same number of patients (equal capacity), that all therapists in a centre are available when a patient is randomised (equal availability), and that therapists remain involved for the entire duration of the trial (no turnover). In practice, these are strong assumptions that if enforced may lead to delayed or suspended randomisation, prolonged recruitment and increased trial costs. We propose three novel randomisation strategies that explicitly account for therapist capacity, availability, and turnover while maintaining desirable statistical properties. In a simulation study, we compared the proposed and existing randomisation strategies across trial scenarios and design parameters, assessing trade‑offs between operational efficiency and statistical performance. The results provide practical guidance for trialists designing trials of therapist-delivered interventions under real‑world constraints. Further work is needed to understand the relative performance of these randomisation strategies under a range of realistic values of the model parameters.

\end{abstract}

\section{Introduction}\label{sec.introduction}

The person who delivers an intervention, whether that is, for example, psychotherapy, physiotherapy, occupational therapy, rehabilitation or surgery, is often central to the treatment a patient receives. They not only deliver the 'mandatory' components of an intervention to patients with potentially different levels of training or experience, but also 'optional' components at their own discretion. The person may be referred to as a 'therapist', 'surgeon', 'nurse', 'coach', 'facilitator', 'professional' or 'practitioner', for example, but we will refer to them generically as therapists here. Randomised controlled trials of therapist-delivered interventions frequently employ an individually-randomised parallel-arm design (also referred to as an individually-randomised group-treatment (IRGT) trial design) in which interventions are simply randomised to patients (\cite{Pals_2008,Walwyn_2025_prep}). This trial design can be used regardless of the relationship between therapists and interventions (also referred to as the complex intervention design). Where every therapist delivers each intervention (the 'crossed' therapist-intervention design), two options are to include the therapists as a stratification factor or as a treatment factor in the randomisation \cite{Walwyn_2025_prep}. The primary objective in these trials is often to estimate the fixed intervention effect, but there may also be interest in estimating the random therapist effect (between-therapist variance in patient outcomes) and therapist-by-intervention interaction (between-therapist variance in the intervention effect).

Recently, Walwyn et al \cite{Walwyn_2026} built on a classical Design of Experiments (DoE) approach \cite{Bailey_2008} to formally set out the statistical design and associated data analysis for trials which regard therapists and interventions as two potentially interacting treatment factors (one random, one fixed) with a factorial structure. The motivation for this was to start to address a longstanding methodological challenge in trials of therapist-delivered interventions \cite{Wilkins_1986, Elkin_1999}: the inherent confounding of therapist and intervention effects typical in individually-randomised parallel-arm trials. This challenge was highlighted by Walwyn and Roberts \cite{Walwyn_2010}, who outlined two potential selection biases in parallel-arm trials, relating to the non-random allocation of (i) interventions to therapists and (ii) therapists to patients. These biases result in difficulties interpreting intervention effects and therapist effects as causal effects of the intervention and therapist, respectively. Walwyn et al \cite{Walwyn_2026} set out a family of three orthogonal factorial designs. The first assumed patients formed a single 'batch' and therapists were affiliated with a single 'centre' (completely randomised factorial design). The second assumed that patients formed multiple batches over time, allowing for the possibility of therapist learning (the randomised block factorial design). The third further assumed that therapists were nested in multiple centres, allowing for the possibility of contextual differences (the multicentre randomised block factorial design).

A number of assumptions were made by Walwyn et al \cite{Walwyn_2026} that present logistical challenges for trial delivery. First, it was assumed that each therapist is randomly allocated the same number of patients in each batch and over the course of the trial. In practice, therapists will commonly have variable capacities to take on trial patients. This may be because some are 'research therapists' employed to deliver the trial, while others are embedded into clinical services and work part-time on the trial. Or because some 'embedded therapists' have more additional responsibilities outside of the trial than others. We refer to whether a therapist is full-time or embedded in a service as the 'status' of the therapist. In routine clinical practice, it is possible that therapist capacity is also associated with their level of seniority and is therefore informative. The impact of varying capacities is that therapists will have unequal caseloads or cluster sizes, indicating the need to consider non-orthogonal factorial designs. Second, it was assumed that all therapists in a centre are available to accept a new patient at the point of randomisation. In practice, not every therapist will have space in their caseload to accept new patients in a particular batch. This may be because they have already reached their limit for that batch, or because of scheduled leave or unexpected additional responsibilities outside of the trial. The impact of varying therapist availability is that only a subset of the therapists in a centre can be considered when randomising each patient in each batch, with the subset varying across time and not being under the control of the research team. This indicates the need to consider highly flexible non-orthogonal factorial designs. Third, it was assumed that all therapists remain involved for the duration of the trial. In practice, centres may come on board incrementally over time and not every therapist will start treating patients from the beginning of a trial. Also, therapists may stop treating trial patients mid-way through a trial, a phenomenon known as therapist turnover. Typically, new therapists are recruited to replace them. The impact of this is that the sample of therapists should be regarded as a dynamic or open cohort in the trial design.

Related problems have been studied in relation to (i) cluster-randomised trials (CRTs) and (ii) the Design of Experiments in agriculture and industry. First, in relation to CRTs, work on design challenges arising from unequal cluster sizes (see \cite{Candel_2016}; \cite{Girling_2018}) has focused on optimal design and methods for calculating sample size, particularly quantifying the efficiency loss relative to equal cluster sizes using a 'coefficient of variation' (CV) correction (refer to \cite{Candel_2009}). Unequal cluster sizes have thus been treated largely as a nuisance, and not a natural constraint to accommodate in the randomisation. Second, also in relation to CRTs, sample size determination for open cohorts has been studied \cite{Kasza_2020}, and analytical strategies have been proposed, using modelling frameworks such as multiple membership models \citep{Fielding_2006, Marsden_2022}. However, the open cohorts here consist of clusters (therapists) in blocks (batches) rather than experimental units (patients) in clusters (therapists). This suggests new inferential procedures may be required. Third, in the Design of Experiments, when the number of treatments exceeds the number of experimental units (patients) in a block (batch), the inability to randomise all the treatments to the units in each block is referred to as `incomplete blocking'. Incomplete Block Designs (IBDs) were introduced to address this (see Supplementary file A). Their construction, statistical properties and optimality have been widely studied \cite{John_1977, Hinkelmann_2005, Mead_2012}. Originally proposed by Yates \cite{Yates_1936}, IBDs aim to improve the precision of estimates of treatment effects by reducing unit (patient) heterogeneity while ensuring balanced treatment comparisons through controlled replication and co-occurrence of treatments across blocks (batches). In classical IBDs, the total number of treatments is fixed throughout the experiment. Where therapists are open cohorts in trials of therapist-delivered interventions, the total number of therapist-intervention combinations (treatments) is not fixed. So, although the conceptual framework of IBDs is informative, it is not possible to directly apply these designs here.

In this paper, we treat the practical constraints resulting from variable therapist capacity, availability and turnover as core components of the trial design. We define therapist capacity as the number of patients a therapist can treat in a batch. We allow therapist capacity to differ depending on their status, with full-time and embedded therapists forming two distinct therapist populations. We define therapist availability as a therapist’s readiness to accept new patients. If availability is operationalised as whether each therapist is available or not in a specific batch, constraints resulting from variable capacity and availability can be decided on a batch-by-batch basis (i.e., at discrete time points throughout a trial). If availability is operationalised as whether each therapist is available or not on a patient-by-patient basis, the constraints will need to apply continuously throughout a trial. Finally, we define therapist turnover as changes to the therapist pool over time. Recognising that turnover will depend on the length of the trial recruitment and treatment periods, we will treat turnover as known at the same point as availability (i.e., before patient randomisation at discrete time points or on a patient-by-patient basis). The aim of this paper is to propose and explore the properties of a family of non-orthogonal factorial designs. These designs aim to relax the assumptions made by Walwyn et al \cite{Walwyn_2026} to increase the feasibility and operational efficiency of delivering factorial designs in future trials. We start by outlining two motivating examples in Section \ref{sec.motivating.examples}. In Section \ref{Sec.Rand.Methods}, we then describe three existing randomisation strategies [\cite{Walwyn_2026, Shapiro_1994, Lamb_2015, Lambert_1989, Lutz_2007}], and propose three more flexible alternatives. In Section \ref{sec.models}, we set out two statistical models possible for data analysis. In Section \ref{sec.simulation.studies}, we present a simulation study exploring the impact of the randomisation strategy on the ability to randomise patients to time and target, and on the robust estimation of the treatment effects. We end with a discussion and some conclusions in Section \ref{sec.discussion}. Appendix~\ref{appendix.examples} provides brief examples illustrating each randomisation strategy considered. Appendix~\ref{appendix.recommendations} summarises the key considerations and recommendations for clinical trialists when selecting among these strategies.

\section{Motivating Examples}\label{sec.motivating.examples}

\subsection{Example A}
In 1994, \cite{Shapiro_1994} reported the results of a landmark psychotherapy trial, the Second Sheffield Psychotherapy Project. The trial compared two psychotherapy modalities (cognitive behavioral and psychodynamic interpersonal psychotherapy) and two durations (8 or 16 sessions) for major depressive disorder in a $2\times$2 factorial trial. One of five `research therapists' were randomised to patients, accounting for therapist availability (\cite{Shapiro_1990}), so that the therapist was a third treatment factor with five levels. The five therapists were all clinical psychologists affiliated to a single centre. A target of 120 patients was set, with patients either being self-referred or referred by general practitioners, psychiatrists or occupational health services over a 6-year recruitment period. The randomisation was stratified by initial symptom severity (low, moderate, high). It was planned that each therapist would treat 24 patients (equal capacity was assumed), two in each modality, duration and severity combination. As such, subject to availability of suitable patients, equal replication was planned across combinations of the levels of the stratification factor and three treatment factors. That is, an orthogonal factorial design was planned.

\subsection{Example B}
In 2014, \cite{Lamb_2015} reported the results of the SARAH trial, a trial that compared tailored hand exercises in addition to usual care to usual care alone for patients with rheumatoid arthritis of the hand. Both interventions were delivered by 48 physio- and occupational-therapists 'embedded' in 17 NHS sites. This gives an average of 2.8 therapists per centre. One of the two interventions was individually-randomised to a total of 490 patients, stratified by NHS site, in a pragmatic, multicentre, parallel-arm trial. This gives an average of 28.8 patients per centre and 10.2 patients per therapist. The recruitment period was 19 months. Stratified block randomisation was used, with variable block sizes of between 2-8 patients to facilitate concealment of allocations prior to randomisation. As a result, batches were of size 2-8 patients per centre. So across the trial as a whole, batches were of 34-136 patients (85 patients on average), giving around 5-6 batches. Unequal replication of therapists per centre, patients per therapist, and patients per batch-centre combination makes the trial design non-orthogonal. This is typical in a pragmatic trial.

\section{Randomisation Strategies} \label{Sec.Rand.Methods}

We first set out our notation for the parameters whose values must be decided at the design stage to precisely specify the trial design. We refer to these parameters collectively as the \emph{design parameters}. A brief summary of all important notation is given in Table \ref{Table.notation}. We then summarise three existing randomisation strategies and three novel randomisation strategies. A combined flowchart of all the randomisation strategies is given in Figure \ref{fig.flowchart.all} and the individual flowcharts are given in Supplementary file I. Each strategy is illustrated with a hypothetical example, building on the motivating examples, in the Appendix \ref{appendix.examples}. Until stated otherwise, we assume that patients are available to be randomised if a therapist is selected to treat them. 

\subsection{Notation}
We denote numbers with normal fonts and sets with calligraphic fonts. Let $N$ denote the overall target number of patients. Let $n_I$ be the number of \emph{interventions} (i.e., therapies or procedures, including any controls), $n_B$ be the target number of \emph{batches} (i.e., cohorts of patients formed across a recruitment period), $n_C$ be the target number of \emph{centres} (i.e., sites, services or regions by which therapists are grouped), and $n_T^{(C)}$ be the planned number of \emph{therapists} in centre $C$. Where $\mathcal{T}^{(C)}$ is the set of therapists in centre $C$, and $|\mathcal{T}^{(C)}|=n_T^{(C)}$, we assume $\mathcal{T}^{(C)} = \{1^{(C)},2^{(C)},\ldots,n_T^{(C)}\}$. As a result, we allow each centre to have a different number of therapists, recognising that, in part due to turnover, the total number of therapists in a centre across a trial may be unknown at the design stage, but will be predictable. Where $n_{Rj}^{(C)}$ denotes the number of patients therapist $j$ nested in centre $C$ treats with each intervention in each batch, we denote the vector of these therapist \emph{capacities} as $\boldsymbol{n_R^{(C)}}=(n_{R1}^{(C)},\ldots,n_{Rn_T}^{(C)})$. The `R' in the subscript refers to \emph{replication}. We use this to differentiate the letter `C' for centres from the letter `R' for capacities. We allow therapist capacities to vary across therapists in each centre, but assume they are set at the design stage, remain fixed for each therapist across the trial, and remain equal across batches. For centre $C$, we let $n_E^{(C)}$ be the number of embedded therapists and $n_F^{(C)}$ be the number of full-time research therapists, such that $n_T^{(C)}=n_E^{(C)}+n_F^{(C)}$. Until stated otherwise, we assume that therapist capacities only differ based on their status (i.e., whether they are classified as an 'embedded therapist' or a 'research therapist'). We denote the expected capacity of an embedded and a research therapist as $E_E$ and $E_F$, respectively. For simplicity, for a single-centre trial, we omit the superscript $C$ from all notations. Finally, we denote the set of therapists that are available to treat patients in batch $B$ of centre $C$ as $\mathcal{T}_{B,A}^{(C)}= \{1_{B,A}^{(C)},2_{B,A}^{(C)},\ldots,n_{T_{B,A}}^{(C)}\}$. It is anticipated that this will be a subset of $\mathcal{T}^{(C)}$.

\subsection{Existing Randomisation Strategies}

Although one randomisation strategy is unlikely to work well across all therapist-delivered intervention trials, we suggest possible strategies can be classified into three broad types: (i) \emph{single-time}, where all the treatments to be randomised are determined at the start of the trial, (ii) \emph{discrete-time}, where the subset of treatments to be randomised is determined at predefined specific time intervals (e.g., at the start of each batch), and (iii) \emph{continuous-time}, where the subset of treatments to be randomised is determined on a patient-by-patient basis. We assume in each case that each patient will be randomised, and will enter the trial, in continuous time. 

\subsubsection{Strategy E1: Single-Time Equal-Replicate Randomisation of $IT$ Combinations}\label{sec.walwyn.randomisation}

The orthogonal factorial design proposed by Walwyn et al \cite{Walwyn_2026} can be classified as having a one-step, single-time, equal-replicate randomisation strategy in which intervention-therapist combinations are the treatments, and are fully determined at the start of the trial. We focus on the multicentre design \cite{Walwyn_2026}, where each centre has $n_T^{(C)} = n_T$ therapists for centres $C=1,\ldots,n_C$, and the number of treatments is fixed at $n_I n_C n_T$ throughout the trial. Here, it was assumed that the number of patients (units) in every centre-batch combination (block) is identical. The randomisation is then performed so that, independently in every block, every intervention-therapist combination (treatment) for that centre is randomly allocated to $n_R$ patients (units). This ensures that the trial design has equal replication. As a consequence, each therapist will treat $n_I n_R$ patients in every centre-batch combination. Therapists have equal capacities, treating the same number of patients with each intervention, not only overall but in every batch. As such, every therapist is available in every batch, with no turnover. The randomisation strategy suggested in \cite{Walwyn_2026} is equivalent to stratified block randomisation with $n_C n_B$ blocks, each of fixed size $n_I n_T n_R$.

For this paper, we made a small adaptation to the randomisation strategy proposed by \cite{Walwyn_2026}. Instead of requiring an equal number of therapists to be affiliated with every centre, i.e., $n_T^{(C)} = n_T$, we allowed the number of therapists $n_T^{(C)}$ to differ in each centre. As in \cite{Walwyn_2026}, we assumed the number of treatments is fixed throughout the trial, but here at $n_I\sum_{C=1}^{n_C}n_T^{(C)}$ rather than $n_I n_C n_T$. The resulting strategy is equivalent to stratified block randomisation with $n_C n_B$ blocks of variable size $n_I n_T^{(C)} n_R$ for $C=1,\ldots, n_C$. The inclusion of $n_R$ here retains the property of equal replication.

A consequence of adopting Strategy E1 is that not all therapists can operate at their full capacity. Each therapist must treat $n_I n_R$ patients, where $n_R$ is the minimum therapist capacity across centres, $$n_R=min(\boldsymbol{n_R^{(1)}},\boldsymbol{n_R^{(2)}},\ldots,\boldsymbol{n_R^{(n_C)}}).$$ A second consequence, arising from every therapist treating patients in all $n_B$ batches, is that if a therapist is unavailable, randomisation must pause. This has the potential to substantially increase the trial duration and, therefore, the associated running costs. It is just not feasible if there is therapist turnover. As such, Strategy E1 was expected to be advantageous for robustly estimating treatment effects, but perform poorly in terms of feasibility of delivery, resource utilisation and randomisation of patients to time to target.

\subsubsection{Strategy E2: Continuous-Time Equal-Replicate Randomisation of $T$ then $I$} \label{sec.E2.rand}

The orthogonal factorial design used by Shapiro et al \cite{Shapiro_1994} can be classified as having a two-step, continuous-time, equal-replicate randomisation strategy. The trial had only one centre, but if multiple centres had been involved, the strategy could have been performed independently in each of $n_C$ centres. As in Strategy E1, the number of treatments is fixed at $n_I\sum_{C=1}^{n_C}n_T^{(C)}$ throughout the trial, allowing the number of therapists per centre to vary. As a result, the number of patients (units) in each centre (block) also varies. 

The randomisation is performed in two steps. To ensure concealment of intervention allocations, the first step is to randomise a therapist to a patient. If the therapist allocated has capacity and is available, the assignment is confirmed. Otherwise, the assignment is ignored, and randomisation is repeated until a therapist is successfully assigned to the patient. This procedure ensures that each therapist is assigned a total of $n_I n_P$ patients, where $n_P$ denotes the equal number of patients treated by each therapist throughout the trial with an intervention. Since randomisation occurs continuously over time, the notion of a `batch' is no longer relevant. However, to facilitate comparison with the other strategies, we have assumed $n_P=n_B n_R$. In the second step, an intervention is randomised to the patient, and therapists in centres are the blocks. The randomisation of $n_I$ interventions is performed using stratified block randomisation, with $\sum_{C=1}^{n_C}n_T^{(C)}$ blocks of size $n_P$.

Step one of Strategy E2 incorporates variable therapist availability and turnover but not variable therapist capacity. As a result, one advantage of Strategy E2 is that randomisation need no longer pause if a therapist is unavailable. However, to preserve equal replication in the number of patients per therapist, $n_P$ is the minimum therapist capacity across centres. As such, Strategy E2 is expected to be advantageous in robustly estimating treatment effects and in feasibility of delivery, but perform poorly in terms of resource utilisation and randomisation of patients to time and target (i.e., operational efficiency). 

\subsubsection{Strategy E3: Single-Time Equal-Replicate Randomisation of $I$ Only}\label{sec.Non-random}

Currently, the most commonly used design is a parallel-arm design (\cite{Walwyn_2025_prep}) in which interventions are randomised to patients, but therapists are non-randomly allocated to patients. The precise methods adopted for allocating therapists to patients are unclear, as they are rarely (if ever) reported in detail in trial protocols or results papers. Our Strategy E3 was therefore based on what we consider to be a statistically sensible method. Strategy E3 can be classified as a two-step, single-time, equal-replicate randomisation strategy, where the number of treatments is fixed at $n_I$ throughout the trial. Batches, and therapists in centres, are stratification factors, with randomisation performed using stratified block randomisation, using $n_B\sum_{C=1}^{n_C}n_T^{(C)}$ blocks of size $n_I n_R$. Despite the simplicity of the randomisation, therapists must also be assigned to patients. In our experience, this is often done after informing centres of the intervention assigned. However, randomising interventions to patients first can increase covariate imbalance among patients subsequently assigned to therapists, potentially leading to inflated type-I error rates (\cite{Walwyn_2026}). So, for comparison, we assumed that therapists are non-randomly assigned to patients in step one, prior to randomisation of interventions to patients in step two. To mimic non-random allocation of therapists to patients, we assumed that more complex patients are matched to more experienced therapists, using a baseline patient covariate (see \cite{Walwyn_2026} for an initial evaluation of other Strategy E3 methods). Specifically, in step one, for each batch-centre combination, each available therapist $j\in \mathcal{T}_{B,A}^{(C)}$ in centre $C$ is assigned a total of $n_I n_{Rj}^{(C)}$ patients. This serves to spread patients out across therapists in a centre, intuitively to minimise unequal cluster sizes. Then in step two, for each batch-therapist combination, $n_I$ interventions are randomised equally to $n_{Rj}^{(C)}$ patients.

Step one of our Strategy E3 incorporates variable therapist capacity, availability and turnover. As a result, Strategy E3 was expected to perform better than Strategies E1 and E2 in terms of feasibility of delivery and operational efficiency. Strategy E3 was expected to perform less well in terms of robustly estimating treatment effects, however.

\subsection{Novel Randomisation Strategies}

Next we propose three alternative strategies, which all incorporate therapist capacity, availability and turnover, while randomising both interventions and therapists to patients. Each novel strategy aimed to ensure unbiased and more precise estimation of treatment effects than Strategy E3, while also performing better than Strategies E1 and E2 on feasibility of delivery and operational efficiency.

\subsubsection{Strategy N1: Discrete-Time Incomplete Block Randomisation of $T$ then $I$}\label{sec.IBD.rand}

First, an Incomplete Block Design (IBD) (see \cite{Yates_1936} and Supplementary file A) was adapted, to give a novel non-orthogonal factorial design. Strategy N1 is classified as a two-step, discrete-time, incomplete block randomisation of therapists then interventions to patients. Here, the \emph{number} of treatments in each centre $C$, not the precise treatments, is fixed at $n_I n_{T_{min}}^{(C)}$ at the beginning of the trial. Here, $$n_{T_{min}}^{(C)}= \min \big\{ n_{T_{1,A}}^{(C)}, n_{T_{2,A}}^{(C)}, \ldots, n_{T_{n_B,A}}^{(C)} \big\},$$ where $n_{T_{B,A}}^{(C)}$ denotes the \emph{number} of available therapists in each block (batch-centre combination). As such, the individual therapists are allowed to be unknown at the design stage, but the number of therapists per block is fixed. While this is a restriction, it is not as great a restriction as requiring all therapists to be available or known. Under Strategy N1, the specific subset of $n_{T_{min}}^{(C)}$ available therapists selected to be randomised in each block is determined in discrete-time at the start of each block. This permits the set of available therapists in centre $C$, denoted $\mathcal{T}_{B,A}^{(C)}$, to change from one batch to the next, as long as at least $n_{T_{min}}^{(C)}$ are available to be randomised. Under Strategy N1, the number of patients in each block is allowed to vary, and is determined at the start of each block.

Under Strategy N1, independently at the start of each block, $n_{T-IBD}^{(C)} \leq n_{T_{min}}^{(C)}$ therapists are randomly selected from the set $\mathcal{T}_{B,A}^{(C)}$ of available therapists in each block. Note that we denote the set of \emph{selected} therapists for batch $B$ in centre $C$ as $$\mathcal{T}_{B,S}^{(C)} = \big\{1_{B,S}^{(C)}, 2_{B,S}^{(C)}, \ldots, n_{T-IBD,B,S}^{(C)} \big\}.$$ This incorporates the availability and turnover constraints. Randomisation is then performed in two steps. In step one, to ensure full utilisation of therapist capacity across all $n_B$ batches, therapist $j$ is randomised to $n_I n_{Rj}^{(C)}$ patients in each batch, where $n_{Rj}^{(C)}$ is each therapist's maximum capacity, determined at the start of the trial. This incorporates the capacity constraint. Consequently, for each block, the total number of patients randomised to the selected therapists is $$N_{B,S}^{(C)} = \sum_{j \in \mathcal{T}_{B,S}^{(C)}} n_I n_{Rj}^{(C)},$$ with the size of each block depending not only on the number of interventions and the fixed number of therapists, but also on the precise set $\mathcal{T}_{B,S}^{(C)}$ of selected therapists and their respective capacities.

In step one, patients are initially listed as $1,2,\ldots,N_{B,S}^{(C)}$, and for $j \in \mathcal{T}_{B,S}^{(C)}$, each therapist $j$ is linked to $n_I n_{Rj}^{(C)}$ patients in a systematic design. A random permutation of $\{1,\ldots,N_{B,S}^{(C)}\}$ is generated to shuffle the order, followed by renumbering of the list from $1$ to $N_{B,S}^{(C)}$, thereby randomising therapists to patients. In step two, each intervention is randomised to $n_{Rj}^{(C)}$ patients. This is performed using stratified block randomisation, with $n_C n_T$ blocks of variable size $n_I n_{Rj}^{(C)}$.

\subsubsection{Strategy N2: Discrete-Time Flexible Randomisation of $IT$ Combinations}\label{sec.discflex.rand}

Next, a more flexible non-orthogonal factorial design is proposed, relaxing the restriction that only $n_{T-IBD}^{(C)}$ therapists can be selected from those available, which led to equal replication of the number of therapists across blocks within centres. Strategy N2 is classified as an iterative, discrete-time, flexible randomisation of intervention-therapist combinations to patients. Under Strategy N2, the ideal target block size should be around $N/n_C n_B$ to minimise unequal block sizes. Recognising that pragmatic trials may include a mix of `embedded therapists' and `full-time research therapists', under Strategy N2, therapist capacities not only differ across centres, but also depend on the expected capacities $E_E$ and $E_F$ of embedded and full-time therapists, respectively. Consequently, the target block size for batch $B$ and centre $C$ is
\begin{equation}\label{eq.target.sample}
   N_B^{(C)} = \frac{N}{n_B}\frac{E_E n_E^{(C)}+E_F n_F^{(C)}}{\sum_{i=1}^{n_C}E_E n_E^{(i)}+E_F n_F^{(i)}} ,
\end{equation}
where $n_E^{(C)}$ and $n_F^{(C)}$ are number of embedded and full-time therapists in centre $C$. Note that the target block size $N_B^{(C)}$ depends on the collective capacity of all the therapists in centre $C$.

Like Strategy N1, an available therapist in each block treats $n_I n_{Rj}^{(C)}$ patients, if selected. Each selected therapist will also treat an equal number of patients with each intervention in that block. However here, the total number of patients that can be treated in each block if all available therapists work at their full capacity is
$$N_{B,A}^{(C)} = \sum_{j \in \mathcal{T}_{B,A}^{(C)}} n_I n_{Rj}^{(C)}.$$ The size of the block depends on the number of interventions, the variable number of therapists available, and their precise capacities. 
Assuming there are always patients available to be randomised, under Strategy N2, all available therapists can be randomised to $N_{B,A}^{(C)}$ patients in a block, ensuring their full capacity is utilised. Thus, after each time period (e.g., a month) has passed (i.e., at the end of each block):

\begin{itemize}
\item[(1)] Create a systematic design for the current block with the set of available therapists $\mathcal{T}_{B,A}^{(C)}$ and the patients available to be randomised (i.e., the patients recruited in a block) numbered as $1,2,\ldots,N_{B,A}^{(C)}$. Do the following:
\begin{itemize}
    \item[(1.1)] \textbf{Under-recruitment:} If the number of patients recruited in a block is lower than the target number of patients, i.e., $N_{B,A}^{(C)} < N_B^{(C)}$, where the target $N_{B}^C$ is defined in \eqref{eq.target.sample}, use the same systematic design created in step (1) and go to Step (1.4) to proceed with randomisation. Here, the set of selected therapists equals the set of available therapists, i.e., $\mathcal{T}_{B,S}^{(C)} = \mathcal{T}_{B,A}^{(C)}$, and the number of patients randomised equals the number of patients recruited in the block, i.e., $N_{B,S}^{(C)} = N_{B,A}^{(C)}$.
   
    \item[(1.2)] \textbf{Mild over-recruitment:} If the number of patients recruited in a block is over the target defined in \eqref{eq.target.sample} by an acceptable fraction, i.e., $N_B^{(C)} < N_{B,A}^{(C)} < (1+frac) N_B^{(C)}$, use the same systematic design created in step (1) and go to Step (1.4) to proceed with randomisation. Here, $frac\in [0,1]$ is the parameter for the fraction of permissible over-randomisation of patients in each block. As in step (1.1), $\mathcal{T}_{B,S}^{(C)} = \mathcal{T}_{B,A}^{(C)}$, and $N_{B,S}^{(C)} = N_{B,A}^{(C)}$. 
    
    \item[(1.3)] \textbf{Excessive over-recruitment:} If the number of patients recruited in a block is over the target defined in \eqref{eq.target.sample} by an unacceptable fraction, i.e., $N_{B,A}^{(C)} > (1+frac) N_B^{(C)}$, a subset of the available therapists $\mathcal{T}_{B,S}^{(C)} \subseteq \mathcal{T}_{B,A}^{(C)}$ is selected so that the total number of patients assigned to these selected therapists in the systematic design created in step (1) approximately agrees with the target block size. That is, the set of therapists $\mathcal{T}_{B,S}^{(C)}$ and associated number of patients $N_{B,S}^{(C)} = \sum_{j \in \mathcal{T}_{B,S}^{(C)}}n_I n_{Rj}^{(C)}$ are selected such that $$N_{B,S}^{(C)} = \min_{\mathcal{T}_{B,S}^{(C)} \subseteq \mathcal{T}_{B,A}^{(C)}}^{}\left| N_{B,S}^{(C)}-(1+frac)N_B^{(C)} \right|.$$
    Create a new systematic design for the current block with the set of selected therapists $\mathcal{T}_{B,S}^{(C)}$ and the selected patients renumbered as $1,2,\ldots,N_{B,S}^{(C)}$. Go to Step (1.4) to proceed with randomisation.    
    \item[(1.4)] Randomise the selected $n_I |\mathcal{T}_{B,S}^{(C)}|$ intervention-therapist combinations to the selected patients numbered as $1, 2, \ldots , N_{B,S}^{(C)}$, according to each therapist's respective capacities ($n_I n_{Rj}^{(C)}$). The randomisation is performed by randomly permuting the rows of the systematic design and then renumbering the patients from $1, 2, \ldots , N_{B,S}^{(C)}$. 

\end{itemize}
       
    \item[(2)] \textbf{Sufficient total randomisation:} If after randomisation of a block, the total number of patients randomised equals or exceeds the overall target, i.e., $\sum_{C=1}^{n_C} \sum_{B=1}^{n_B} N_{B,S}^{(C)} \geq N$, then stop.
    
    \item[(3)] \textbf{Insufficient total randomisation:} If after randomisation of a block, the overall target number of patients randomised has not been reached, add a new batch indexed by $B + 1$, with a set of available therapists $\mathcal{T}_{B+1,A}^{(C)}$ and go back to Step (1).
\end{itemize}

Unlike Strategies E1, E2 and N1, Strategy N2 allows over-randomisation of patients in a block when additional therapists are available, thereby using any surplus capacity. Such over-randomisation serves as a compensatory mechanism for blocks where randomisation is suboptimal due to limited therapist availability. The permissible threshold is defined as `$frac$' above the target block size $N_B^{(C)}$. The value of this threshold is not an inherent property of the design. It is specified for each trial by the research team.

\subsubsection{Strategy N3: Continuous-Time Flexible Randomisation of $T$ then $I$}

Finally, an alternative more flexible non-orthogonal factorial design is proposed, adapting Strategy E2 to incorporate variable therapist capacity constraints. In contrast to Strategy N2, Strategy N3 can be implemented on a patient-by-patient basis, with no requirement to wait until the end of a block to randomise patients. Strategy N3 is classified as a two-step, continuous-time, flexible randomisation of therapists then interventions to patients. Unlike Strategy E2, where each therapist is randomised an equal number of patients $n_In_P=n_In_Bn_R$, Strategy N3 allocates $n_In_{Pj}^{(C)}=n_I n_B n_{Rj}^{(C)}$ patients according to the full capacity of therapist $j$ in centre $C$.

As with Strategy E2, the randomisation is performed in two steps. In step one, a therapist $j$ is randomised to each patient within a centre as they enter the trial. If the selected therapist has been randomised fewer than $n_In_{Pj}^{(C)}$ patients and is available to accept the patient, the randomisation is confirmed. Otherwise, the therapist is removed from the candidate list, and the procedure is repeated until a therapist is successfully randomised to the patient. In step two, each intervention is randomised equally to patients associated with each therapist using stratified block randomisation, with $\sum_{C=1}^{n_C} n_T^{(C)}$ blocks of different sizes (i.e., $n_{Pj}^{(C)}$ for therapist $j$ in centre `C'). This ensures full utilisation of therapist capacity and availability.

\section{Analysis Models}\label{sec.models}

Two linear regression models are given below, one for trials with a single-time or discrete-time randomisation strategy and one for trials with a continuous-time randomisation strategy. In the first model, time is a discrete factor, represented by batches. In the second model, time is a continuous variable, allowing estimation of therapist learning.

\subsection{Model-1 - Discrete-Time Model}

The first model was proposed by Walwyn et al \cite{Walwyn_2026} for analysing data from the orthogonal multicentre randomised block factorial design using Strategy E1 (Section~\ref{sec.walwyn.randomisation}). A property of this design is that it enables orthogonal estimation of the main effects of all treatment and block effects (Intervention ($I$), Therapist ($T$), Batch ($B$), and Centre ($C$)) and their interactions. The intervention effect is assumed to contribute fixed effects, while all other effects and interactions are assumed to be random. 

Specifically, for a continuous outcome $Y_{ijkl}$, for patient $i$ ($i=1,\ldots,N$), nested in a combination of therapist $j$ and batch $k$, where therapist $j$ is nested in centre $l$ and batch $k$ is crossed with centre $l$, a cross-classified model can be written as:
\begin{align}\label{model.1}
 Y_{ijkl} &= \delta_{0} + \delta_{1}I_{ijkl}+u_{1j}+ u_{2k}+ u_{3jk}+u_{4l}+u_{5kl}+v_{1j}I_{ijkl}\\ \nonumber
    & \qquad +v_{2k}I_{ijkl}+ v_{3jk} I_{ijkl}+v_{4l} I_{ijkl}+v_{5kl} I_{ijkl} + e_{ijkl}, 
\end{align}
where $I_{ijkl}$ is an indicator for the intervention given to patient $i$ nested in therapist $j$, batch $k$, and centre $l$, representing a treatment contrast coded as $1$ and $-1$ for levels 1 and 2 of the intervention. Here, $\delta_{0}$ is the overall mean, and $\delta_{1}$ is the mean intervention effect, defined as the difference between the mean of all outcomes under an intervention and the overall mean. The random effect $u_{1j}$ represents deviation around $\delta_{0}$ for therapist $j$, $u_{2k}$ deviation around $\delta_{0}$ for batch $k$, $u_{3jk}$ deviation around $\delta_{0}+u_{2k}$ for therapist $j$ in batch $k$, $u_{4l}$ deviation around $\delta_{0}$ for centre $l$, and $u_{5kl}$ deviation around $\delta_{0}+u_{2k}$ for centre $l$ in batch $k$. Similarly, $v_{1j}$ represents deviation around $\delta_{1}$ for therapist $j$, $v_{2k}$ deviation around $\delta_{1}$ for batch $k$, $v_{3jk}$ deviation around $\delta_{1}+v_{2k}$ for therapist $j$ in batch $k$, $v_{4l}$ deviation around $\delta_{1}$ for centre $l$, and $v_{5kl}$ deviation around $\delta_{1}+v_{2k}$ for centre $l$ in batch $k$. Finally, $e_{ijkl}$ denotes the patient-level residual error. The random effects are assumed to follow normal distributions: 
$$
u_{1j} \sim \mathcal{N}(0, \sigma_{u_{1}}^2), 
u_{2k} \sim \mathcal{N}(0, \sigma_{u_{2}}^2), 
u_{3jk} \sim \mathcal{N}(0, \sigma_{u_{3}}^2), 
u_{4l} \sim \mathcal{N}(0, \sigma_{u_{4}}^2), 
u_{5kl} \sim \mathcal{N}(0, \sigma_{u_{5}}^2),
$$
$$
v_{1j} \sim \mathcal{N}(0, \sigma_{v_{1}}^2), 
v_{2k} \sim \mathcal{N}(0, \sigma_{v_{2}}^2), 
v_{3jk} \sim \mathcal{N}(0, \sigma_{v_{3}}^2), 
v_{4l} \sim \mathcal{N}(0, \sigma_{v_{4}}^2), 
v_{5kl} \sim \mathcal{N}(0, \sigma_{v_{5}}^2), 
e_{ijkl} \sim \mathcal{N}(0, \sigma_{e}^2).
$$

Here, $\sigma_{u_{1}}^2$ represents the between-therapist variance, $\sigma_{u_{2}}^2$ the between-batch variance, $\sigma_{u_{3}}^2$ the therapist-by-batch variance, $\sigma_{u_{4}}^2$ the between-centre variance, and $\sigma_{u_{5}}^2$ the centre-by-batch variance. Similarly, $\sigma_{v_{1}}^2$ represents the between-therapist variance in the intervention effect, $\sigma_{v_{2}}^2$ the between-batch variance in the intervention effect, $\sigma_{v_{3}}^2$ the therapist-by-batch variance in the intervention effect, $\sigma_{v_{4}}^2$ the between-centre variance in the intervention effect, and $\sigma_{v_{5}}^2$ the centre-by-batch variance in the intervention effect. Finally, $\sigma_{e}^2$ denotes the patient-level residual variance.

\subsection{Model-2 - Continuous-Time Model}

We consider a second model for analysing data from trials with a continuous-time randomisation strategy. Time is here represented by a continuous \emph{time-in-therapist} variable, contributing a fixed effect. This replaces the Batch factor from Model~1, and more directly represents therapist learning. Although this need not be linear, for simplicity, we assume a constant learning rate here. As such, the intervention-by-time interaction also contributes a fixed effect. All other effects and interactions remain random. For a continuous outcome $Y_{ijl}$, for patient $i$ ($i=1,\ldots,N$), nested in therapist $j$, nested in centre $l$, Model~2 can be written as
\begin{align}\label{model.2}
    Y_{ijl} &= \delta_{0} + \delta_{1} I_{ijl} + \delta_2 t_{ijl} + \delta_3 t_{ijl} I_{ijl} + w_{1j} + w_{2j} t_{ijl} + w_{3l} + w_{4l} t_{ijl} \nonumber \\
    &\qquad + z_{1j} I_{ijl} + z_{2j} t_{ijl} I_{ijl} + z_{3l} I_{ijl} + z_{4l} t_{ijl} I_{ijl} + e_{ijl},
\end{align}
where $t_{ijl}$ denotes the number of patients treated by therapist $j$ in centre $l$ prior to patient $i$. Here, $I_{ijl}$ is an indicator for the intervention given to patient $i$ nested within therapist $j$ nested in centre $l$, $\delta_{0}$ is the overall mean, $\delta_{1}$ is the fixed intervention effect, $\delta_2$ is the fixed time effect, and $\delta_3$ is the fixed intervention-by-time interaction effect. The random effect $w_{1j}$ represents deviation around $\delta_{0}$ for therapist $j$, $w_{2j}$ deviation around $\delta_{0}+\delta_2 t_{ijl}$ for therapist $j$ at time $t_{ijl}$, $w_{3l}$ deviation around $\delta_{0}$ for centre $l$, and $w_{4l}$ deviation around $\delta_{0}+\delta_2 t_{ijl}$ for centre $l$ at time $t_{ijl}$. Similarly, $z_{1j}$ represents deviation around $\delta_{1}$ for therapist $j$, $z_{2j}$ deviation around $\delta_{1}+(\delta_2+\delta_3) t_{ijl}$ for therapist $j$ at time $t_{ijl}$, $z_{3l}$ deviation around $\delta_{1}$ for centre $l$, and $z_{4l}$ deviation around $\delta_{1}+(\delta_2+\delta_3) t_{ijl}$ for centre $l$ at time $t_{ijl}$. Finally, $e_{ijl}$ is the patient-level residual error. The random effects are again assumed to follow normal distributions:
$$
w_{1j} \sim \mathcal{N}(0, \sigma_{w_{1}}^2), \quad
w_{2j} \sim \mathcal{N}(0, \sigma_{w_{2}}^2), \quad
w_{3l} \sim \mathcal{N}(0, \sigma_{w_{3}}^2), \quad
w_{4l} \sim \mathcal{N}(0, \sigma_{w_{4}}^2),
$$
$$
z_{1j} \sim \mathcal{N}(0, \sigma_{z_{1}}^2), \quad
z_{2j} \sim \mathcal{N}(0, \sigma_{z_{2}}^2), \quad
z_{3l} \sim \mathcal{N}(0, \sigma_{z_{3}}^2), \quad
z_{4l} \sim \mathcal{N}(0, \sigma_{z_{4}}^2), \quad
e_{ijl} \sim \mathcal{N}(0, \sigma_{e}^2).
$$

In Model~2, $\sigma_{w_{1}}^2$ represents the between-therapist variance, $\sigma_{w_{2}}^2$ the between-therapist variance in the time effect, $\sigma_{w_{3}}^2$ the between-centre variance, and $\sigma_{w_{4}}^2$ the between-centre variance in the time effect. Similarly, $\sigma_{z_{1}}^2$ represents the between-therapist variance in the intervention effect, $\sigma_{z_{2}}^2$ the between-therapist variance in the intervention-by-time interaction effect, $\sigma_{z_{3}}^2$ the between-centre variance in the intervention effect, and $\sigma_{z_{4}}^2$ the between-centre variance in the intervention-by-time interaction effect. Finally, $\sigma_{e}^2$ represents patient-level residual variance.

\section{Simulation Study}\label{sec.simulation.studies}

Next we summarise the methods and results of a simulation study that was designed to explore the six randomisation strategies considered in this paper. Specifically, we investigated how departures from orthogonality arising from variable therapist capacity, availability and turnover affect the performance of these randomisation strategies. Given the current lack of real data from trials implementing the six randomisation strategies, the trial scenarios and model parameter values considered in the simulations are hypothetical. However, the trial scenarios were motivated by the real trial examples presented in Section \ref{sec.motivating.examples} and the authors' prior experience of conducting trials of therapist-delivered interventions. The model parameter values were chosen for illustration purposes, informed by point-estimates reported in the literature, including \cite{Pals_2008}, \cite{Baldwin_2008}, \cite{Herzog_2003}, \cite{Imel_2008}, and \cite{Roberts_2005}, recognising their wide confidence intervals.

We considered three broad trial scenarios representing different trial phases. Scenario 1 corresponds to a single-centre proof of concept trial, where the total number of patients recruited is typically relatively small, often fewer than 200. Scenario 2 represents a medium-sized multi-centre explanatory trial, in which several hundred patients across a few centres may be recruited to evaluate efficacy (i.e., intervention effects with limited interest in generalisability). Scenario 3 represents a large multi-centre pragmatic trial, in which more than 1000 patients are recruited across many centres to evaluate clinical and cost effectiveness, providing generalisable evidence to change standard clinical practice.

\subsection{Simulation Methods}\label{sec.simulation.methods}

The protocol for this simulation study is available as Supplementary Material B. The code used to generate data was developed in \textsf{R} version 4.5.3. The simulated datasets were subsequently analysed using linear mixed-effects models, with parameter estimation performed via residual maximum likelihood (REML) using the \texttt{lmerTest} package (\cite{Kuznetsova_2017}), following \cite{Walwyn_2026}. All the code developed is available at \url{https://github.com/arpan52/TDI-Randomisations}. Note that a covariate with parameter ($\delta_4$) was incorporated during data generation but deliberately omitted during analysis to allow bias introduced in parameter estimation to be quantified, where interventions are randomised to patients but therapist allocation is systematic.

\subsubsection{Data Generated and Model Parameters}\label{sec.data.generating.mechanisms}

We generated data under both Model~1 and Model~2 separately for a single-centre (for trial scenario 1) and a multi-centre trial (for trial scenarios 2 and 3). The parameter values for both the models are given in Table \ref{Table.parameter.scenarios}(b). In all cases, the overall mean $\delta_{0}$ was fixed to be zero, the intervention effect $\delta_{1}$ was fixed to be 0.4, and the between-therapist variance ($\sigma_{u_{1}}^2$ or $\sigma_{w_{1}}^2$) was fixed to be 0.1. In the single-centre trial, the between-therapist variance in the intervention effect ($\sigma_{v_{1}}^2$ or $\sigma_{z_{1}}^2$) was fixed to be 0.15; it was fixed to be 0.1 in the multi-centre trial. In the multi-centre trial, the between-centre variance ($\sigma_{u_{4}}^2$ or $\sigma_{w_{3}}^2$) and the between-centre variance in the intervention effect ($\sigma_{v_{4}}^2$ or $\sigma_{z_{3}}^2$) were both fixed to be 0.1; they were fixed to be zero in the single-centre trial. 

Under Model~1, the between batch variance ($\sigma_{u_{2}}^2$) and the between-batch variance in the intervention effect ($\sigma_{v_{2}}^2$) were both fixed to be 0.1 for both the single- and multi-centre trials. The therapist-by-batch variance ($\sigma_{u_{3}}^2$) and the therapist-by-batch variance in the intervention effect ($\sigma_{v_{3}}^2$) were fixed to be 0.1 in the single- and 0.05 in the multi-centre trial. The centre-by-batch variance ($\sigma_{u_{5}}^2$) and the centre-by-batch variance in the intervention effect ($\sigma_{v_{5}}^2$) were both fixed to be 0.05 in the multi-centre trial and zero in the single-centre trial. 

Under Model~2, the time effect ($\delta_2$) and the intervention-by-time effect ($\delta_3$) were both fixed to be 0.1. The between-therapist variance in the time effect ($\sigma_{w_{2}}^2$) was fixed to be 0.1 in the single- and 0.05 in the multi-centre trial. The between-therapist variance in the intervention-by-time interaction ($\sigma_{z_{2}}^2$) was fixed to be 0.1 in the single- and 0.05 in the multi-centre trial. The between-centre variance in the time effect ($\sigma_{w_{4}}^2$) and the between-centre variance in the intervention-by-time interaction ($\sigma_{z_{4}}^2$) were both fixed to be 0.05 in both the single- and multi-centre trials. 

Finally, under Model~1 and Model~2, the patient-level variance ($\sigma_{e}^2$) was fixed to be 0.35 in the single- and 0.2 in the multi-centre trial to allow each of the the variance components to be interpreted as a proportion of a total variance of 1.0. It is worth noting that existing literature reporting values of variance components has focused solely on the therapist effect. As such, there is greater uncertainty about the other variance components. We expect the relative performance of the randomisation strategies to be robust to moderate changes in the individual variance components.

\begin{table}[ht]
\centering
\caption{Parameter values used in the simulations}
\setlength{\tabcolsep}{3pt}
\renewcommand{\arraystretch}{1.15}
\scriptsize

\textbf{(a) Parameters used in the randomisation strategies}

\vspace{2mm}

\begin{tabular}{c|c|c|c|c|c|c|c|c|c|c|c}
\hline
Scenario & $N$ & $n_C$ & $E(n_T^{(C)})$ & $n_B$ & $a_E$
& $E_E$ & $E_F$ & $\sigma_E^2$ & $\sigma_F^2$ & $a$
& $n_{T\text{-IBD}}^{(C)}$ \\
\hline
Small & 240 & 1 & $12 \& 6$ & $\{5,10\}$
& $\{0,0.25,0.5,0.75,1\}$ & $18 \& 35$ & $24 \& 40$
& 8 & 2 & $\{0.5,0.75,1\}$ & $a n_T^{(C)}$ \\

Medium & 640 & 6 & $8 \& 4$ & $\{5,8\}$
& $\{0,0.25,0.5,0.75,1\}$ & $12 \& 25$ & $30 \& 35$
& 6 & 2 & $\{0.5,0.75,1\}$ & $a n_T^{(C)}$ \\

Large & 1200 & 15 & $8 \& 4$ & $\{5,8\}$
& $\{0,0.25,0.5,0.75,1\}$ & $8 \& 15$ & $15 \& 20$
& 6 & 2 & $\{0.5,0.75,1\}$ & $a n_T^{(C)}$ \\
\hline
\end{tabular}

\vspace{4mm}

\textbf{(b) True parameter values used in Model 1 and Model 2}

\vspace{2mm}

\begin{tabular}{c|c|c|c|c|c|c|c|c|c|c|c|c|c|c|c}
\hline
Model & Scenario & $\delta_0$ & $\delta_1$ & $\delta_4$
& $\sigma_{u_1}^2$ & $\sigma_{u_2}^2$ & $\sigma_{u_3}^2$
& $\sigma_{u_4}^2$ & $\sigma_{u_5}^2$
& $\sigma_{v_1}^2$ & $\sigma_{v_2}^2$ & $\sigma_{v_3}^2$
& $\sigma_{v_4}^2$ & $\sigma_{v_5}^2$ & $\sigma_e^2$ \\
\hline
Model-1 & Small & 0 & 0.4 & 0.3 & 0.1 & 0.1 & 0.1 & 0 & 0 & 0.15 & 0.1 & 0.1 & 0 & 0 & 0.35 \\
& Medium & 0 & 0.4 & 0.3 & 0.1 & 0.1 & 0.05 & 0.1 & 0.05 & 0.1 & 0.1 & 0.05 & 0.1 & 0.05 & 0.2 \\
& Large & 0 & 0.4 & 0.3 & 0.1 & 0.1 & 0.05 & 0.1 & 0.05 & 0.1 & 0.1 & 0.05 & 0.1 & 0.05 & 0.2 \\
\hline
Model-2 & Small & 0 & 0.4 & 0.1 & 0.1 & 0.1 & 0.1 & 0.1 & 0 & 0 & 0.15 & 0.1 & 0 & 0 & 0.35 \\
& Medium & 0 & 0.4 & 0.1 & 0.1 & 0.3 & 0.1 & 0.05 & 0.1 & 0.05 & 0.1 & 0.05 & 0.1 & 0.05 & 0.2 \\
& Large & 0 & 0.4 & 0.1 & 0.1 & 0.3 & 0.1 & 0.05 & 0.1 & 0.05 & 0.1 & 0.05 & 0.1 & 0.05 & 0.2 \\
\hline
\end{tabular}

\vspace{2mm}

\parbox{0.95\textwidth}{%
\footnotesize
\textit{Note:} Table 2(a) corresponds to the parameters used in the randomisation strategies for `Small', `Medium' and `Large' trial scenarios given in Section \ref{sec.simulation.scenarios}. $E(n_T^{(C)})$ denotes expected number of therapists in centre C. A brief definition of the rest of the parameters is listed in Table \ref{Table.notation}. For the parameters $E_E$ and $E_F$, the symbol `\&' indicates that one of the two values is selected conditional on the corresponding value of $E(n_T^{(C)})$. For example, in the single‑centre trial scenario, when $E(n_T^{(C)})=12$, the associated values $E_E=18$ and $E_F=24$ are used. The parameters $a_E$ and $a$ are varied independently over their specified ranges. Table 2(b) corresponds to parameter used in Model-1 and Model-2 given in \eqref{model.1} and \eqref{model.2}, respectively. }\label{Table.parameter.scenarios}
\end{table}

\subsubsection{Designs Generated}\label{sec.simulation.scenarios}

For each of the three trial scenarios, the parameters necessary to specify the trial design are given below and summarised in Table \ref{Table.parameter.scenarios}(a) based on the assumed target size of the trial. \\

\textbf{(i) Small single-centre trial:}
First, for our proof of concept trial, we assumed that a new intervention was to be compared against a control intervention in a single centre. We fixed the target number of patients to be $N=240$. We considered two options for the number of therapists, assuming $n_T$ follows a \textit{Poisson} distribution, i.e., $n_T\sim Poisson(E(n_T))$, where $E(n_T)\in \{6,12\}$. When $E(n_T)=6$, the associated values $E(E)=35$ and $E(F)=40$. When $E(n_T)=12$, the associated values $E(E)=18$ and $E(F)=24$. In both cases, $\sigma_E^2=8$ and $\sigma_F^2=2$. We independently considered five options for the proportion of embedded therapists out of all therapists, using the parameter $a_E\in \{0,0.25,0.5,0.75,1\}$. The remaining proportion $1-a_E$ gives the proportion of full-time therapists. We also independently considered two options for the number of batches, assuming $n_B\in \{5,10\}$, and three options for the minimum proportion of therapists available to treat patients in a batch, parameterised by $a$, assuming $a\in \{0.5,0.75,1\}$. As a result, the fixed number of therapists in each batch for Strategy N1 was calculated to be $n_{T-IBD}=n_T a$. \\

\textbf{(ii) Medium multi-centre trial:} 
Next, for our efficacy trial, we assumed that a new intervention was to be compared against a control intervention in multiple centres. We fixed the target number of patients to be $N = 640$. We assumed that multiple centres would be required to achieve this sample size. We fixed the number of centres to be $n_C = 6$. We assumed the therapists were nested in centres. We also assumed the number of therapists across centres might vary substantially. We considered two options for the number of therapists per centre, assuming $n_T^{(C)}\sim Poisson(E(n_T^{(C)}))$, with $E(n_T^{(C)})\in \{4,8\}$. When $E(n_T)=4$, the associated values $E(E)=25$ and $E(F)=35$. When $E(n_T)=8$, the associated values $E(E)=12$ and $E(F)=30$. In both cases, $\sigma_E^2=6$ and $\sigma_F^2=2$. As before, we considered $a_E\in \{0,0.25,0.5,0.75,1\}$. Here, $n_B\in \{5,8\}$. And finally, $a\in \{0.5,0.75,1\}$, allowing the fixed number of therapists in each centre-batch combination (block) for Strategy N1 to be $n_{T-IBD}^{(C)}=n_T^{(C)} a$. \\

\textbf{(iii) Large multi-centre trial:}
Finally, for our pragmatic trial, we assumed that a promising intervention was to be compared with a control intervention in multiple centres. We fixed the target number of patients to be $N = 1200$ and the number of centres at $n_C = 15$. Similar to our medium multi-centre trial, we assumed $n_T^{(C)}\sim Poisson(E(n_T^{(C)}))$, where $E(n_T^{(C)})\in \{4,8\}$. However, when $E(n_T)=4$, the associated values $E(E)=15$ and $E(F)=20$, and when $E(n_T)=8$, the associated values $E(E)=8$ and $E(F)=15$. Again, $\sigma_E^2=6$ and $\sigma_F^2=2$. As such, the main change was the number of centres. However, the distribution of therapists across centres was also assumed to change to reflect the resulting reduction in the number of patients required per centre. As before, we considered $a_E\in \{0,0.25,0.5,0.75,1\}$, $n_B\in \{5,8\}$ and $a\in \{0.5,0.75,1\}$. Also $n_{T-IBD}^{(C)}=n_T^{(C)} a$. \\

In each trial scenario, we then defined a distribution of total capacities per therapist across the trial. This gave the vector of total capacities for all therapists per centre, denoted $\boldsymbol{c^{(C)}}=(c_{1}^{(C)},\ldots,c_{n_T}^{(C)})$. We then calculated the capacities per centre-batch combination (i.e., the replications per block) as $\boldsymbol{n_R^{(C)}}=(n_{R1}^{(C)},\ldots,n_{Rn_T}^{(C)})$, where $$\boldsymbol{n_R^{(C)}}=\boldsymbol{c^{(C)}}/n_I n_B.$$ The total capacities of embedded and full-time therapists were assumed to follow \textit{truncated normal} distributions with range $(0,\infty)$, i.e., $\bc_E^{(C)} \sim \mathcal{TN}(0,\infty,E_E,\sigma_e^2)$ and $\bc_F^{(C)} \sim \mathcal{TN}(0,\infty,E_F,\sigma_f^2)$. 

\subsubsection{Performance Indicators}\label{sec.performance.measures}

The two performance indicators were the ability of each randomisation strategy to (i) enable patients to be randomised to time and target and (ii) enable robust estimation of the treatment effects (i.e., $\hat{\delta}_{1}$, $\hat{\sigma}_{u_{1}}^2$ and $\hat{\sigma}_{v_{1}}^2$ under Model~1 and $\hat{\delta}_{1}$, $\hat{\sigma}_{w_{1}}^2$ and $\hat{\sigma}_{z_{1}}^2$ under Model~2). Randomisation to time and target was measured by the actual number of batches required to hit the target sample size (i.e., the target number of patients randomised). It is desirable that this is no higher than the number of batches in the original design (i.e, $n_B$). Larger values indicate higher trial costs due to prolonged randomisation. Robust estimation of the treatment effects was indicated by (a) \textit{model convergence}, measured by the proportion of REML fits that failed to converge; (b) \textit{numerical stability}, measured by the proportion of REML fits with boundary estimates defined as zero estimates of variance components whose true values are non-zero; (c) \textit{unbiasedness}, measured by the average estimated values of the treatment effects; (d) \textit{precision}, measured by the average estimated value of the standard error of the intervention effect $SE(\hat{\delta}_1)$; (e) \textit{inferential accuracy}, measured by the Type~II error rate for testing the intervention effect; and (f) \textit{predictive accuracy}, measured by the root mean squared error (RMSE) of prediction. Note that RMSE is computed using predictions based on the fixed effects only, treating the fitted model as population-averaged. Performance indicators are given over 10{,}000 simulation replicates.

\subsection{Simulation Results}

\subsubsection{Randomisation to Time and Target}
We first report the impact of limited therapist capacity and availability on the randomisation of patients to time and target. Table~\ref{Table.time.to.target.single.centre} gives the number of batches actually required to achieve the target sample size under different randomisation strategies for our single-centre trial. As a reminder, in the absence of availability constraints, the target number of batches $n_B \in \{5,10\}$. As the results hold for our medium and large multi-centre trials, their results are provided in Supplementary Files D and E. 

Across all six randomisation strategies, the number of batches required was lowest when all therapists are full-time (i.e., $a_E = 0$), compared to where therapists embedded in services are included ($a_E > 0$). For example, under Strategy E2, when the target number of batches is $n_B = 5$ and all therapists are available in every batch ($a = 1$), the actual number of batches equals $5$ for $a_E = 0$ but increases to $8.41$ when $a_E = 0.25$ (i.e., $25\%$ of embedded therapists). This is a consequence of a reduction in the average collective capacity of the therapists, as we have assumed embedded therapists have less capacity than full-time therapists. Also across all six randomisation strategies, the actual number of batches was lowest when therapist availability ($a$) was higher. For example, under Strategy N1, with $25\%$ embedded therapists ($a_E = 0.25$) and $n_B = 5$, the actual number of batches is $10.73$ when $50\%$ of therapists are available in each batch ($a = 0.5$) but reduces to $8.41$ when all therapists are available ($a = 1$). As therapist availability reduces, waiting times for patients increase, unless the number of therapists is increased.

As expected, differences were apparent across the three existing randomisation strategies. As Strategy E1 requires the availability of all therapists, we only report results for $a=1$. Even where all therapists are available in every batch, Strategy E1 consistently requires the largest actual number of batches, implying that, even under ideal circumstances, Strategy E1 is not an efficient choice. In contrast, Strategy E2 can be implemented even where some therapists are unavailable. Despite this, Strategy E2 requires a similar number of batches to Strategy E1 when $a=1$. This is because Strategies E1 and E2 enforce strict balance in therapist caseloads by using the minimum capacity $c = (c_{1}^{(C)},\ldots,c_{n_T}^{(C)})$ across all therapists (see sections \ref{sec.walwyn.randomisation} and \ref{sec.E2.rand}). This leads to some of the collective therapist capacity not being utilised. Strategy E3 is the most flexible of the existing randomisation strategies, allowing variable therapist capacity and availability. As such, across all of the designs considered, Strategy E3 required the smallest number of batches, and was thus the most efficient.

Differences were also apparent between Strategy N1 and the remaining five randomisation strategies. Strategy N1 was the most sensitive to variable therapist availability of all six strategies. When only half the therapists were available in each batch ($a = 0.5$), Strategy N1 required approximately $65\%$ more batches than Strategy N2. Conversely, when all therapists are available in every batch ($a = 1$), Strategy N1 became the best-performing strategy overall, requiring the smallest number of batches. These differences are explained with the parameter $n_{T\text{-IBD}}^{(C)}$. As we set $n_{T\text{-IBD}}^{(C)}= a n_T^{(C)}$, when only $50\%$ availability is guaranteed ($a=0.5$), only half the therapists can treat patients randomised in each batch ($n_{T\text{-IBD}}^{(C)} = n_T^{(C)}/2$). This substantially increases the actual number of batches required compared to other strategies. Conversely, when full availability is guaranteed ($a = 1$), the number of batches required by Strategy N1 becomes comparable to Strategies N2 and N3, as all therapists can participate in every batch, i.e., $n_{T\text{-IBD}}^{(C)} = n_T^{(C)}$.

Existing randomisation strategies were not universally inferior to Strategy N1. Specifically, although Strategy E2 was inferior in the actual number of batches required overall, it outperformed Strategy N1, and requiring comparable numbers of batches to Strategies E1, E3, N2 and N3, when the number of therapists $n_T = 6$. This is because, for $n_T = 6$, we assumed greater control over the therapists than for $n_T = 12$, with most therapists being full‑time and having approximately similar capacities. So, therapists tended to operate near their maximum capacity, limiting inefficiencies caused by including embedded therapists. This reduced the penalty typically associated with less flexible randomisation strategies, and implies that there will be a tipping point at which Strategy N1 is superior to Strategy E2.

Finally, the flexible randomisation strategies (E3, N2 and N3) consistently required at least $60\%$ fewer actual batches compared to restrictive strategies (E1, E2, and N1) when the target number of batches was low and embedded therapists are present. For example, when $(n_B, a_E, a) = (5, 0.25, 0.5)$, Strategy N2 required $6.20$ batches, whereas Strategy E2 required $10.73$ batches. Conversely, when $(n_B, a_E, a) = (10, 0.25, 0.5)$, Strategy N2 required $12.34$ batches, comparable to Strategy E2 at $12.59$ batches. This occurs because Strategy N2 randomises approximately $2n_{Rj}^{(C)} = c_j^{(C)} / n_B$ patients to therapist $j$ per batch. As $n_B$ increases from $5$ to $10$ with the total therapist capacity remaining fixed at $c_j^{(C)}$, the batch size halves, increasing the actual number of batches required. This increases the recruitment period if the time taken to randomise a batch of patients remains constant.

In summary, across the trial scenarios and designs considered, regardless of the model, Strategies E3, N2 and N3 consistently outperform the other randomisation strategies on this performance indicator, performing comparably to each other. This is expected, as these strategies are specifically designed to assign patients according to therapist capacities to the maximum number of available therapists.

\begin{sidewaystable}
\centering
\caption{Randomisation to time and target against the fraction of therapist availability for the single-centre trial}
\setlength{\tabcolsep}{2pt}  
\begin{tabular}{p{2cm}|c|ccc|ccc|ccc|ccc|ccc}
\hline
&&\multicolumn{15}{c}{Randomisation to Time and Target: Number of batches actually required to meet target} \\
\cline{2-17}
 & $a_E$ & \multicolumn{3}{c|}{$0$}  
  & \multicolumn{3}{c|}{$0.25$} 
 & \multicolumn{3}{c|}{$0.5$} 
 & \multicolumn{3}{c|}{$0.75$} 
 & \multicolumn{3}{c}{$1$} \\
\cline{2-17}
&Design $\downarrow$ a $\rightarrow$ & $0.5$ & $0.75$ & $1$ & $0.5$ & $0.75$ & $1$ & $0.5$ & $0.75$ & $1$ & $0.5$ & $0.75$ & $1$&$0.5$ & $0.75$ & $1$\\
\hline
\multirow{6}{*}{
  \parbox[c]{2cm}{
    \centering
    $N=240$\\
    $n_B=5$\\ 
$n_T = 12$  }}  
& E1        & -&-&5&-&-&8.41&-&-&9.52&-&-&9.85&-&-&9.96\\
 & E2      & 6.79&5.99&5&10.73&9.42&8.41&11.87&10.55&9.52&12.43&10.86&9.85&13.03&11.25&9.96\\
& E3    & 6.18&5.37&4.97&6.20&5.45&5.09&6.38&5.68&5.30&6.71&5.91&5.53&7.30&6.38&5.80\\
\cline{2-17}

 & N1          & 9.45&6.48&4.97&9.90&6.76&5.09&10.28&7.03&5.30&10.70&7.31&5.53&11.26&7.67&5.80\\
 & N2     & 6.18&5.37&4.98&6.20&5.46&5.09&6.38&5.68&5.30&6.71&5.91&5.53&7.30&6.38&5.80\\
 & N3 & 6.24&5.40&4.97&6.30&5.48&5.09&6.55&5.73&5.30&6.96&5.99&5.53&7.49&6.46&5.80\\
\hline 
\multirow{6}{*}{
  \parbox[c]{2cm}{
    \centering
    $N=240$\\
    $n_B=5$\\
$n_T = 6$  }}
& E1        &-&-&5.04&-&-&6.35& -&-&-& -&-&-& -&-&-\\
 & E2      &6.44&5.90&5.04&7.60&6.88&6.35& -&-&-& -&-&-& -&-&-\\
& E3    &6.40&5.87&5.02&6.37&5.79&5.47& -&-&-& -&-&-& -&-&- \\
\cline{2-17}

 & N1          &10.02 &8&5.02&10.54&7.09&5.47& -&-&-& -&-&-& -&-&-\\
 & N2     &6.40&5.87&5.02&6.37&5.79&5.47& -&-&-& -&-&-& -&-&- \\
 & N3 &6.41&5.87&5.02&6.37&5.79&5.47& -&-&-& -&-&-& -&-&-\\
 \hline
 \multirow{6}{*}{
  \parbox[c]{2cm}{
    \centering
    $N=240$\\
    $n_B=10$\\
$n_T = 12$  }}
& E1       & -&-&10 &-&-&10&-&-&10&-&-&10&-&-&10\\
 & E2    & 13.09&11.29&10  &12.59&11.01&10&12.44&11.01&10&12.59&11.01&10&13.09&11.29&10\\
& E3  & 13.06&11.26&10  &12.34&10.77&9.98&12.02&10.6&9.92&11.98&10.45&9.82&12.27&10.65&9.72 \\
\cline{2-17}

 & N1      & 19.96&13.96&10    &19.69&13.73&9.98&19.45&13.51&9.92&19.21&13.29&9.82&18.96&13.09&9.72\\
 & N2   & 13.06&11.26&10  & 12.34&10.77&9.98&12.02&10.6&9.92&11.98&10.46&9.82&12.27&10.65&9.72\\
 & N3& 13.07&11.26&10 &12.35&10.77&9.98&12.02&10.6&9.92&11.99&10.45&9.82&12.39&10.69&9.72\\
\hline  
\multirow{6}{*}{
  \parbox[c]{2cm}{
    \centering
    $N=240$\\
    $n_B=10$\\
$n_T = 6$  }}
& E1 &-&-&10&12.40&12.50&12.44& -&-&-& -&-&-& -&-&- \\
 & E2 &12.42&11.02&10&14.72&13.45&12.44& -&-&-& -&-&-& -&-&-\\
& E3 &12.42&11.02&10&12.24&11.13 &10.24& -&-&-& -&-&-& -&-&- \\
\cline{2-17}

 & N1 &20&15&10&20.59&15.45&10.24 & -&-&-& -&-&-& -&-&-\\
 & N2 &12.42&11.02&10&12.24 &11.13&10.24& -&-&-& -&-&-& -&-&-\\
 & N3 &12.42&11.02&10&12.24&11.13&10.24& -&-&-& -&-&-& -&-&-\\
 \hline
\end{tabular}
\parbox{0.95\textwidth}{%
 \footnotesize
\textit{Note:} Here $N, n_B$ and $n_T$ denote the total sample size, number of batches and number of therapists, respectively.  The proportion of embedded therapists is $a_E$. Parameter $a$ denotes that at least $100a\%$ of the total therapists available in each batch. The expected capacities of embedded and full-time research therapists $(E_E,E_F)$ is fixed to (18,24) or (35,40) for $n_T=12$ or $n_T=6$, respectively. The standard deviation of therapist capacities is fixed to $(\sigma_E,\sigma_F)= (8,2)$. Parameter $frac$ is fixed at $0.25$.  }\label{Table.time.to.target.single.centre}
\end{sidewaystable}

\subsubsection{Robust Estimation of Treatment Effects}\label{sec.comp.of.rand}

We next report the impact of the six randomisation strategies on the statistical properties of the analysis using Model~1 and Model~2. 
We are primarily interested in the estimates of therapist effect and therapist-by-intervention interaction effect, i.e., $\sigma^2_{u_1}$ and $\sigma^2_{v_1}$ for model-1 and $\sigma^2_{w_1}$ and $\sigma^2_{z_1}$ for model-2. The estimates of the remaining parameters are listed in supplementary files F, G and H provided in the GitHub repository \url{https://github.com/arpan52/TDI-Randomisations}. 

Results of the parameters of interest are summarised in Tables \ref{Table.estimation.model.1.single.1} and \ref{Table.estimation.model.2.single.1}, respectively, for the single-centre trial scenario where the target number of batches $n_B = 5$, varying the proportion of embedded therapists $a_E$ and therapists available $a$ (see also Supplementary File C). Results for our multi-centre trial scenarios are given Supplementary Files D and E. \\

\begin{sidewaystable}[ht]
\centering
\caption{Estimation results for the single centre trial under Model-1}
\setlength{\tabcolsep}{3pt}  
    \begin{tabular}{ p{2cm}| p{2cm}| c| c  c c  c| c  c c| c c| c c| c c c}
  \hline
&& Design&$\hat{\delta_1}$& $SE(\hat{\delta_1})$ & $\hat{\sigma}_{u_1}^2$ & $\hat{\sigma}_{v_1}^2$ &T-2 Error & Boundary& Fail& RMSE\\
\hline
&True Values&&0.4& -& 0.1& 0.15& -&-&-&-\\
\hline
\multirow{36}{*}{  \parbox[c]{2cm}{
    \centering
$N=240$,\\
$n_T = 12$
  }}
  & \multirow{6}{*}{
  \parbox[c]{2cm}{
    \centering
    $n_B=5$\\
    $a_E=0$\\
    $a=0.5$
  }}
  & E1 & - & - &- & - & - & - & -& -  \\
&& E2 & 0.399 & 0.121 & 0.11 & 0.15 & 0.17 & 0.36 & 0.01 &0.97\\

&& E3 & 0.399 & 0.141 & 1.1 & 0.15 & 0.33 & 0.75 & 0.01&1.68 \\
\cline{3-11}

&& N1 & 0.398 & 0.115 & 0.11 & 0.15 & 0.12 & 0.37 & 0.01&0.97 \\

&& N2 & 0.4 & 0.124 & 0.11 & 0.14 & 0.19 & 0.37 & 0.01&0.97 \\
&& N3 & 0.398 & 0.123 & 0.11 & 0.14 & 0.19 & 0.38 & 0.01&0.97 \\
\cline{2-11}
& \multirow{6}{*}{
  \parbox[c]{2cm}{
    \centering
    $n_B=5$\\
    $a_E=0$\\
    $a=0.75$
 }}    & E1 & - & - &- & - & - & - & -& -  \\
    && E2 & 0.399 & 0.123 & 0.11 & 0.14 & 0.19 & 0.36 & 0.01 & 0.97 \\
    && E3 & 0.399 & 0.144 & 1.09 & 0.15 & 0.36 & 0.74 & 0.01 & 1.68 \\
    \cline{3-11}

  && N1 & 0.4 & 0.122 & 0.11 & 0.14 & 0.17 & 0.37 & 0.01 & 0.97 \\
&& N2 & 0.4 & 0.127 & 0.11 & 0.15 & 0.23 & 0.39 & 0.01 & 0.96 \\
&& N3 & 0.399 & 0.126 & 0.11 & 0.14 & 0.23 & 0.38 & 0.01 & 0.96 \\
     \cline{2-11}
& \multirow{6}{*}{   
  \parbox[c]{2cm}{
    \centering
    $n_B=5$\\
    $a_E=0$\\
    $a=1$
  }}& E1 & 0.399 & 0.129 & 0.11 & 0.15 & 0.25 & 0.41 & 0.01 & 0.96 \\
  && E2 & 0.398 & 0.129 & 0.11 & 0.14 & 0.26 & 0.41 & 0.01 & 0.96 \\

&& E3 & 0.401 & 0.145 & 1.1 & 0.16 & 0.37 & 0.74 & 0.01 & 1.68 \\
\cline{3-11}

&& N1 & 0.402 & 0.128 & 0.11 & 0.14 & 0.24 & 0.38 & 0.01 & 0.96 \\
&& N2 & 0.401 & 0.129 & 0.11 & 0.14 & 0.25 & 0.38 & 0.01 & 0.96 \\
&& N3 & 0.4 & 0.129 & 0.11 & 0.14 & 0.26 & 0.38 & 0.01 & 0.96 \\
     \cline{2-11}  
     & \multirow{6}{*}{
  \parbox[c]{2cm}{
    \centering
    $n_B=5$\\
    $a_E=0.25$\\
    $a=0.5$
  }}
  & E1 & - & - &- & - & - & - & - & - \\
&& E2 & 0.398 & 0.122 & 0.11 & 0.14 & 0.18 & 0.36 & 0.7&0.97  \\

 && E3 & 0.398 & 0.141 & 1.09 & 0.15 & 0.33 & 0.75 & 0.01  &1.68\\
 \cline{3-11}

&& N1 & 0.402 & 0.114 & 0.11 & 0.14 & 0.11 & 0.36 & 0.01&0.97  \\
&& N2 & 0.396 & 0.123 & 0.11 & 0.14 & 0.2 & 0.37 & 0.01  &0.97\\
&& N3 & 0.403 & 0.123 & 0.11 & 0.14 & 0.18 & 0.37 & 0.01 &0.97 \\
\cline{2-11}
& \multirow{6}{*}{
  \parbox[c]{2cm}{
    \centering
    $n_B=5$\\
    $a_E=0.25$\\
    $a=0.75$
  }}& E1 & - & - &- & - & - & - & -& -  \\
  && E2 & 0.401 & 0.123 & 0.11 & 0.14 & 0.19 & 0.35 & 0.69&0.97 \\

&& E3 & 0.402 & 0.143& 1.09& 0.15& 0.35& 0.73& 0.02&1.68 \\
\cline{3-11}

&& N1 & 0.401 & 0.121 & 0.11 & 0.14 & 0.17 & 0.37 & 0.01&0.97 \\

&& N2 & 0.401 & 0.126 & 0.11 & 0.14 & 0.21 & 0.38 & 0.01 &0.96\\
&& N3 & 0.4 & 0.126 & 0.11 & 0.14 & 0.22 & 0.38 & 0.01&0.96 \\
     \cline{2-11}
& \multirow{6}{*}{
  \parbox[c]{2cm}{
    \centering
    $n_B=5$\\
    $a_E=0.25$\\
    $a=1$
  }}& E1 & 0.395 & 0.129 & 0.11 & 0.14 & 0.25 & 0.41 & 0.68&0.96 \\
  && E2 & 0.398 & 0.129 & 0.11 & 0.14 & 0.25 & 0.41 & 0.69&0.97 \\

&& E3 & 0.399 & 0.144& 1.11& 0.15&  0.35& 0.75& 0.01&1.68 \\
\cline{3-11}

&& N1 & 0.4 & 0.128 & 0.11 & 0.14 & 0.24 & 0.38 & 0.01&0.96 \\

&& N2 & 0.4 & 0.128 & 0.11 & 0.14 & 0.25 & 0.39 & 0.01&0.96 \\
&& N3 & 0.403 & 0.128 & 0.11 & 0.14 & 0.24 & 0.39 & 0.01&0.96 \\
\hline
\end{tabular}
\parbox{0.95\textwidth}{%
 \footnotesize
\textit{Note:} Here $N, n_B$ and $n_T$ denote the total sample size, number of batches and number of therapists, respectively. The proportion of embedded therapists is $a_E$. Parameter $a$ denotes that at least $100a\%$ of the total therapists available in each batch. The true values of the model-1 parameters are also given. $\hat{\delta_1}$ and $SE(\hat{\delta_1})$ denote the average estimate and standard error of the intervention effect $\delta_1$. $\hat{\sigma}_{u_1}^2$ and $\hat{\sigma}_{v_1}^2$, T-2 Error, Boundary, Fail and RMSE denote the average estimates of $\sigma_{u_1}^2$, $\sigma_{v_1}^2$, type-II error rate, boundary estimate rate, REML failure rate and root mean square error, respectively. The value of the expected capacities $(E_E,E_F)$ is fixed to (18,24) or (35,40) for $n_T=12$ or $n_T=6$, respectively. The standard deviation of therapist capacities is fixed to $(\sigma_E,\sigma_F)= (8,2)$. Parameter $frac$ is fixed at $0.25$.  }\label{Table.estimation.model.1.single.1}
\end{sidewaystable}

\clearpage

\noindent \textbf{(1) Model~1}

\paragraph{(a) Model Convergence:} 
The primary criterion for evaluating a design, of which the randomisation strategy is part, is whether it reliably produces sufficient data to enable the intended analysis model to be fitted. For a model to be estimable, each factor level must have at least two replications; failure to achieve this leads to incomplete data for estimating the effect associated with that factor.

In Table~\ref{Table.estimation.model.1.single.1}, when $a_E = 0$ (all therapists full-time), the failure rate is negligible ($\approx 0.01$) across all randomisation strategies. However, as the proportion of embedded therapists increases, replication becomes a severe problem for rigid Strategies E1 and E2. For example, when $a_E = 0.25$ ($25\%$ embedded therapists), the failure rate for Strategies E1 and E2 jumps to $70\%$. In Table C.2, where all therapists are embedded ($a_E = 1$), Strategies E1 and E2 fail in nearly all simulations due to the increase in batches required. The small $c$ when embedded therapists are included (see Sections~\ref{sec.walwyn.randomisation} and~\ref{sec.E2.rand}) means each therapist is randomised approximately $c/(2n_B)\approx 1$ patient per intervention per batch. This lack of replication at the intervention-by-therapist-by-batch level leads to frequent REML fitting failures, particularly when estimating variance $\sigma^2_{v_3}$. Strategy E3 and novel strategies (N1, N2 and N3) mitigate these issues, provided $n_B$ is not excessively large, by utilising the full capacity of available therapists. Exclusively using full‑time therapists also alleviates convergence problems, as illustrated in Table C.3, where for $n_T = 6$ fitting failure rates are negligible.

If the target number of batches $n_B$ is large, $c_j/(2n_B) \approx 1$ for all therapists $j = 1,\ldots,n_T$. Here, all strategies randomising therapists to patients exhibit high failure rates (often exceeding $90\%$; see Tables C.4-C.5). The exception is Strategy E3, which maintains low failure rates even for large $n_B$. This robustness stems from its two-step randomisation. This introduces imbalance of patients in intervention-therapist combinations in a batch, ensuring that some combinations receive multiple patients while others receive none. Although this imbalance provides necessary replication for model fitting, it should be used cautiously as it also inflates Type~I error rates, as shown in \cite{Walwyn_2026}.

Further problems arise across all six randomisation strategies when there is insufficient replication of therapists in batches and/or centres. When, in a batch and/or centre, only a single therapist is available to treat patients, the model’s ability to estimate the therapist-by-batch interaction ($\sigma_{u_3}^2$) and/or intervention-by-therapist-by-batch interaction ($\sigma_{v_3}^2$) is limited. This problem did not occur in our single-centre trial, where we assumed at least $50\%$ therapist availability ($a \geq 0.5$). It was common in our multi-centre trials, particularly when the number of therapists per centre was small (e.g., in Table E.4, where $N = 1200$, $n_C=15$ and $E(n_T^{(C)})=4$). This problem was unrelated to therapist capacity.

\paragraph{(b) Numerical Stability:}

Boundary estimates signal weak identifiability and likelihood degeneracy. They affect inference because in packages such as \texttt{lmerTest} in R, an analysis of the fixed effects is equivalent to an analysis based on the assumption that the relevant variance components are known to be zero (see \cite{Gilmour_2009}).

Table~\ref{Table.estimation.model.1.single.1} shows that Strategy E3 yields an extremely high proportion of boundary estimates ($\approx 75\%$), indicating its unreliability in providing estimates of variance components. In contrast, the remaining randomisation strategies give a lower but notable proportion of boundary estimates ($\approx 35\%$). Table C.3 and C.4 show that boundary estimates increase by more than $10\%$ overall when the number of therapists decreases from $n_T = 12$ to $n_T = 6$, reflecting the reduced information available for estimating the between-therapist variance (see also Table C.17). For example, when $(n_T,a_E,a) = (12,0,0.75)$, $16\%$ of between-therapist variance estimates under Strategy N1 are boundary estimates, whereas for $(n_T,a_E,a) = (6,0,0.75)$, the proportion rises to $27\%$. Boundary estimates arise from an insufficient number of levels for a given random factor, resulting in zero or near-zero degrees of freedom for estimating the corresponding variance components. Notably, Strategy E3 yields high proportions of boundary estimates across all variance components except the between-therapist variance, while strategies randomising therapists to patients yield boundary estimates primarily for the between-therapist and between-batch variances.

\paragraph{(c) Unbiasedness:} 

Across all trial scenarios, designs, and randomisation strategies considered, the intervention effect and intervention-by-therapist interaction are estimated without bias: $\hat{\delta}_1 \approx 0.4$ and $\hat{\sigma}_{v_1}^2 \approx 0.15$. The between-therapist variance is also unbiasedly estimated ($\hat{\sigma}_{u_1}^2 \approx 0.1$), provided therapists are randomised to patients, except when the number of therapists is small (e.g., $n_T=6$ in Table C.3). As expected, Strategy E3 leads to severely biased estimates of the between-therapist variance $\sigma_{u_1}^2$ (see Walwyn et al. \cite{Walwyn_2026} for further details).

\paragraph{(d) Precision:}
Across all $a_E$ in Table~\ref{Table.estimation.model.1.single.1} (and Table C.2), Strategy N1 consistently yielded the smallest $SE(\hat{\delta}_1)$ when therapist availability was limited ($a < 1$). For example, in Table~\ref{Table.estimation.model.1.single.1}, with $(a_E,a) = (0,0.5)$, $SE(\hat{\delta}_1) = 0.115$ under Strategy N1, compared to $SE(\hat{\delta}_1) \geq 0.121$ for the remaining strategies. As availability improves ($a \geq 0.75$), differences in $SE(\hat{\delta}_1)$ across strategies randomising therapists to patients became negligible. This aligns with the results in Table~\ref{Table.time.to.target.single.centre}. Under limited availability ($a=0.5$), Strategy N1 requires more batches, improving estimation of $\sigma_{u_2}^2$ (see Tables F.1-F.3). This marginal gain reduces $SE(\hat{\delta}_1)$. When availability is high ($a>0.5$), Strategy N1 requires a similar number of batches as other strategies, eliminating its precision advantage.

In contrast, across trial scenarios and designs, Strategy E3 consistently produced the largest $SE(\hat{\delta}_1)$ as a bi-product of the inflated between-therapist variance. Flexible Strategies N2 and N3 yielded precision equivalent to the rigid Strategy E2. When $(a_E, a) = (0, 0.5)$, $SE(\hat{\delta}_1) = 0.124$ for Strategy N2 with $n_T = 12$ (Table~\ref{Table.estimation.model.1.single.1}), whereas $SE(\hat{\delta}_1) = 0.144$ for $n_T = 6$ (see Table C.3). This suggests a trade-off between operational efficiency and statistical precision when reducing the number of therapists.

\paragraph{(e) Inferential Accuracy:}
The Type~II error associated with the significance of $\delta_1$ follows the same pattern as $SE(\hat{\delta}_1)$. From Table~\ref{Table.estimation.model.1.single.1} (and Tables C.2-C.3), Strategy E3 consistently produces the highest Type~II error across all trial scenarios and designs considered. In contrast, similar to (d), Strategy N1 yielded the lowest Type~II error when therapist availability is limited, performing similarly to other strategies that randomise therapists to patients otherwise. 

Type~II error inflated significantly for rigid Strategies E1 and E2 when a higher proportion of embedded therapists were included. This occurs due to high REML failures. For instance, in Table~\ref{Table.estimation.model.1.single.1}, Strategy E2 for $(a_E,a)=(0,0.5)$ had a Type~II error of $0.17$ with $99\%$ successful REML fits. However, in Table C.2, Strategy E2 for $(a_E,a)=(0.75,0.75)$ had a Type~II error of $0.33$, equivalent to Strategy E3, because REML fits fail in $97\%$ of simulations. 

As expected, across all six randomisation strategies, Type~II error increased if fewer therapists ($n_T=6$) compared to more therapists ($n_T=12$) were included (see Table~\ref{Table.estimation.model.1.single.1}, and Table C.3). 

\paragraph{(f) Predictive Accuracy:}

Across Table~\ref{Table.estimation.model.1.single.1} and Tables C.2-C.3, RMSE under Strategy E3 is substantially larger than under the remaining strategies. This reduction in performance is consistent with the severe bias observed in estimation of the between-therapist variance $\sigma_{u_1}^2$, which, if implicitly treated as correct during model fitting, degrades marginal predictions. In contrast, strategies randomising therapists to patients yield RMSE values that are practically indistinguishable across designs varying $a_E$, $a$ and $n_T$.\\

In summary, replication in the design is crucial for model convergence. This replication can be improved by greater control over therapist capacity and availability and planning the number of batches carefully. The number of therapists and batches included affect numerical stability. Randomising therapists to patients will avoid inflating the between-therapist variance, which will have knock on effects on the standard error and Type~II error for estimating ${\delta}_1$ and on predictive accuracy. \\ 

\noindent \textbf{(2) Model~2}

\paragraph{(a) Model Convergence:}  
In contrast to Model~1, all randomisation strategies yielded sufficient data to estimate all the parameters in Model~2 regardless of $a_E$, $a$, or $n_B$. Table~\ref{Table.estimation.model.2.single.1} (and Tables C.8-C.9) show that the failure rate does not exceed $5\%$ for Strategy E3 and remained below $3\%$ for all other strategies. Using continuous variable $t_{ijl}$ in Model~2 eliminated the problem of insufficient replication at the intervention-by-therapist-by-batch level. However, analogous to model-1, therapist replication in centres became problematic in large multi-centre trials because the intercept ($\sigma^2_{w_1}$) and slope ($\sigma^2_{w_2}$) for therapist learning effects must be estimated from few therapists (see Table E.10). 

\paragraph{(b) Numerical Stability:}  
Across all designs considered, Model~2 yielded substantially fewer boundary estimates compared to Model~1. A comparison of Table \ref{Table.estimation.model.2.single.1} and Table C.9 shows that boundary estimates increased by more than $17\%$ when the number of therapists decreased from $n_T = 12$ to $n_T = 6$. Most problems observed with boundary estimates in Model~2 arose from correlated estimates produced by REML for the intercept ($\sigma^2_{w_1}$) and slope ($\sigma^2_{w_2}$) therapist learning effects (see Table C.18). More than $25\%$ of simulations had correlated estimates across all randomisation strategies.

\paragraph{(c) Unbiasedness:} 
Performance was the same whether Model~1 or Model~2 was fitted. 

\paragraph{(d) Precision:} 
Performance was similar whether Model~1 or Model~2 was fitted. The only notable difference was that Strategy N1 did not dominate the other strategies randomising therapists to patients; rather, its performance was comparable to Strategies E1, E2, N2 and N3, even when therapist availability was limited.

\paragraph{(e) Inferential Accuracy:} 
Consistent with Model~1, Strategy E3 yielded the highest Type~II error rates. However, in contrast to Model~1, Strategy N1 consistently yielded Type~II error rates slightly higher than other strategies randomising therapists to patients. Among all strategies, Strategy E2 exhibited the lowest Type~II error across designs, although its advantage over Strategies N2 and N3 was minimal. Consistent with Model~1, reducing the number of therapists increased the frequency of Type~II errors (see Table~\ref{Table.estimation.model.2.single.1} and Table C.9). 

\paragraph{(f) Predictive Accuracy:}  
In contrast to Model~1, under Model~2 the randomisation strategies yielded clearly distinct RMSE values, reflecting differences in predictive accuracy arising from the underlying randomisation strategies and design parameters.\\

In summary, replication was less problematic, except for therapists in centres. The cause of numerical instability was correlated estimates produced by REML for the therapist learning effects. Strategies randomising therapists to patients were comparable in terms of unbiasedness, precision and inferential accuracy but not predictive accuracy.

\begin{sidewaystable}
\centering
\caption{Estimation results for the single centre trial for model-2}
 \setlength{\tabcolsep}{3pt}  
   \begin{tabular}{ p{2cm}| p{2cm}| c| c  c c  c| c  c c| c c| c c| c c c}
  \hline
&& Design&$\hat{\delta_1}$& $SE(\hat{\delta_1})$ & $\hat{\sigma}_{w_1}^2$ & $\hat{\sigma}_{z_1}^2$ &T-2 Error & Boundary& Fail&RMSE\\
\hline
&True Values&&0.4& -& 0.1& 0.15& -&-&-&-\\
\hline
\multirow{36}{*}{  \parbox[c]{2cm}{
    \centering
$N=240$,\\
$n_T = 12$
  }}& \multirow{6}{*}{
  \parbox[c]{2cm}{
    \centering
    $n_B=5$\\
    $a_E=0$\\
    $a=0.5$
  }}& E1 & - & - & - & - & - & - &-&- \\
&& E2 & 0.401 & 0.086 & 0.11 & 0.14 & 0.01 & 0.28 & 0.03 & 2.95 \\

&& E3 & 0.4 & 0.115 & 1.1 & 0.16 & 0.13 & 0.21 & 0.05 & 3.54 \\
\cline{3-11}
&& N1 & 0.4 & 0.09 & 0.11 & 0.14 & 0.03 & 0.3 & 0.03 & 3.53 \\
&& N2 & 0.4 & 0.086 & 0.11 & 0.14 & 0.01 & 0.28 & 0.03 & 3.06 \\
&& N3 & 0.4 & 0.086 & 0.1 & 0.14 & 0.02 & 0.28 & 0.03 & 3.1 \\ 
\cline{2-11}
& \multirow{6}{*}{
  \parbox[c]{2cm}{
    \centering
    $n_B=5$\\
    $a_E=0$\\
    $a=0.75$
  }}& E1 & - & - & - & - & - & - &-&- \\
  && E2 & 0.401 & 0.086 & 0.11 & 0.14 & 0.01 & 0.28 & 0.03 & 2.95 \\
&& E3 & 0.4 & 0.115 & 1.1 & 0.16 & 0.13 & 0.21 & 0.05 & 3.54 \\
\cline{3-11}

&& N1 & 0.4 & 0.09 & 0.11 & 0.14 & 0.03 & 0.3 & 0.03 & 3.53 \\
&& N2 & 0.4 & 0.086 & 0.11 & 0.14 & 0.01 & 0.28 & 0.03 & 3.06 \\
&& N3 & 0.4 & 0.086 & 0.1 & 0.14 & 0.02 & 0.28 & 0.03 & 3.1 \\
\cline{2-11}
& \multirow{6}{*}{
  \parbox[c]{2cm}{
    \centering
    $n_B=5$\\
    $a_E=0$\\
    $a=1$
  }}& E1 & 0.402 & 0.085 & 0.11 & 0.15 & 0.01 & 0.27 & 0.03 & 2.55 \\
  && E2 & 0.399 & 0.085 & 0.11 & 0.14 & 0.01 & 0.28 & 0.03 & 2.55 \\
&& E3 & 0.399 & 0.111 & 1.1 & 0.16 & 0.1 & 0.19 & 0.05 & 3.36 \\
\cline{3-11}

&& N1 & 0.4 & 0.085 & 0.1 & 0.15 & 0.01 & 0.27 & 0.03 & 3.04 \\
&& N2 & 0.401 & 0.085 & 0.1 & 0.15 & 0.01 & 0.28 & 0.03 & 2.91 \\
&& N3 & 0.401 & 0.085 & 0.11 & 0.15 & 0.01 & 0.27 & 0.04 & 3.06 \\
     \cline{2-11}     
& \multirow{6}{*}{
  \parbox[c]{2cm}{
    \centering
    $n_B=5$\\
    $a_E=0.25$\\
    $a=0.5$
  }}
  & E1 & - & - & - & - & - & - &-&- \\
  && E2 & 0.4 & 0.085 & 0.1 & 0.15 & 0.01 & 0.28 & 0.03 & 2.81 \\
&& E3 & 0.4 & 0.117 & 1.09 & 0.16 & 0.15 & 0.23 & 0.06 & 3.63 \\
\cline{3-11}

&& N1 & 0.4 & 0.091 & 0.11 & 0.14 & 0.03 & 0.31 & 0.04 & 3.64 \\
&& N2 & 0.401 & 0.086 & 0.11 & 0.14 & 0.02 & 0.27 & 0.03 & 3.09 \\
&& N3 & 0.4 & 0.087 & 0.11 & 0.15 & 0.02 & 0.29 & 0.03 & 3.2 \\ 
\cline{2-11}
& \multirow{6}{*}{
  \parbox[c]{2cm}{
    \centering
    $n_B=5$\\
    $a_E=0.25$\\
    $a=0.75$
  }}& E1 & - & - & - & - & - & - &-&- \\
  && E2 & 0.399 & 0.085 & 0.1 & 0.15 & 0.01 & 0.27 & 0.03 & 2.77 \\

&& E3 & 0.4 & 0.114 & 1.1 & 0.16 & 0.13 & 0.21 & 0.05 & 3.53 \\
\cline{3-11}

&& N1 & 0.4 & 0.089 & 0.11 & 0.14 & 0.02 & 0.3 & 0.03 & 3.44 \\

&& N2 & 0.398 & 0.085 & 0.11 & 0.14 & 0.02 & 0.28 & 0.03 & 3 \\
&& N3 & 0.4 & 0.086 & 0.11 & 0.14 & 0.02 & 0.28 & 0.03 & 3.17 \\
\cline{2-11}
& \multirow{6}{*}{
  \parbox[c]{2cm}{
    \centering
    $n_B=5$\\
    $a_E=0.25$\\
    $a=1$
  }}& E1 & 0.4 & 0.085 & 0.1 & 0.15 & 0.01 & 0.27 & 0.03 & 2.55 \\
  && E2 & 0.402 & 0.084 & 0.11 & 0.14 & 0.01 & 0.27 & 0.03 & 2.55 \\

&& E3 & 0.401 & 0.113 & 1.1 & 0.16 & 0.12 & 0.21 & 0.05 & 3.44 \\
\cline{3-11}

&& N1 & 0.4 & 0.086 & 0.11 & 0.14 & 0.02 & 0.29 & 0.03 & 3.13 \\

&& N2 & 0.399 & 0.086 & 0.1 & 0.15 & 0.01 & 0.28 & 0.03 & 2.96 \\
&& N3 & 0.399 & 0.086 & 0.11 & 0.14 & 0.01 & 0.29 & 0.03 & 3.14 \\
 \hline
\end{tabular}
\parbox{0.95\textwidth}{%
 \footnotesize
\textit{Note:} Here $N, n_B$ and $n_T$ denote the total sample size, number of batches and number of therapists, respectively.  The proportion of embedded therapists is $a_E$. Parameter $a$ denotes that at least $100a\%$ of the total therapists available in each batch. The true values of the model-1 parameters are also given. $\hat{\delta_1}$ and $SE(\hat{\delta_1})$ denote the average estimate and standard error of the intervention effect $\delta_1$. $\hat{\sigma}_{w_1}^2$ and $\hat{\sigma}_{z_1}^2$, T-2 Error, Boundary, Fail and RMSE denote the average estimates of $\sigma_{w_1}^2$, $\sigma_{z_1}^2$, type-II error rate, boundary estimate rate, REML failure rate and root mean square error, respectively. The value of the expected capacities $(E_E,E_F)$ is fixed to (18,24) or (35,40) for $n_T=12$ or $n_T=6$, respectively. The standard deviation of therapist capacities is fixed to $(\sigma_E,\sigma_F)= (8,2)$. Parameter $frac$ is fixed at $0.25$.}\label{Table.estimation.model.2.single.1}
\end{sidewaystable}
\clearpage

\subsubsection{Trade-Offs in Performance}\label{sec.tradeoffs.rand}

\begin{sidewaysfigure}[ht]
\caption{Power and randomisation time to target against percentage of therapists available}
\subfigure[$(n_T,a_E) = (12,0)$]{
  \centering
  \includegraphics[width=0.3\linewidth, height=0.2\textheight]{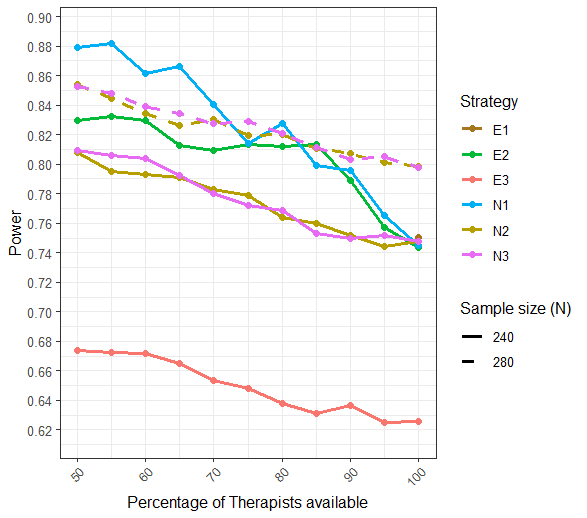}  
}
\subfigure[$(n_T,a_E) = (12,0.5)$]{
  \centering
  \includegraphics[width=0.3\linewidth, height=0.2\textheight]{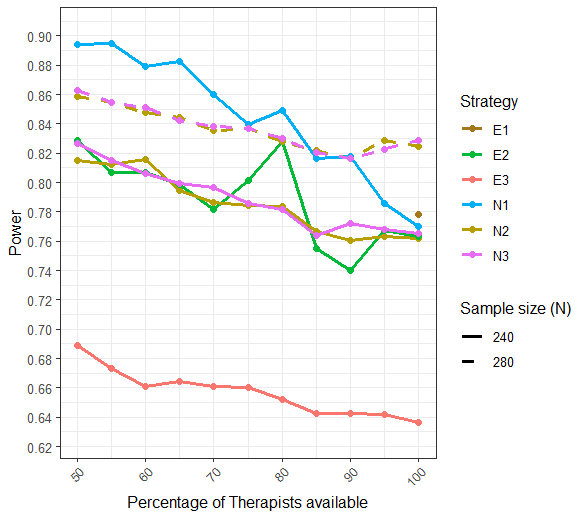}  
}
\subfigure[$(n_T,a_E) = (12,1)$]{
  \centering
  \includegraphics[width=0.3\linewidth, height=0.2\textheight]{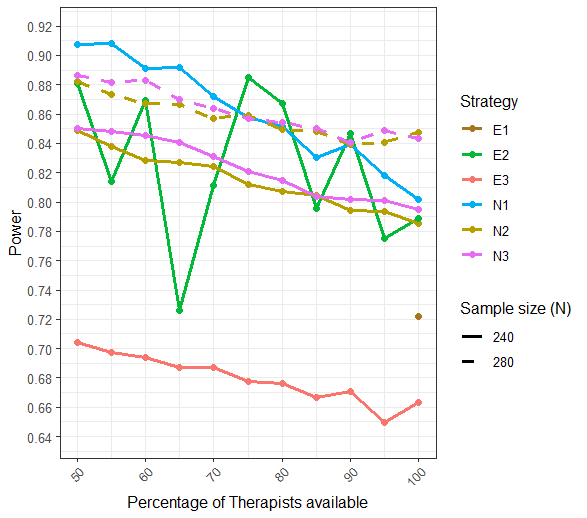}  
}
\subfigure[$(n_T,a_E) = (12,0)$]{
  \centering
  \includegraphics[width=0.3\linewidth, height=0.2\textheight]{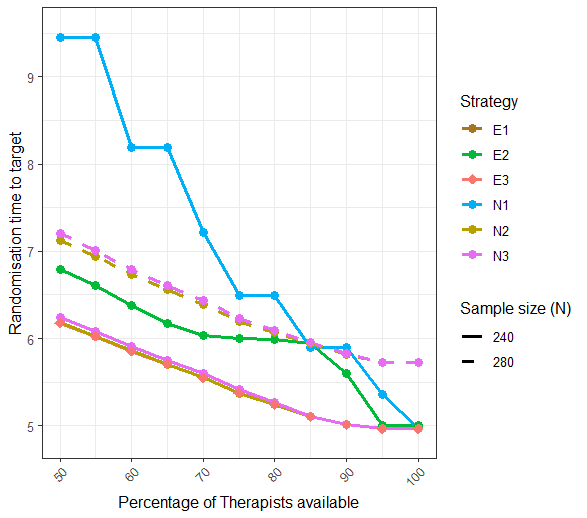}  
}
\subfigure[$(n_T,a_E) = (12,0.5)$]{
  \centering
  \includegraphics[width=0.3\linewidth, height=0.2\textheight]{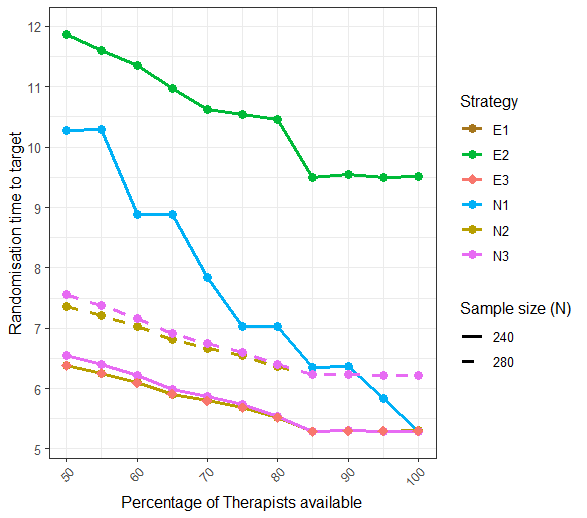}  
}
\subfigure[$(n_T,a_E) = (12,1)$]{
  \centering
  \includegraphics[width=0.3\linewidth, height=0.2\textheight]{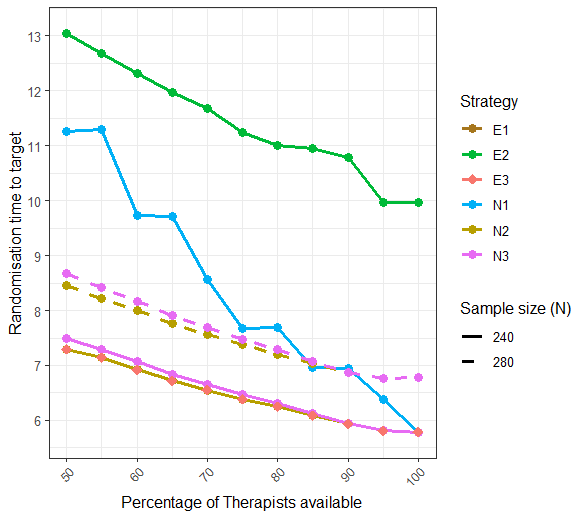}  
}
\parbox{0.95\textwidth}{%
 \footnotesize
\textit{Note:} Here $N$ and $n_T$ denote the total sample size and number of therapists, respectively.  The proportion of embedded therapists is $a_E$. Plots (a)-(c) represents the power curves, whereas, plots (d)-(f) represents the randomisation time to target curves of various randomisation strategies against the percentage of therapists available for model-1 for the single centre trial. The parameter $a_E$ represents the proportion of embedded therapists in the centre.}
\label{Fig.power.Model.1}
\end{sidewaysfigure}

We now explore the trade-off between \emph{inferential accuracy} and \emph{operational efficiency} across randomisation strategies for Model~1. Specifically, we compare the power to detect the intervention effect $\delta_1$ with randomising to time and target by levels of therapist availability (see Figures~\ref{Fig.power.Model.1}(a)-(c)) Figure~\ref{Fig.power.Model.1}(a) shows that when $(n_T, a_E) = (12, 0)$, power decreases as therapist availability decreases, with Strategy E3 consistently yielding the lowest power. Strategy N1 yields the highest power when therapist availability is constrained below approximately $80\%$, whereas all strategies perform similarly as availability exceeds this threshold. Flexible Strategies N2 and N3 yield intermediate power under limited therapist availability for $N=240$; remaining stable and reliable. When the sample size increases from $240$ to $280$, flexible Strategies N2 and N3 show improved power, becoming comparable to Strategy N1 and surpassing Strategy E2 (as shown by the dashed lines).

The corresponding actual randomisation periods for each strategy are shown in Figures~\ref{Fig.power.Model.1}(d)--(e). As expected, randomisation time to target decreases for all strategies as therapist availability increases. Moreover, under constant recruitment rates, the time required by Strategies N2 and N3 to reach $N=280$ is substantially shorter than that required by rigid Strategies N1 and E2 to reach their original target of $N=240$. A similar pattern is observed in Figures~\ref{Fig.power.Model.1}(b)--(c), which consider designs with a higher proportion of embedded therapists ($a_P\geq0.5$). As the proportion of embedded therapists increases, Strategy E2 exhibits increasingly unstable behaviour due to more frequent REML failures. Similar plots for $N=240$ and $N=300$ are provided in Figure~C.1. In summary, these results suggest that a modest increase in sample size can allow Strategies N2 and N3 to achieve power comparable to N1 and E2 while requiring shorter randomisation periods.

For Model~2, the power to detect the intervention effect $\delta$ is similar across all six randomisation strategies. We therefore explored the trade-off between \emph{predictive accuracy} and \emph{operational efficiency} (see Figures C.2(a)--(f) for boxplots of RMSE distributions for $n_T=12$ across combinations of $(a_E,a)$). Figure C.2(a) shows that when only $50\%$ of therapists are available ($a=0.5$), Strategy N1 performs comparably to Strategy E3, both exhibiting the highest RMSE. Although the RMSE for Strategy N1 decreases as therapist availability increases (see Figures C.2(b)--(c)), it remains suboptimal relative to Strategies E1, E2, N2 and N3. When combined with its substantially longer randomisation period, these results indicate that strategy N1 is not a good choice under constrained therapist availability when Model~2 is fitted. Across all scenarios, Strategies E1 and E2 achieve the smallest RMSE when they are feasible. Figures C.2(d)--(f) show that Strategies E1 and E2 incur substantially longer randomisation periods compared with their flexible counterparts (Strategies N2 and N3), particularly when therapist status varies ($a_E>0$). 

In summary, for Model~2, despite their superior predictive accuracy, Strategies E1 and E2 are inefficient from an operational perspective. Strategies N2 and N3 occupy an intermediate position in terms of RMSE but dominate in operational efficiency.

\clearpage

\section{Discussion and Conclusions}\label{sec.discussion}

Disentangling therapist from intervention effects has been a longstanding methodological challenge in trials of therapist-delivered interventions. A sizeable proportion already include the therapists as a factor when generating the randomisation sequence \cite{Walwyn_2025_prep}. However, the associated methodological literature has been limited. Walwyn et al \cite{Walwyn_2026} set out a statistically efficient but operationally rigid family of factorial designs, each of which randomises intervention-therapist combinations to patients. We have extended this work to simultaneously incorporate several practical constraints specific to clinical trials, setting out three existing randomisation strategies and proposing a further three novel more flexible strategies. As a result, we have begun to address many of the practical considerations identified by \cite{Walwyn_2025_prep} explicitly within the trial design. The novel randomisation strategies were intended to achieve a better balance between statistical and operational efficiency. We have summarised the results of a detailed simulation study, exploring the properties of the six randomisation strategies across three broad trial scenarios, several design parameters, and two analysis models. Notably, we have allowed research teams to include any mix of full-time research therapists and therapists embedded in services, and better understand the implications of doing so. We have also given research teams a choice between setting up the randomisation at the start of the trial, at discrete intervals throughout the trial, or on a patient-by-patient basis, as therapist availability becomes known. 

As we expected, there was no universally preferred randomisation strategy. Rather, the choice depended on the proportion of therapists embedded in services, their capacity, availability, the analysis model, and a trade-off between statistical efficiency and the length of the randomisation period. We have confirmed the benefit of randomising therapists to patients on unbiased estimation of the between-therapist variance and a consequent reduction in the standard error of the intervention effect and increase in power. We have shown the benefits of incorporating therapist availability in the randomisation strategy on feasibility of trial delivery, and of better utilisation of therapist capacity on operational efficiency. We have highlighted the importance of replication of treatments in blocks and of therapists in centres on model convergence, and the need to include sufficient therapists (and batches) for numerical stability, when fitting the planned analysis model. Our two motivating examples were chosen to display these design features. Overall, greater experimental control led to more robust estimation of treatment effects, but where this control is not possible, as will be the case in large pragmatic trials, our more flexible randomisation strategies provide a good balance between statistical and operational performance. The choice can then be made between setting up the randomisation at discrete intervals or on a patient-by-patient basis. If discrete intervals are preferred, then care is needed in determining the number of intervals (batches) to ensure sufficient replication in the design, otherwise model convergence will not be guaranteed. Reducing the analysis model assumes the effects omitted are zero. We found that replacing the random factor Batches with a fixed continuous variable Time-in-Therapist had benefits. However, the analysis then depends on modeling assumptions, and specifically the assumed shape of the therapist learning curve.

There are a number of challenges that we have not addressed. First, in our alternative model, we assumed therapist learning was linear across time, primarily for illustration. Further work is needed to provide detailed guidance on trial design and its associated analysis, where there is specific interest in robustly estimating therapist learning effects. Second, we fixed the overall sample size in the three trial scenarios considered in our simulation study. Currently, sample size is often calculated under idealised assumptions. Research is currently ongoing to propose and evaluate methods for calculating sample size in non-orthogonal factorial trials of therapist-delivered interventions. Third, we assumed that therapist capacities are predictable at the start of the trial and constant across the trial. We also assumed that therapist turnover can be incorporated simply by classifying therapists as unavailable for each block or on a patient-by-patient basis. As a result, we have not addressed the possibility that cluster sizes are informative. Further work is needed to set out the implications of informative cluster sizes for trial design and its associated analysis. Fourth, we have explored the performance of the randomisation strategy on the length of the randomisation period. While this is closely related to the cost of running a trial, it is not the only factor that is. Further work is planned to develop methods for finding optimal designs that balance the trade-off between increased information and increased cost, when estimating fixed and random treatment effects is of interest. Fifth, given the early stage of the methods considered in this paper, there was a lot of uncertainty around realistic sizes of the model parameters. Further work is needed to obtain realistic estimates, not only of the between-therapist variance but of all the random effects considered here. Following this, more extensive simulation studies will be needed to fully evaluate the performance of the six randomisation strategies in realistic scenarios. Sixth, given the uncertainty in the size of the model parameters, and the potential sensitivity to mis-specification, Bayesian or minimax design strategies may increase feasibility. Further work is needed to propose and evaluate Bayesian and/or minimax approaches. Seventh, last minute changes to therapist availability may lead to violations of the randomisation protocol. When interventions are delivered over many sessions, therapist turnover might change the therapist allocated to a patient mid-intervention. While our randomisation strategies are more flexible, some non-compliance is likely and methods are needed for handling this in the design and analysis. Eighth, we have focused on a single continuous outcome. In practice, longitudinal or multivariate outcomes may be available and/or binary or survival outcomes. Further research is needed to generalise our work to these outcome types.

There are also some limitations to note. First, although we have considered six randomisation strategies in some detail here, others are also possible. We would recommend better reporting of the randomisation process used in trials of therapist-delivered interventions, including sufficient details of how the therapists are allocated to patients, whether this allocation is random or not. In all our strategies, we have assumed that therapists are allocated to patients before interventions or at the same time. This was to avoid the risk of inflating the between-therapist variance in the intervention effect through lack of concealment. Where this is not the case (see \cite{Walwyn_2026}) the statistical properties of the trial design are anticipated to be worse. Second, we assumed the rate-limiting factor in randomising to time and target is the collective capacity of the available therapists. Further work is needed to explore the additional impact of problems in patient recruitment, as in practice either or both could be rate-limiting factors. Third, the scenarios that we considered did not include pilot or feasibility trials. Small-sample properties of the six randomisation strategies were therefore not explored. It is expected that the number of centres, therapists, batches and replications would all be small in such studies. The point at which centres, therapists and batches should be considered as fixed rather than as random factors therefore remains unknown. Fourth, incorporating the complexity resulting from considering therapists as a treatment factor in the trial design is likely to have implications for the difficulty in communicating the design in grant applications and in conducting the trial. Experience is needed of both to demonstrate feasibility and share the lessons learnt. Finally, although we have made the programming code used to implement the randomisation strategies available on GitHub, the next step will be to incorporate these options into central randomisation systems and the software used there.

To conclude, purposeful trial design can not only be more statistically efficient but more operationally feasible if key practical constraints are incorporated in the randomisation at the design stage.

\newpage
\bibliographystyle{unsrtnat}

\bibliography{Bibfile}

\newpage

\begin{sidewaysfigure}
    \centering
    \caption{A flowchart of randomisation strategies}
    \includegraphics[width=0.9\textheight]{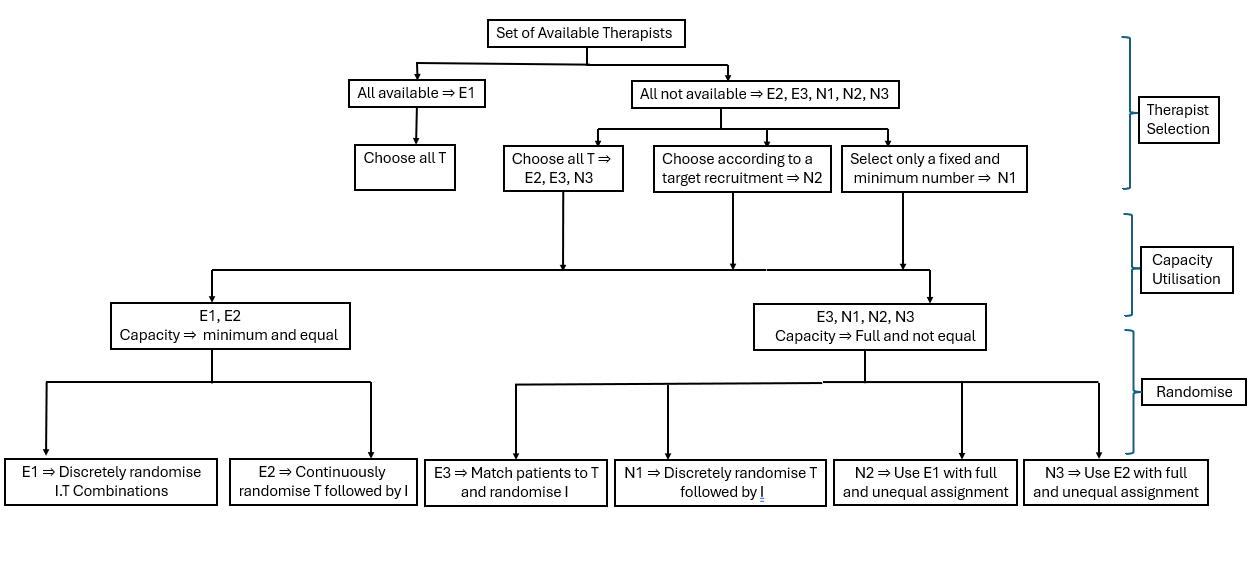}

    \parbox{0.9\textheight}{%
        \footnotesize
        \textit{Note:} `T' denotes the therapist and `I' denotes the intervention. Strategy E1 is single-time equal-replicate randomisation
        of $IT$ combinations; Strategy E2 is continuous-time equal-replicate
        randomisation of $T$ then $I$; Strategy E3 is single-time
        equal-replicate randomisation of $I$ only; Strategy N1 is
        discrete-time incomplete block randomisation of $T$ then $I$;
        Strategy N2 is discrete-time flexible randomisation of $IT$
        combinations; and Strategy N3 is continuous-time flexible
        randomisation of $T$ then $I$. The randomisation strategies differ in:
        (1) the selection of therapists from those available;
        (2) the utilisation of capacity of selected therapists; and
        (3) how patients are randomised to selected therapists in each block.
    }
    \label{fig.flowchart.all}
\end{sidewaysfigure}

\begin{sidewaystable}
\centering
\caption{Summary of notation}
\label{Table.notation}

\setlength{\tabcolsep}{3pt}
\renewcommand{\arraystretch}{1.15}

\begin{tabular}{c|@{\hspace{4pt}}p{0.78\textheight}}
\hline
Notation & Description \\
\hline
$(I,C,T,B)$ & (Intervention, Centre, Therapist, Batch)\\
$N$ & Target total number of patients\\
$(n_I, n_C, n_T^{(C)},n_B)$ & Number of interventions, centres, therapists in centre $C$, and batches\\
$a_E$& Proportion of embedded therapists in a centre\\
$(n_E^{(C)},n_F^{(C)})$ & Number of embedded and research therapists in centre C, i.e., $(a_E n_T^{(C)}, n_T^{(C)}-a_E n_T^{(C)})$\\
$\mathcal{T}^{(C)}$ & Set $\{1^{(C)},2^{(C)},\ldots,n_T^{(C)}\}$ of $n_T^{(C)}$ therapists in centre $C$\\
$n_{R_j}^{(C)}$ & Capacity of therapist $j$ in centre $C$ in each batch\\
$\boldsymbol{n_R^{(C)}}$ & Collection of capacities of therapists in centre C, $(n_{R1}^{(C)},\ldots,n_{Rn_T}^{(C)})$\\
$n_R$ & $\min(\boldsymbol{n_R^{(1)}},\boldsymbol{n_R^{(2)}},\ldots,\boldsymbol{n_R^{(n_C)}})$\\
$(E_E,E_F)$ & Expected capacity of embedded and research therapists throughout a trial\\
$(\sigma^2_E,\sigma^2_F)$ & Variance of the capacity of embedded and research therapists throughout the trial\\
$\mathcal{T}_{B,A}^{(C)}$ & Set $\{1_{B,A}^{(C)},2_{B,A}^{(C)},\ldots,n_{T_{B,A}}^{(C)}\}\subseteq\mathcal{T}^{(C)}$ of available therapists in batch $B$ of centre $C$\\
$n_{T_{B,A}}^{(C)}$ & $|\mathcal{T}_{B,A}^{(C)}|$, i.e. number of available therapists in batch $B$ of centre $C$\\
$N_{B,A}^{(C)}$ & $\sum_{j\in\mathcal{T}_{B,A}^{(C)}}n_I n_{R_j}^{(C)}$, total number of patients that can be treated by the available therapists\\
$\mathcal{T}_{B,S}^{(C)}$ & Subset of $\mathcal{T}_{B,A}^{(C)}$ of selected therapists in batch $B$ of centre $C$\\
$n_{T_{B,S}}^{(C)}$ & $|\mathcal{T}_{B,S}^{(C)}|$, i.e. number of selected therapists in batch $B$ of centre $C$\\
$N_{B,S}^{(C)}$ & $\sum_{j\in\mathcal{T}_{B,S}^{(C)}}n_I n_{R_j}^{(C)}$, total number of patients randomised to the selected therapists\\
$n_{T_{\min}}^{(C)}$ & $\min\{n_{T_{1,A}}^{(C)},n_{T_{2,A}}^{(C)},\ldots,n_{T_{n_B,A}}^{(C)}\}$, minimum number of available therapists across batches\\
$n_{T\text{-IBD}}^{(C)}$ & $\leq n_{T_{\min}}^{(C)}$, number of therapists randomly selected in centre $C$ from $\mathcal{T}_{B,A}^{(C)}$ for strategy N1\\
$N_B^{(C)}$ & $\displaystyle\frac{N}{n_B}\frac{E_E n_E^{(C)}+E_F n_F^{(C)}}{\sum_{i=1}^{n_C}(E_E n_E^{(i)}+E_F n_F^{(i)})}$, target size of a batch in centre $C$ for strategy N2\\
$\text{frac}$ & Permissible over-recruitment above the target batch size\\
$a$ & Parameter of the distribution of therapist availability\\
$\text{frac}_P,\text{frac}_F\sim\text{Uniform}[a,1]$ & Fraction of therapists available out of $(n_E^{(C)},n_F^{(C)})$ in a centre-batch combination\\
\hline
\end{tabular}\label{Table.notation}
\end{sidewaystable}

\newpage

\appendix

\section{Appendix}
\subsection{Brief examples}\label{appendix.examples}
In the following, we shall briefly explain the existing and novel randomisation strategies through hypothetical examples. The names of the examples are self explanatory, for instance, Example A.1.E2 denotes the example for strategy E2. The examples are motivated by Example B presented in Section \ref{sec.motivating.examples}. For all the examples, we shall assume that we are comparing a control therapy with a novel therapy, i.e., $n_I=2$. We assume that $n_C=17$ centres, $n_T^{C}=n_T=3$ therapists within each centre, and $n_B=5$ batches. For simplicity, let us assume that

$$
\boldsymbol{n_R^{(1)}}=\boldsymbol{n_R^{(2)}}=\ldots=\boldsymbol{n_R^{(n_C)}}=(2,3,4).
$$

That is, therapist 1 in every centre will treat $2$ patients in each batch using each intervention. Similarly, therapists 2 and 3 in every centre will treat $3$ and $4$ patients in each batch using each intervention, respectively. Thus, we have variable therapist capacity in each centre.

\medskip

\noindent\textbf{Example A.1.E1:} 
\medskip

For strategy E1, it is necessary to have balancing in all factors. The number of treatments is fixed at $n_In_Cn_T=102$. Note that this strategy is applicable only when every therapist is available to treat patients in every batch and has equal capacity associated with them. The randomisation is performed so that, in each of the $n_Cn_B=85$ centre--batch combinations (blocks), $n_R=\min(\boldsymbol{n_R^{(1)}},\boldsymbol{n_R^{(2)}},\ldots,\boldsymbol{n_R^{(n_C)}})=2$ patients are randomly allocated to every intervention--therapist ($n_In_T=6$) combination. Thus, $n_R=2$ patients are allocated to each intervention within each therapist, and each of the $n_Cn_T=51$ therapists shall treat $n_In_R=4$ patients in each block. The randomisation strategy is equivalent to stratified block randomisation with $n_Cn_B=85$ blocks, each of fixed size $n_In_Tn_R=12$. Therefore, the total number of patients in the trial is $ N=n_Cn_Tn_Bn_Rn_I=1020.$

\medskip

\noindent\textbf{Example A.1.E2:}
\medskip

As in Example 3.E1, the number of treatments is fixed at $n_In_Cn_T=102$. Since the batch structure is not defined for this strategy, the capacity of a therapist throughout the trial is $n_In_P = n_In_Bn_R = 20$. The randomisation is performed in two steps. In the first step, as patients enter the trial, they are randomised to a therapist. If the selected therapist is available at that time and has capacity left (i.e., fewer than $n_In_P=20$ patients have been randomised to the therapist so far), the assignment is confirmed. Otherwise, the allocation is ignored and the patient is re-randomised until a therapist is allocated. In the second step, using therapist as a blocking factor, $10$ patients are randomised to intervention 1 and the other $10$ to intervention 2.

\medskip

\noindent\textbf{Example A.1.E3:}
\medskip

This is a two-step randomisation. In the first step, patients are non-randomly assigned to available therapists using a patient--therapist matching procedure based on a covariate (e.g., severity of the patient). For instance, let the set of therapists available to treat patients in the first batch and first centre be $\mathcal{T}_{1,A}^{(1)}=\{1,3\}$. Then, in the first batch, $n_In_{R_1}^{(1)}=2\times2=4$ patients in centre 1 will be assigned to therapist 1, while $n_In_{R_3}^{(1)}=2\times4=8$ patients will be assigned to therapist 3, based on the patient severity covariate. Similar assignments hold for other centres, with the availability and capacity constraints of therapists taken into account.

In the second step, taking therapist 1 as a block, $n_{R_1}^{(1)}=2$ out of the $4$ patients assigned to therapist 1 are randomised to each of the $n_I=2$ interventions. Similarly, taking therapist 3 as a block, $n_{R_3}^{(1)}=4$ out of the $8$ patients assigned to therapist 3 are randomised to each of the $n_I=2$ interventions. The same procedure is followed for the other patients assigned to each therapist within each centre.

\medskip

\noindent\textbf{Example A.1.N1:} 
\medskip

We illustrate how the randomisation operates for $C=2$ and $B=1$. Let us assume that a minimum of two therapists are always available to treat patients in centre 2. That is, $
n_{T_{min}}^{(2)}
=
\min\left\{
n_{T_{1,A}}^{(2)},n_{T_{2,A}}^{(2)},\ldots,n_{T_{n_B,A}}^{(2)}
\right\}
=2,
$ where $n_{T_{1,A}}^{(2)}=|\mathcal{T}_{1,A}^{(2)}|$ denotes the number of therapists available in batch $B$ in centre $2$. This is a two-step randomisation strategy. We randomly select $
n_{T\text{-}IBD}^{(2)}=n_{T_{min}}^{(2)}=2
$ therapists from the set $\mathcal{T}_{1,A}^{(2)}$ of all available therapists in batch 1 centre 2 to treat patients in that batch. Suppose that, for batch 1, $
\mathcal{T}_{1,A}^{(2)}=\{1,2,3\}.
$ Then we randomly select $
\mathcal{T}_{1,S}^{(2)}=\{1,3\}
$ (say) as the set of therapists to whom patients will be allocated.

In Step 1, for centre 2 and batch 1, $n_In_{R_1}^{(2)}=2\times2=4$ patients are allocated to therapist 1, while $n_In_{R_3}^{(2)}=2\times4=8$ patients are allocated to therapist 3. Thus, the total number of patients randomised to selected therapists $\mathcal{T}_{1,S}^{(2)}=\{1,3\}$ in batch 1 of centre 2 is $
N_{1,S}^{(2)}=4+8=12.
$ A systematic design is then constructed by considering patients labelled $1,2,\ldots,12$ and assigning them to the selected therapists according to the above allocation. A random permutation of ${1,2,\ldots,12}$ is subsequently generated to shuffle the order of the patients, after which the patients are renumbered from 1 to 12. The same procedure is followed independently for all other centres and batches.

In Step 2, $n_{R_1}^{(2)}=2$ of the $4$ patients allocated to therapist 1 are randomised to each of the $n_I=2$ interventions. Similarly, $n_{R_3}^{(2)}=4$ of the $8$ patients allocated to therapist 3 are randomised to each of the $n_I=2$ interventions.
\medskip

\noindent\textbf{Example A.1.N2:}
\medskip

For simplicity, let the number of embedded and full-time therapists in centre C be $n_E^{(C)}=2$ and $n_F^{(C)}=1$, respectively, for $C=1,\ldots,n_C$. Further, let the expected capacity of an embedded and full-time therapist is $E_E=2$ and $E_F=4$, respectively. From Example E.1, the target total sample size is $N=1020$ patients. Therefore, the target batch size in centre $C$ is $$N_B^{(C)} = \frac{N}{n_B}\frac{E_E n_E^{(C)}+E_F n_F^{(C)}}{\sum_{i=1}^{n_C}(E_E n_E^{(i)}+E_F n_F^{(i)})}=12.$$ We illustrate how the randomisation procedure operates for centre 1. 

Suppose that the set of available therapists in centre 1 during batch 1 be $\mathcal{T}_{1,A}^{(1)}=\{1,2\}$. The total number of patients that can be treated in batch 1 if all available therapists work at their full capacity in centre 1 is $$N_{1,A}^{(1)} = n_I( n_{R_1}^{(1)}+n_{R_2}^{(1)}) = 2\times(2+3)  = 10.$$ 

\begin{itemize}
    \item[(1)] A systematic design for this block (centre 1 batch 1) with set $\mathcal{T}_{1,A}^{(1)}=\{1,2\}$ of available therapists and their associated $N_{1,A}^{(1)} =10$ patients is created. 
    \begin{itemize}
        \item[(1.1)] Since $N_{1,A}^{(1)} < N_B^{(1)}$, we have a situation of under-recruitment. Therefore, all available therapists in this block are selected, i.e., $\mathcal{T}_{1,S}^{(1)}=\mathcal{T}_{1,A}^{(1)}=\{1,2\}$. Randomise the selected $n_I |\mathcal{T}_{1,S}^{(1)}| = 2\times 2 = 4$ intervention-therapist combinations by randomly permuting the rows of the systematic design and renumbering the patients from $1,2, \ldots,10$.  
    \end{itemize}

\end{itemize}

In contrast, suppose that the set of available therapists in centre 2 during batch 1 is $\mathcal{T}_{1,A}^{(2)}=\{1,2,3\}$. The total number of patients that can be treated in this block if all available therapists work at their full capacity is $N_{1,A}^{(2)} = n_I( n_{R_1}^{(2)}+n_{R_2}^{(2)}+n_{R_3}^{(2)}) = 18$.

\begin{itemize}
    \item[(1)] A systematic design for this block with set $\mathcal{T}_{1,A}^{(2)}=\{1,2,3\}$ of available therapists and their associated $N_{1,A}^{(2)} =18$ patients is created. 
    \begin{itemize}
        \item[(1.1)] For $\mathrm{frac}=0.25$, since $N_{1,A}^{(2)} > (1+\mathrm{frac})N_B^{(2)}=15$, we have a situation of excessive over-recruitment. Therefore, we select a subset $\mathcal{T}_{1,S}^{(2)}\subseteq\mathcal{T}_{1,A}^{(2)}$ with selected batch size $N_{1,S}^{(2)}=\sum_{j\in\mathcal{T}_{1,S}^{(2)}}n_I n_{R_j}^{(2)}$ such that $\left|N_{1,S}^{(2)}-(1+\mathrm{frac})N_B^{(2)}\right|$ is minimised. In particular, if $\mathcal{T}_{1,S}^{(2)}=\{1,2\}$, then $N_{1,S}^{(2)}=2\times2+2\times3=10$. Similarly, for $\mathcal{T}_{1,S}^{(2)}=\{1,3\}$ and $\mathcal{T}_{1,S}^{(2)}=\{2,3\}$, we obtain $N_{1,S}^{(2)}=12$ and $N_{1,S}^{(2)}=14$, respectively. Clearly, $\mathcal{T}_{1,S}^{(2)}=\{2,3\}$ produces the minimum value of $\left|N_{1,S}^{(2)}-(1+\mathrm{frac})N_B^{(2)}\right|$. Therefore, therapists 2 and 3 are selected.   
 Finally, randomise the selected $n_I |\mathcal{T}_{1,S}^{(1)}| = 2\times 2 = 4$ intervention-therapist combinations by randomly permuting the rows of the systematic design and renumbering the patients from $1,2, \ldots,14$.  
    \end{itemize}

\end{itemize}

\medskip

\noindent\textbf{Example A.1.N3:}

\medskip

This is an extension of Example A.1.E2. The randomisation is again performed in two steps: therapist allocation followed by intervention randomisation. The only difference is that, instead of imposing a minimum capacity of $n_P=10$, the full capacity of every therapist is utilised. For instance, let $\mathcal{T}_{1,A}^{(1)}=\{1,3\}$ denote the set of therapists available in centre 1 during batch 1. As a patient enters the trial in centre 1 during batch 1, suppose that the patient is initially randomised to therapist 2. Since $2\notin\mathcal{T}_{1,A}^{(1)}$, therapist 2 is not available to treat the patient, and the patient needs to be re-randomised to another therapist. Suppose that the patient is subsequently randomised to therapist 3. Since $1\in\mathcal{T}_{1,A}^{(1)}$ and therapist 3 has sufficient capacity $$n_In_{P_3}^{(1)} = n_In_Bn_{R_3}^{(1)} = 2\times 5\times 4 = 40,$$ to treat the patient, the patient is allocated to therapist 3. The same procedure is followed for all patients throughout the trial, with the randomisation continuing until each patient is allocated to an available therapist with remaining capacity.

In the second step, patients are randomised equally to the interventions within each centre, taking therapist as a blocking factor. For instance, therapist 3 in centre 1 will have $n_In_{P_3}^{(1)}=40$ patients allocated to them, of which $20$ patients are randomised to each of the two interventions.

\subsection{Cautions and Recommendations}\label{appendix.recommendations}

We compared these existing and novel randomisation strategies on two key objectives, \emph{operational} efficiency and \emph{inferential} efficiency. The results reveal that there is no universally preferred randomisation; rather, the choice depends on the proportion of therapists embedded in services (therapist status), their capacity, availability, the analysis model, and the trade-off between cost reduction and estimation precision. To enhance the practical utility of these findings, we summarize below the key cautions and recommendations for trialists conducting such complex trials. A brief summary of the recommended strategies in different combinations of therapist availability and capacity is given in Tables \ref{Table.recommendations}(a) and \ref{Table.recommendations}(b) for model-1 and 2, respectively, below.  

\begin{enumerate}

\item \textbf{Strategies E1 \& E2:}

\textbf{Cautions:}
\begin{enumerate}
    \item Both randomisation strategies enforcing strict balancing of patients across therapists results in each therapist not being used to their maximum capacity. This substantially increases randomisation time to target, sometimes more than doubling the expected duration, leading to significant cost escalation due to operational costs. Therefore, if not all therapists are working full-time, Strategies E1 and E2 pose a high risk in terms of operational efficiency.
    
    \item If embedded therapists are included, it is critical to ensure they have sufficient capacity to treat at least two patients from each intervention within a time period. If this condition is not met, i.e., the trialists lack control over therapist capacities, both randomisation schemes fail to generate adequate replication for estimating some of the effects that might be of interest. Consequently, these methods should only be used when nearly all therapists have high treatment capacity.
\end{enumerate}

\textbf{Recommendations:}  
When the trial team has full control over therapists, meaning all therapists are recruited and trained prior to trial initiation and are expected to treat a high number of patients, Strategies E1 and E2 are ideal choices both in terms of randomisation time to target and estimation precision.

\item \textbf{Strategy E3:}  

\textbf{Cautions:}  
Across all scenarios of therapist status, capacity, and availability, Strategy E3 performs poorly in terms of parameter estimation. It severely inflates the therapist variance, leading to adverse consequences such as increased Type II error, higher RMSE and, more frequent boundary estimates. Due to these issues, E3 assignment should be avoided if possible.  

\textbf{Recommendations:}  
E3 assignment is highly flexible with respect to therapist status, capacity, and availability. Consequently, it achieves the shortest randomisation time to target. It also generates sufficient data to estimate all random effects (though estimates may be biased). Thus, while the estimates may lack validity, convergence of the model is virtually guaranteed.

\item \textbf{Strategy N1:} 

\textbf{Cautions:} 
\begin{enumerate}
\item Strategy N1 only selects $n_{T-IBD}^{(C)}$ therapists per batch from the pool of available therapists even if more therapists are available during a batch. Conversely, if fewer than $n_{T-IBD}^{(C)}$ therapists are available in a given batch, randomisation cannot proceed until the required number is met. This underscores the critical importance of choosing an appropriate $n_{T-IBD}^{(C)}$ at the design stage. An incorrect choice can dramatically increase randomisation time to target (highest among all novel strategies). 
    
\item If $n_{T\text{-IBD}}^{(C)} \leq 2$, Strategy N1 can produce model failures at the analysis stage, thus, to avoid that the number of therapists per centre should be managed such that at least two therapists are always available to treat patients.
\end{enumerate}

\textbf{Recommendations:}
\begin{enumerate}
    \item For Model~1, Strategy N1 yields the highest power and the smallest standard error for the intervention effect among all randomisation strategies when therapist availability is constrained, irrespective of therapist status. Therefore, Strategy N1 should be preferred over Strategies E1 and E2 when there is limited control over therapist availability and status across centres.

    \item For Model~2, when therapist availability is constrained, Strategy N1 exhibits one of the highest root mean squared errors for prediction. Thus, Strategy N1 should be avoided if the primary objective is prediction rather than estimation of the intervention effect.
\end{enumerate}

\item \textbf{Strategies N2 and N3:} 

\textbf{Cautions:} 
\begin{enumerate}
\item Similar to other strategies, strategies N2 and N3 can produce insufficient data for estimating therapist-specific parameters if fewer than two therapists are available in a centre at any time. Therefore, if a centre recruits only a small number of therapists, this number should exceed two, and the trial team must maintain strong control over therapist calendars.

\item Due to their non-orthogonal nature, both N2 and N3 perform slightly worse than more rigid strategies such as N1, E1, and E2 in terms of precision and statistical inference. Consequently, to achieve comparable precision, the sample size may need to be modestly inflated. Nevertheless, both are substantially better than Strategy E3.
\end{enumerate}

\textbf{Recommendations:}
\begin{enumerate}
 \item N2 and N3 achieve the shortest randomisation time to target across all designs of therapist status and availability. In some cases, randomisation time is reduced by more than half compared to other strategies.
 
\item When the trial team has strong control over therapist status and scheduling, N2 and N3 yield parameter estimates that are comparable in precision to rigid strategies such as N1, E1, and E2. 
  \item Although N2 and N3 are not the most precise when therapist capacity and availability are constrained, as shown in Figure \ref{Fig.power.Model.1}, a slight increase in sample size achieves precision comparable to rigid strategies N1, E1, and E2, while still managing a faster randomisation time to target. Taken together, these findings indicate that N2 and N3 are a practical and cost-efficient alternative when full control over therapists is not feasible.
\end{enumerate}

\end{enumerate}

\clearpage

\begin{sidewaystable}[ht]
\renewcommand{\thetable}{A1}   
\centering
\caption{Parameter values used in the simulations}\label{Table.recommendations}
\addtocounter{table}{-1}
\centering
\setlength{\tabcolsep}{3pt}
\renewcommand{\arraystretch}{1.15}
\scriptsize

\textbf{(a) Recommended randomisation strategies for model-2}

\hspace{2mm}

\setlength{\tabcolsep}{0.3pt}  
\begin{tabular}{c|c|c|c|c|c|c|c|c}
\hline
Capacity&Availability&\multicolumn{7}{c}{Efficiency Measure}\\
\hline
&&Trial Cost&Convergence&Stability&Unbiasedness&Precision&Inference&Prediction\\
\hline
 \multirow{2}{*}{Low}& Low &E3, N2, N3&E3, N1, N2, N3 &N1, N2, N3 &N1, N2, N3&N1&N1&N1, N2, N3 \\ 
 \cline{2-9}
  & High &E3, N1, N2, N3&E3, N1, N2, N3 &N1, N2, N3 &N1, N2, N3&N1, N2, N3&N1, N2, N3&N1, N2, N3\\
\cline{2-9}
 \hline
 \multirow{2}{*}{High}& Low  &E3, N2, N3&All$\setminus \{E1\}$ &All$\setminus \{E1, E3\}$&All$\setminus \{E1, E3\}$&N1&N1&All$\setminus \{E1, E3\}$\\ 
  \cline{2-9}
 & High &E3, N1, N2, N3&All&All$\setminus \{E3\}$&All$\setminus \{E3\}$&All$\setminus \{E3\}$&All$\setminus \{E3\}$&All$\setminus \{E3\}$\\ 
 \cline{2-9}
 \hline
\end{tabular}

\hspace{4mm}

\textbf{(b) Recommended randomisation strategies for model-2}

\hspace{2mm}

\setlength{\tabcolsep}{0.4pt}  
\begin{tabular}{c|c|c|c|c|c|c|c|c}
\hline
Capacity&Availability&\multicolumn{7}{c}{Efficiency Measure}\\
\hline
&&Trial Cost&Convergence&Stability&Unbiasedness&Precision&Inference&Prediction\\
\hline
 \multirow{2}{*}{Low}& Low &E3, N2, N3&All$\setminus \{E1\}$&All$\setminus \{E1\}$&All$\setminus \{E1, E3\}$&E2, N2, N3&E2, N2, N3&E2 \\ 
 \cline{2-9}
  & High &E3, N1, N2, N3&All &All&All$\setminus \{E3\}$&All$\setminus \{E3\}$&All$\setminus \{E3\}$&E1, E2\\
\cline{2-9}
 \hline
 \multirow{2}{*}{High}& Low  &E3, N2, N3&All$\setminus \{E1\}$&All$\setminus \{E1\}$&All$\setminus \{E1, E3\}$&E2, N2, N3&E2, N2, N3&E2 \\ 
  \cline{2-9}
 & High &E3, N1, N2, N3&All &All&All$\setminus \{E3\}$&All$\setminus \{E3\}$&All$\setminus \{E3\}$&E1, E2\\
 \cline{2-9}
 \hline
\end{tabular}

\hspace{2mm}

\parbox{0.95\textwidth}{%
\footnotesize
\textit{Note:} Table A1(a) presents a brief summary of recommended randomisation strategies for `Low' and `High' configurations of therapist capacity and availability against different performance indicators for Model-1 given in \ref{sec.performance.measures}. Similarly, Table A1(b) presents a brief summary of recommended randomisation strategies for Model-2}
\end{sidewaystable}

\end{document}



\title{Supplementary File A: Incomplete Block Design}
\date{}

\maketitle

Here we present a brief overview of \emph{Incomplete Block Designs} (IBDs), outlining their origin and practical relevance. 

\section{Incomplete Block Design}

IBDs originated in agricultural experimentation and have since been extensively studied, with a rich body of literature addressing both their construction and statistical properties and optimality (refer to books \cite{Hinkelmann_2005} and \cite{Mead_2012} and paper \cite{John_1977}).

When there are no restrictions on the randomisation of treatments to experimental units, the \emph{completely randomised design (CRD)} is the most straightforward choice, in which treatments are randomly assigned to plots. However, in classical agricultural experiments, plots are typically arranged within \emph{blocks}, each representing a relatively homogeneous portion of the experimental field. Owing to physical constraints and the desire to control local sources of variability, the number of plots within a block is often limited, while the number of treatments under comparison may be large. Consequently, not all treatments can be accommodated within every block. This scenario gives rise to \emph{incomplete blocking}, and designs that address such situations are known as \emph{Incomplete Block Designs}.

\cite{Yates_1936} introduced incomplete block designs to achieve greater precision in treatment comparisons by reducing field heterogeneity beyond what was possible with completely randomised or randomised complete block designs when the number of treatments was large. The precision of a treatment effect estimate increases with the number of replications of that treatment. Similarly, the precision of estimating differences between two treatment effects depends on how frequently the pair of treatments occurs together within the same block. To ensure equal (or nearly equal) precision across all treatment comparisons, treatments are allocated to blocks of equal size such that (i) each treatment appears at most once per block, (ii) each treatment is replicated the same number of times across blocks, and (iii) each pair of treatments occurs together the same (or nearly the same) number of times.

Designs satisfying these conditions exactly are known as \emph{Balanced Incomplete Block Designs } (BIBD), whereas those where the number of joint occurrences varies slightly between treatment pairs are referred to as \emph{Partially Balanced Incomplete Block Designs} (PBIBD).

Formally, consider a design with $b$ blocks, each containing $k$ plots, and $t$ treatments, each replicated $r$ times. The design is incomplete whenever $k < t$, ensuring that not all treatments appear in every block and that each treatment appears at most once per block. The total number of treatment assignments satisfies the relationship
$$
bk = tr.
$$
For instance, if $(b, k, t, r) = (4, 3, 6, 2)$, one possible design is
$$
\{1, 2, 3\}, \ \{1, 4, 5\}, \ \{2, 4, 6\}, \ \{3, 5, 6\}.
$$
The use of equal block sizes and equal replication across treatments facilitates separate estimation of treatment and block effects, although these effects are generally \emph{not orthogonal} in incomplete block designs. Consequently, treatment effects must be estimated after adjusting for block effects (yielding \emph{adjusted treatment effects}), typically using the method of least squares. Similarly, one may obtain adjusted block effects by accounting for treatment differences.

\bigskip

\bibliographystyle{apalike}
\bibliography{Bibfile}


\title{Supplementary File B: Simulation protocol}
\date{}

\maketitle
\section{Background}

This paper deals with designing trials of therapist-delivered interventions taking into account the practical constraints associated with involving therapists which lead to non-orthogonal trial designs. In particular, variable therapist capacity, availability and turnover which are defined as follows.
\begin{enumerate}
    \item \emph{Capacity} of a therapist is defined as the number of patients a therapist can treat in a batch (eg., a fixed time period).
    \item \emph{Availability} of a therapist is defined as a therapist’s readiness to accept new patients.
    \item \textit{Turnover} is defined as the dynamic change in the therapist pool over the course of the trial, resulting in a different set of therapists than those initially engaged.
\end{enumerate}

We assume a crossed therapist-intervention design in which each therapist delivers every intervention (to different patients). We then assume an individually-randomised, factorial design in which the intervention and the therapist are treatment factors (one fixed and one random).

In this paper we propose three novel randomisation strategies (N1, N2, N3) and compare them to three existing ones (E1, E2, E3). For a through understanding of these randomisation strategies, refer to the main article. A brief description is as follows. 
\begin{enumerate}
    \item The existing strategy E1 is an orthogonal factorial design proposed by Walwyn et al \cite{Walwyn_2026}. It is a one-step, single-time, equal-replicate randomisation strategy in which intervention-therapist combinations are the treatments, and are fully determined at the start of the trial.
    \item The existing strategy E2 was used by Shapiro et al \cite{Shapiro_1994}. It can be classified as a two-step, continuous-time, equal-replicate randomisation strategy.
    \item The existing strategy E3 is a parallel-arm design in which patients are randomised to interventions, but therapists are non-randomly allocated to patients.
    \item The novel strategy N1 is inspired from Incomplete Block Design (IBD) (\cite{Yates_1936}). It is classified as a two-step, discrete-time, incomplete block randomisation of therapists then interventions to patients.
    \item The novel strategy N2 is a more flexible non-orthogonal extension of Strategy E2. It is classified as an iterative, discrete-time, flexible randomisation of intervention-therapist combinations to patients.
    \item The novel strategy N3 is a more flexible non-orthogonal extension of Strategy E2, incorporating variable therapist capacity constraints. It is classified as a two-step, continuous-time, flexible randomisation of therapists then interventions to patients.
\end{enumerate}

\subsection{Aim of the Simulations}\label{sec.aims.of.sim}

The aim of this simulation study is to develop evidence-based recommendations for trials of therapist-delivered interventions. In particular, we will try to answer the following research questions,
\begin{enumerate}
    \item Is randomisation of therapists to patients necessary in pragmatic trials of therapist-delivered interventions?
    \item What is the effect of adopting six possible randomisation strategies on randomisation time to target?
    \item How do these randomisation strategies affect the primary analysis and estimation of treatment effects?
\end{enumerate}

\subsection{Notation}\label{sec.notation}

The following notation will be used for the parameters used in the six randomisation strategies. We shall define all the notations in terms of a multi-centre trial, for a trial with a single centre, we shall drop $(C)$ from the superscript of the symbols and stick to a normal font instead of a bold font. A brief definition of all the notations is given in Table 5 in the main article as well.
\begin{itemize}
\item $N$ - number of patients required
\item $(n_I,n_B,n_C)$ - number of (Interventions, Batches, Centres)
\item $a_E$ - proportion of embedded therapists in a centre
\item $(n_T^{(C)},n_E^{(C)},n_F^{(C)}) = (n_T^{(C)},n_E^{(C)},n_T^{(C)}-n_E^{(C)})$ - number of (total therapists, embedded therapists, full-time therapists) in Centre C
\item $\boldsymbol{n_T^{C}} = (n_T^{(1)},\ldots,n_T^{(n_C)})$ - $n_C\times 1$ vectors of total number of therapists in each centre
\item $\boldsymbol{n_E^{C}}=a_E(n_T^{(1)},\ldots,n_T^{(n_C)}) = (n_E^{(1)},\ldots,n_E^{(n_C)})$ and $\boldsymbol{n_F^{C}}=(1-a_E)(n_T^{(1)},\ldots,n_T^{(n_C)})=(n_F^{(1)},\ldots,n_F^{(n_C)})$ denote number of embedded therapists and number of full-time therapists, across all $n_C$ centres, respectively
\item $n_{R_j}^{(C)}$ - Capacity of therapist `j' in centre `C' in each batch
\item $\boldsymbol{n_R^{(C)}}$= $(n_{R1}^{(C)},\ldots,n_{Rn_T}^{(C)})$
\item $n_R$ = $min(\boldsymbol{n_R^{(1)}},\boldsymbol{n_R^{(2)}},\ldots,\boldsymbol{n_R^{(n_C)}})$
\item  $\bc^{(C)} = n_B \boldsymbol{n_R^{(C)}} = (\bc_E^{(C)},\bc_F^{(C)}) = (c_{1}^{(C)},\ldots,c_{n_T^{(C)}}^{(C)})$ - $n_T^{(C)} \times 1$ vector of capacities of embedded followed by full-time therapists throughout the trial in Centre C
\item $(E_E,E_F)$ - Expected capacity of (Embedded, Research) therapists throughout the trial
\item $(\sigma^2_E,\sigma^2_F)$ - Variance of the capacity of (Embedded, Research) therapists throughout the trial
\item $N_B^{(C)} =\frac{N}{n_B}\frac{E_E n_E^{(C)}+E_F n_F^{(C)}}{\sum_{i=1}^{n_C}E_E n_E^{(i)}+E_F n_F^{(i)}} $ - target number of patients in a Centre-Batch combination based on the capacities of the therapists in that centre
\item $\boldsymbol{n_{T-IBD}} = (n_{T-IBD}^{(1)},\ldots,n_{T-IBD}^{(n_C)})$  - $n_C \times 1$ vector of number of therapists selected in a Centre-Batch combination in strategy N1. The bold font is dropped if $n_C=1$
\item $frac$ -  Permissible over-recruitment above the target batch size in strategy N2
\item $a$ - parameter of the distribution of therapist availability.
\item $(frac_E, frac_F)$ - both follows $\text{Uniform}[a, 1]$ and denote the fraction of the therapists available out of $(n_E^{(C)},n_F^{(C)})$ in a Centre-Batch combination
\end{itemize}

\subsection{Statistical Models}

Let $(i,j,k,l)$ be the (patient, therapist, batch, centre). The following two models will be used in the simulations to generate and analyse the data.
The first model is taken from \cite{Walwyn_2026}. The second model is proposed here. 

\begin{itemize}
\item[Model-1:]  \begin{align*}
 Y_{ijkl} &= \delta_{0} + \delta_{1}I_{ijkl}+u_{1j}+ u_{2k}+ u_{3jk}+u_{4l}+u_{5kl}+v_{1j}I_{ijkl}\\\nonumber
    & \qquad +v_{2k}I_{ijkl}+ v_{3jk} I_{ijkl}+v_{4l} I_{ijkl}+v_{5kl} I_{ijkl} + e_{ijkl}, \label{model.1}
\end{align*}

where, \begin{itemize}
\item $\delta_0$ - overall mean (fixed)
\item $\delta_1$ - intervention effect (fixed)  
\item $I_{ijkl}$ - $\{-1,1\}$ indicator for interventions
\item $u_{1j}\sim  \mathcal{N}(0, \sigma_{u_{1}}^2)$ therapist effect (random)  
\item $u_{2k}\sim  \mathcal{N}(0, \sigma_{u_{2}}^2)$ batch effect (random) 
\item $u_{3jk}\sim  \mathcal{N}(0, \sigma_{u_{3}}^2)$ therapist-by-batch effect (random)
\item $u_{4l}\sim  \mathcal{N}(0, \sigma_{u_{4}}^2)$ centre effect (random)
\item $u_{5kl}\sim  \mathcal{N}(0, \sigma_{u_{5}}^2)$ centre-by-batch effect (random)
\item $v_{1j}\sim  \mathcal{N}(0, \sigma_{v_{1}}^2)$ intervention-by-therapist effect (random) 
\item $v_{2k}\sim  \mathcal{N}(0, \sigma_{v_{2}}^2)$ intervention-by-batch effect (random) 
\item $v_{3jk}\sim  \mathcal{N}(0, \sigma_{v_{3}}^2)$ intervention-by-therapist-by-batch effect (random)
\item $v_{4l}\sim  \mathcal{N}(0, \sigma_{v_{4}}^2)$ intervention-by-centre effect (random)
\item $v_{5kl}\sim  \mathcal{N}(0, \sigma_{v_{5}}^2)$ intervention-by-centre-by-batch effect (random)
\item $e_{ijkl}\sim  \mathcal{N}(0, \sigma_{e}^2)$ patient level random error (random)
\end{itemize}

\item[Model-2:] \begin{align*}
    Y_{ijl} &= \delta_{0} + \delta_{1} I_{ijl} + \delta_2 t_{ijl} + \delta_3 t_{ijl} I_{ijl} + w_{1j} + w_{2j} t_{ijl} + w_{3l} + w_{4l} t_{ijl} \nonumber \\
    &\qquad + z_{1j} I_{ijl} + z_{2j} t_{ijl} I_{ijl} + z_{3l} I_{ijl} + z_{4l} t_{ijl} I_{ijl} + e_{ijl},
\end{align*}

where, \begin{itemize}
\item $\delta_0$ - overall mean (fixed)
\item $\delta_1$ - intervention effect (fixed) 
\item $t_{ijl}$ - number of patients treated by therapist $j$ before patient $i$ in centre $l$ (a variable for time in therapist)
\item $\delta_2$ - time in therapist effect (fixed)
\item $\delta_3$ - intervention-by-time in therapist effect (fixed)
\item $w_{1j}\sim  \mathcal{N}(0, \sigma_{w_{1}}^2)$ - therapist effect (random)
\item $w_{2j}\sim  \mathcal{N}(0, \sigma_{w_{2}}^2)$ - therapist learning effect over time $t_{ijl}$ (random)
\item $w_{3l}\sim  \mathcal{N}(0, \sigma_{w_{3}}^2)$- centre effect (random)
\item $w_{4l}\sim  \mathcal{N}(0, \sigma_{w_{4}}^2)$ - centre learning effect over time $t_{ijl}$ (random)
\item $z_{1j}\sim  \mathcal{N}(0, \sigma_{z_{1}}^2)$- intervention-by-therapist effect (random)
\item $z_{2j}\sim  \mathcal{N}(0, \sigma_{z_{2}}^2)$ - intervention-by-therapist learning effect over time $t_{ijl}$ (random)
\item $z_{3l}\sim  \mathcal{N}(0, \sigma_{z_{3}}^2)$ - intervention-by-centre effect (random)
\item $z_{4l}\sim  \mathcal{N}(0, \sigma_{z_{4}}^2)$ - intervention-by-centre learning effect over time $t_{ijl}$ (random)
\item $e_{ijl}\sim  \mathcal{N}(0, \sigma_{e}^2)$ patient level random error (random)
\end{itemize}

\end{itemize}

\section{Data Generating Mechanisms}

 For data generation and analysis, do the following for $n_{sim}$ number of simulations. 
\begin{enumerate}
\item Let $(N,n_C,n_B,a_E,a,E_E,E_F,\sigma^2_E,\sigma^2_F,\boldsymbol{n_{T-IBD}}, frac, a_E,a)$ are given. 
\item Therapist overall capacities ($\bc_E^{(C)}$ and $\bc_F^{(C)}$) are assumed to follow, $\bc_E^{(C)} \sim \mathcal{TN}(0,\infty,E_E,\sigma_E^2)$ and $\bc_F^{(C)} \sim \mathcal{TN}(0,\infty,E_F,\sigma_F^2)$, where $E_E$ and $E_F$ are the expected capacities of a embedded and full-time therapist, respectively. Here, $\mathcal{TN}(0,\infty,\mu,\tau^2)$ denotes a truncated normal distribution, truncated at 0 with underlying (pre-truncation) mean $\mu$ and variance $\tau^2$. We assume $E_E < E_F$ to reflect the practical scenario in which a embedded therapist treats, on average, fewer patients than a full-time therapist. Note that the expected capacities ($E_E,E_F$) are chosen in such a way that the number of therapists recruited in the trial are sufficient to treat the total sample size. Furthermore, the capacity of embedded therapists can vary substantially depending on the time they dedicate to the trial. In some cases, it can be as low as treating only a single patient, whereas in certain special scenarios, it may even exceed the capacity of a full-time therapist. Conversely, since full-time therapists allocate all their working hours to the trial, their capacity remains relatively stable. Therefore, we assume $\sigma_E^2 \gg \sigma_F^2$, indicating substantially greater variability in the treatment capacity of embedded therapists compared to full-time therapists.

\item For a combination of aforementioned parameters, generate a design using each of the six randomisation strategies, $\{E1, E2, e3, N1, N2, N3\}$.
 \item For each simulation do the following;
\begin{enumerate}
  \item Simulate data from model-1 or 2.
  \item Fit model-1 or 2 using \texttt{lmerTest} package in R software.
 \item Extract and store the indicators (1 or 0) of failure in model fitting for the given model, a singular model fit and Type-2 error.
 \item Extract and store the estimates of, 
 \begin{itemize}
     \item the fixed effects $(\delta_0,\delta_1)$ and random effect variances $(\sigma^2_{u_1},\sigma^2_{u_2},\ldots, \sigma^2_{e})$ for model-1
     \item the fixed effects $(\delta_0,\delta_1\delta_2,\delta_3)$ and random effect variances $(\sigma^2_{w_1},\sigma^2_{w_2},\ldots, \sigma^2_{e})$ for model-2.
 \end{itemize}
\item Compute and extract the root mean squared error (RMSE) of the fit for prediction.

\end{enumerate}
\item For every randomisation strategy $\{E1, E2, E3, N1, N2, N3\}$, compute the average values of all the extracted indicators, estimates and RMSE.
\item Present the results using tables and figures.
\end{enumerate}

\noindent For simplicity and a better interpretation of results, we shall not run the simulations for all the possible combinations of the parameter spaces, instead we shall stick to the following practically relevant trial scenarios.
A brief summary of all the parameter values is given in Table \ref{Table.parameter.scenarios}. The tabular represented as `Design' lists the design parameter values and `Analysis' lists the model parameter values.
\medskip

\subsection{Trial Scenarios}

\subsubsection{Single centre trial:}

\begin{figure}[ht]
\centering
\includegraphics[width=12cm]{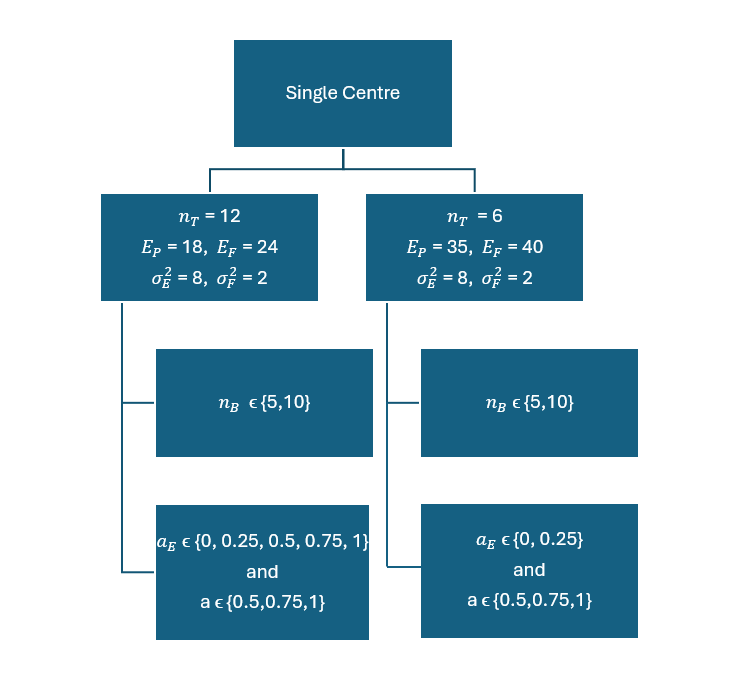}
\caption{Flowchart of the scenarios of the single centre trial}
\label{fig.flow.single.centre} 
\end{figure}

To mimic the situation of a single centre trial we assume; 
\begin{itemize}
  \item The total sample size is fixed at $N = 240$.
    
    \item The number of therapists is fixed to $n_T \in \{12, 6\}$ to mitigate less and more control on therapists.
    
    \begin{enumerate}
        \item $n_T = 12$ denotes that the population of therapists is more variable in terms of their status. That is, not all therapists are recruited full-time in the trial. Therefore, in this case, the status of a therapist (part/full) has five choices, $a_En_T$ where $a_E \in \{0,0.25,0.5,0.75,1\}$. For instance, $a_E = 0.75$ denotes that $75\%$ for the recruited therapists are working embedded and remaining $25\%$ therapists are working full-time. Further, $\bc_E \sim \mathcal{TN}(0,\infty,E_E, 8)$ and $\bc_F \sim \mathcal{TN}(0,\infty,E_F, 2)$. Here, $E_E = 18$, $E_F = 24$ are chosen such that the desired sample size of $240$ can be recruited using these $12$ therapists.
        
        \item $n_T = 6$ denotes that the trial committee has more control over therapists and the population of therapists is fairly similar in terms of their status. That is, almost all therapists are recruited as full-time therapists and work specifically for the trial. Therefore, accordingly $a_E \in \{0,0.25\}$. Further, $\bc_E \sim \mathcal{TN}((0,\infty,E_E, 8)$ and $\bc_F \sim \mathcal{TN}(0,\infty,E_F, 2)$. Here $E_E = 35$, $E_F = 40$ are chosen such that the desired sample size of $240$ can be recruited using these $6$ therapists.
    \end{enumerate}
    
    \item The number of batches is $n_B \in \{5, 10\}$ to mitigate high and low values of the duration of recruitment.
    
    \item To demonstrate the availability of therapists in each batch, it is assumed that $\text{frac}_E, \text{frac}_F \sim \text{Uniform}[a, 1]$. That is, at least $100a\%$ of the embedded and full-time therapist are always available in a batch. We shall assume that $a \in \{0.5, 0.75, 1\}$.
    
    \item For strategy N1, the number of therapists chosen in every batch is $n_T^{IBD} = a n_T$, that is, the minimum number of therapists guaranteed to be available across all batches.
     
    \item Each combination of $a_E$, $n_B$, and $a$ is used for the simulations according to the value of $n_T$, while keeping other parameters fixed. For more details, refer to the flowchart given in Figure~\ref{fig.flow.single.centre}.
    \item To simulate the data:
    \begin{enumerate}
  \item The fixed effects parameters are specified as $(\delta_0,\delta_1,\delta_2,\delta_3,\delta_4) = (0,0.4,0.1,0.1,0.3)$.
\item The random effects parameters for model-1 are fixed at: $$(\sigma_{u_1}^2,\sigma_{u_2}^2,\sigma_{u_3}^2,\sigma_{u_4}^2,\sigma_{u_5}^2,\sigma_{v_1}^2,\sigma_{v_2}^2,\sigma_{v_3}^2,\sigma_{v_4}^2,\sigma_{v_5}^2,\sigma_{e}^2) = (0.1,0.1,0.1,0,0,0.15,0.1,0.1,0,0,0.35).$$
\item The random effects parameters for model-2 are fixed at: $$(\sigma_{w_1}^2,\sigma_{w_2}^2,\sigma_{w_3}^2,\sigma_{w_4}^2,\sigma_{z_1}^2,\sigma_{z_2}^2,\sigma_{z_3}^2,\sigma_{z_4}^2,\sigma_{e}^2) = (0.1,0.1,0,0,0.15,0.1,0,0,0.35).$$
    \end{enumerate}

\end{itemize}
\medskip

\noindent \textbf{Multi centre trial:}
To mimic the situation of a medium and a large multi-centre trial, we assume two different set of parameters.
\medskip

\subsubsection{Medium multi-centre trial}

\begin{itemize}
  \item The total sample size is fixed at $N = 640$.
  \item The number of centres recruited in the trial is $n_C = 6$. 
 \item The number of therapists $n_T^{(C)}$ for $C=1,\ldots,6$, follows $n_T^{(C)} \sim Poisson(8)$ or $n_T^{(C)} \sim Poisson(4)$ to mitigate less and more control on therapists in each centre. Here, $Poisson(\lambda)$ denotes a Poisson Distribution with mean $\lambda$.
    
    \begin{enumerate}
\item $n_T^{(C)} \sim Poisson(8)$ denotes that, on an average there are $8$ therapists per centre. It also indicates that the population of therapists is more variable in terms of their status. That is, in every centre, not all therapists are recruited full-time in the trial. Therefore, the therapist status $a_En_T^{(C)}$ for $C=1,\ldots,6$ has five choices, $a_E \in \{0,0.25,0.5,0.75,1\}$. Further, for $C=1,\ldots,6$, the capacities of a embedded and full-time therapist follows $\bc_E^{(C)} \sim \mathcal{TN}(0,\infty,E_E, 6)$ and $\bc_F^{(C)} \sim \mathcal{TN}(0,\infty,E_F, 2)$, respectively. Here, $E_E = 12$, $E_F = 20$ are chosen such that the desired sample size of $640$ can be recruited assuming average of $8$ therapists in each centre. Note that $\sigma^2_E=6$ here instead of $8$ as compared to the single centre trial. This is reasonably chosen since expected capacity has reduced to $E_E=12$. It ensures that at least 1 patients per batch is assigned to a therapist if they are available to treat patients in that batch.

\item $n_T^{(C)} \sim Poisson(4)$ denotes that on an average there are $4$ therapists per centre and the population of therapists is less variable in terms of their status. That is, in every centre, almost all the therapists are recruited full-time in the trial. Therefore, the therapist status $a_En_T^{(C)}$ for $C=1,\ldots,6$ has two choices, $a_E \in \{0,0.25\}$. Moreover, for $C=1,\ldots,6$, the capacities of a embedded and full-time therapist follows $\boldsymbol{c_E^{(C)}} \sim \mathcal{TN}(0,\infty,E_E, 6)$ and $\boldsymbol{c_F^{(C)}} \sim \mathcal{TN}(0,\infty,E_F, 2)$, respectively. Here $E_E = 25$, $E_F = 30$ are chosen such that the desired sample size of $640$ can be recruited assuming average of $4$ therapists in each centre. 
\end{enumerate}
    
 \item The number of batches is $n_B \in \{5, 8\}$ to mitigate high and low values of the duration of recruitment. Note that $n_B=8$, instead of 10 is chosen to ensure that every therapist has capacity to treat at east 1 patient using each intervention in a batch, i.e., $n_R^{(C)} = c_E^{(C)}/n_In_B \geq 1$.
 
 \item To demonstrate the availability of therapists in each centre-batch combination, it is assumed that $\text{frac}_E, \text{frac}_F \sim \text{Uniform}[a, 1]$ where, $a \in \{0.5, 0.75, 1\}$.
\item For strategy N1, the vector of number of therapists chosen in every batch based on the centre is $ \boldsymbol{n_{T-IBD}}  = a (n_T^{(1)},\ldots,n_T^{(6)})$. That is, the minimum number of therapists guaranteed to be available across all batches in the respective centre.
\item Each combination of $a_E$, $n_B$, and $a$ is used for the simulations according to the distribution of $n_T^{(C)}$, while keeping other parameters fixed. For more details, refer to the flowchart given in Figure \ref{fig.flow.multi.centre}.
\end{itemize}

\subsubsection{Large multi-centre trial}

\begin{figure}[ht]
\centering
\includegraphics[width=17cm]{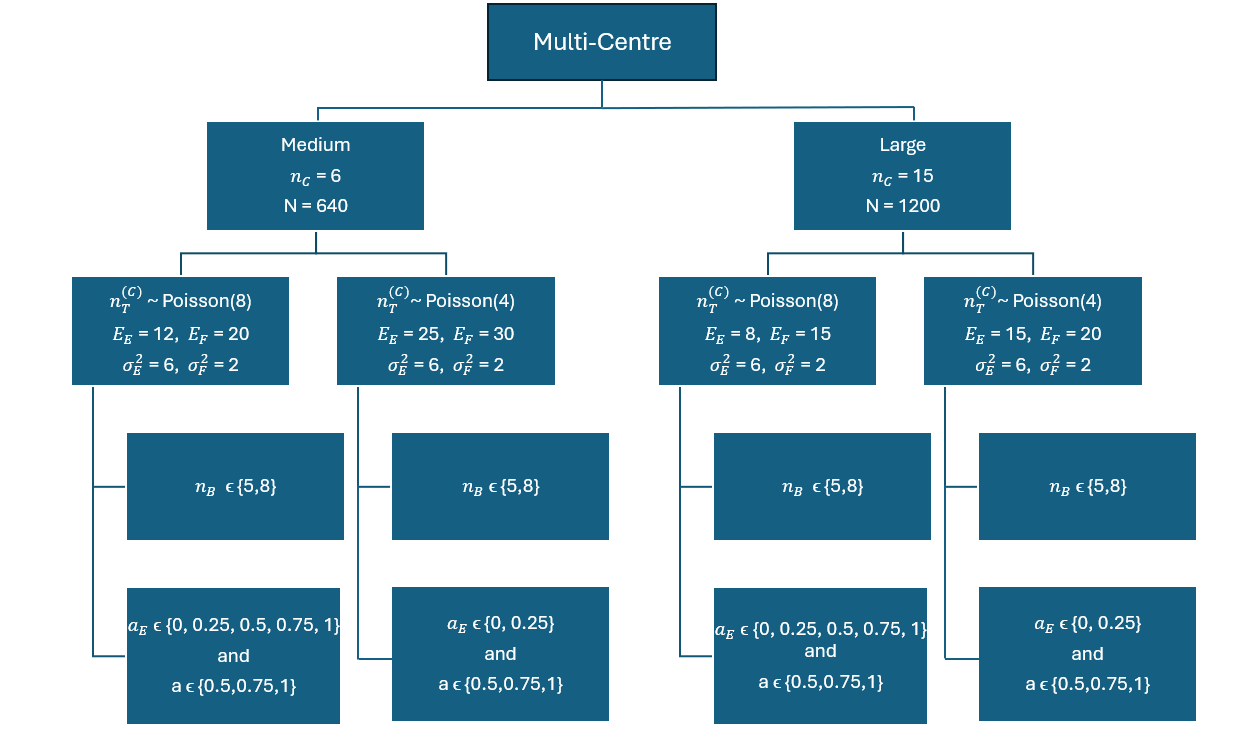}
\caption{Flowchart of the scenarios of a multi-centre trial}
\label{fig.flow.multi.centre} 
\end{figure}

\begin{itemize}
  \item The total sample size is fixed at $N = 1200$.
  \item The number of centres recruited in the trial is $n_C = 15$. 
 \item The number of therapists $n_T^{(C)}$ for $C=1,\ldots,15$, follows $n_T^{(C)} \sim Poisson(8)$ or $n_T^{(C)} \sim Poisson(4)$ to mitigate less and more control on therapists in each centre.
    
\begin{enumerate}
\item $n_T^{(C)} \sim Poisson(8)$ denotes that in every centre, not all therapists are recruited full-time in the trial. Therefore, the therapist status $a_En_T^{(C)}$ for $C=1,\ldots,15$ has four choices, $a_E \in \{0,0.25,0.5,0.75,1\}$. Further, for $C=1,\ldots,15$, the capacities of a embedded and full-time therapist follows $\bc_E \sim \mathcal{TN}(0,\infty,E_E, 6)$ and $\bc_F \sim \mathcal{TN}(0,\infty,E_F, 2)$, respectively, where $E_E = 8$, $E_F = 15$.  The average capacities of embedded and full-time therapists are chosen such that the desired sample size of $1200$ can be recruited assuming average of $8$ therapists in each centre.
\item $n_T^{(C)} \sim Poisson(4)$ denotes that in every centre, almost all the therapists are recruited full-time in the trial. Therefore, the therapist status $a_En_T^{(C)}$ for $C=1,\ldots,n_C$ has two choices, $a_E \in \{0,0.25\}$. Further, the capacities of a embedded and full-time therapist follows $\boldsymbol{c_E^{(C)}} \sim \mathcal{TN}(0,\infty,E_E, 6)$ and $\boldsymbol{c_F^{(C)}} \sim \mathcal{TN}(0,\infty,E_F, 2)$, respectively, where $E_E = 15$, $E_F = 20$.  The average capacities of embedded and full-time therapists are chosen such that the desired sample size of $1200$ can be recruited assuming average of $4$ therapists in each centre.
\end{enumerate}
    
 \item The number of batches is $n_B \in \{5, 8\}$ to mitigate high and low values of the duration of recruitment.
 \item The availability of therapists in each centre-batch combination, it is assumed that $\text{frac}_E, \text{frac}_F \sim \text{Uniform}[a, 1]$ where, $a \in \{0.5, 0.75, 1\}$.
\item For IBD, $ \boldsymbol{n_{T-IBD}}  = a (n_T^{(1)},\ldots,n_T^{(15)})$. 
\item Each combination of $a_E$, $n_B$, and $a$ is used for the simulations according to the distribution of $n_T^{(C)}$, while keeping other parameters fixed. For more details, refer to the flowchart given in Figure \ref{fig.flow.multi.centre}.
  \item To simulate the data, the parameters are identical for both medium and large multi-centre trial. In particular
    \begin{enumerate}
  \item The fixed effects parameters are specified as $(\delta_0,\delta_1,\delta_2,\delta_3,\delta_4) = (0,0.4,0.1,0.1,0.3)$.
\item The random effects parameters for model-1 are fixed at $$(\sigma_{u_1}^2,\sigma_{u_2}^2,\sigma_{u_3}^2,\sigma_{u_4}^2,\sigma_{u_5}^2,\sigma_{v_1}^2,\sigma_{v_2}^2,\sigma_{v_3}^2,\sigma_{v_4}^2,\sigma_{v_5}^2,\sigma_{e}^2) = (0.1,0.1,0.05,0.1,0.05,0.1,0.1,0.05,0.1,0.05,0.2).$$
\item The random effects parameters for model-2 are fixed at $$(\sigma_{w_1}^2,\sigma_{w_2}^2,\sigma_{w_3}^2,\sigma_{w_4}^2,\sigma_{z_1}^2,\sigma_{z_2}^2,\sigma_{z_3}^2,\sigma_{z_4}^2,\sigma_{e}^2) = (0.1,0.05,1,0.05,0.1,0.05,0.1,0.05,0.2).$$
    \end{enumerate}
  
\end{itemize}

\section{Coding and Execution}

\begin{enumerate}
    \item The simulations are coded and implemented in \texttt{R}. The data generated from a simulation is fitted with respective model using residual maximum likelihood (REML) approach. The results are averaged over $n_{\text{sim}} = 10{,}000$ simulations. Since each simulation is independent of each other, to reduce computational time, simulations are executed parallelly using distinct seeds on a high performance computer (HPC) available at University of Leeds.
    
    \item Sequential seeds are used for several job submissions on the HPC cluster. The sequence of seeds always starts from $1001$ and ends at $1000+n_{jobs}$, where $n_{jobs}$ is the number of jobs submitted at a time. For example, when submitting $100$ jobs with $100$ simulations each, assign seeds as $1001, 1002, \dots, 1100$, respectively.
\end{enumerate}

\noindent \textbf{Estimands and other targets:} 
For every randomisation stratergy $\{E1, E2, E3, N1, N2, N3\}$ and every combination of $N$, distribution of $n_T^{(C)}$, $n_B$, $a_E$, $a$ and model (1 or 2) listed in the previous section, we shall compute the average values of the following indicators over $n_{\text{sim}} = 10{,}000$ simulations. Note that the average is carried over the simulations where model fit is successful.
\begin{itemize}
 \item Randomisation time to target - Number of batches (periods) required to achieve the target sample size.
 \item Model failures.
\item Estimates of all the fixed and random effects in both model-1 and 2.
\item Boundary estimates (zero estimates of non-zero random effects).
\item Type-2 error.
\item Root mean squared error (RMSE), where \begin{equation*}
    \mathrm{RMSE}
    \;=\;
    \sqrt{\frac{1}{N}\sum_{i=1}^{N}
    \bigl(Y_{ijkl}-\hat{Y}_{ijkl}\bigr)^2}.
\end{equation*}
\end{itemize}

\noindent \textbf{Performance indicators to compare the strategies:}
\begin{enumerate}
\item Cost efficiency - Randomisation time to target
\item Robust estimation - 
\begin{enumerate}
 \item \textit{model convergence}, measured by the proportion of REML fits that failed to converge; 
 \item \textit{numerical stability}, measured by the proportion of REML fits with boundary estimates defined as zero estimates of variance components whose true values are non-zero; 
 \item \textit{unbiasedness}, measured by the average estimated values of the treatment effects; 
 \item \textit{precision}, measured by the average estimated value of the standard error of the intervention effect $SE(\hat{\delta}_1)$; 
 \item \textit{inferential accuracy}, measured by the Type~II error rate for testing the intervention effect; 
 \item \textit{predictive accuracy}, measured by the RMSE of prediction.
\end{enumerate}

\end{enumerate}

\section{Reporting and Analysis}

In each simulation, a design is generated by selecting one of the parameter combinations listed in the “Design’’ panel of Table~\ref{Table.parameter.scenarios} and applying one of the randomisation strategy. For parameters whose values are presented as pairs separated by `\&’, a single consistent choice must be made across all such parameters: if the first value is selected for one parameter, then the first value must be selected for every parameter containing an `\&’ pair; similarly, if the second value is selected for any such parameter, the second value must be selected for all of them.  
In contrast, parameters presented as sets within curly brackets may be chosen freely and independently from their respective sets.

Once a design is specified using these parameter choices, data are generated from either Model~1 or Model~2 using the corresponding parameter values indicated in the “Analysis’’ panel of Table~\ref{Table.parameter.scenarios}. The same model from which the data were generated is then fitted using the \texttt{lmer} function from the \texttt{lmerTest} package in \textsf{R}, and the resulting estimates are stored.

This process (design generation, data simulation, model fitting, and result storage) is repeated 10{,}000 times for each scenario. The final results are obtained by averaging the target estimands across the 10{,}000 replicates, and all scenarios are reviewed and reported comprehensively.

\begin{table}[ht]
\centering
\setlength{\tabcolsep}{3pt}  
    \begin{tabular}{ c| c| c| c|  c| c|  c| c|  c| c| c| c| c| c| c c c}
  \hline
\multirow{4}{*}{
  \parbox[c]{1.5cm}{
    \centering
    Design }}&Trial& $N$&$n_C$& $E(n_T^{(C)})$ & $n_B$ & $a_E$ &$E_E$ & $E_F$ &$\sigma_E^2$& $\sigma_F^2$& $a$& $n_{T\text{-IBD}}^{(C)}$\\
    \cline{2-13}
& Single &240 & 1 & $12 \& 6$ & $\{5, 10\}$ & $\{0, 0.25, 0.5, 0.75, 1\}$ &$18\&35$ &$24 \& 40$&8&2& $\{0.5, 0.75, 1\}$& $a n_T^{(C)}$ \\
 &Medium &640 & 6 & $8 \& 4$ & $\{5, 8\}$ & $\{0, 0.25, 0.5, 0.75, 1\}$ &$12\&25$ &$30\&35$&6&2& $\{0.5, 0.75, 1\}$ & $a n_T^{(C)}$ \\
&Large &1200 & 15 & $8 \& 4$ & $\{5, 8\}$ & $\{0, 0.25, 0.5, 0.75, 1\}$ &$8\&15$ &$15\&20$&6&2& $\{0.5, 0.75, 1\}$& $a n_T^{(C)}$ \\
\hline
\end{tabular}
\begin{tabular}{ c| c| c| c|  c| c|  c| c|  c| c| c| c| c| c| c| c| c| c| c|}
  \hline
\multirow{8}{*}{
  \parbox[c]{1.5cm}{
    \centering
     Analysis}}&  
&Trial&$\delta_0$&$\delta_1$&$\delta_4$& $\sigma_{u_1}^2$&$\sigma_{u_2}^2$&$\sigma_{u_3}^2$&$\sigma_{u_4}^2$&$\sigma_{u_5}^2$&$\sigma_{v_1}^2$&$\sigma_{v_2}^2$&$\sigma_{v_3}^2$&$\sigma_{v_4}^2$&$\sigma_{v_5}^2$&$\sigma_{e}^2$\\
\cline{2-17}
&\multirow{3}{*}{
  \parbox[c]{1.5cm}{
    \centering
     Model-1}}&Small &0&0.4&0.3&0.1&0.1&0.1&0&0&0.15&0.1&0.1&0&0&0.35\\
 &&Medium&0&0.4&0.3 &0.1&0.1&0.05&0.1&0.05&0.1&0.1&0.05&0.1&0.05&0.2\\
  &&Large&0&0.4&0.3 &0.1&0.1&0.05&0.1&0.05&0.1&0.1&0.05&0.1&0.05&0.2\\
\cline{2-17}
&&&$\delta_0$&$\delta_1$&$\delta_2$&$\delta_3$&$\delta_4$&$\sigma_{w_1}^2$&$\sigma_{w_2}^2$&$\sigma_{w_3}^2$&$\sigma_{w_4}^2$&$\sigma_{z_1}^2$&$\sigma_{z_2}^2$&$\sigma_{z_3}^2$&$\sigma_{z_4}^2$&$\sigma_{e}^2$\\
\cline{2-17}
&\multirow{3}{*}{
  \parbox[c]{1.5cm}{
    \centering
     Model-2}}&Small &0&0.4&0.1&0.1&0.3&0.1&0.1&0&0&0.15&0.1&0&0&0.35\\
 &&Medium &0&0.4&0.1&0.1&0.3&0.1&0.05&0.1&0.05&0.1&0.05&0.1&0.05&0.2\\
  &&Large &0&0.4&0.1&0.1&0.3&0.1&0.05&0.1&0.05&0.1&0.05&0.1&0.05&0.2\\
  \hline
\end{tabular}
\caption{This table lists all the parameter values used in the design for the simulations studies. A comprehensive simulation protocol is given in supplementary file H. The first tabular is corresponding to the parameters used in the randomisation strategies and the second tabular is for the true parameter values in model-1 and 2. For brief definitions of the parameters, refer to Section \ref{sec.notation}. For the parameters $E_E$ and $E_F$, the symbol `\&' indicates that one of the two values is selected conditional on the corresponding value of $E(n_T^{(C)})$. For example, in the single‑centre trial scenario, when $E(n_T^{(C)})=12$, the associated values $E_E=18$ and $E_F=24$ are used. In contrast, the parameters $a_E$ and $a$ are varied independently over their specified ranges; all values in these sets are admissible and need not be paired with any particular value of $E(n_T^{(C)})$.}\label{Table.parameter.scenarios}
\end{table}

The reporting of the findings associated with the three questions raised in Section~\ref{sec.aims.of.sim} is presented in Section~5.2 of the main article. In particular, Section~5.2.1 addresses Question~2 and Section~5.2.2 and 5.2.3 addresses Question~3.  
The answer to Question~1 is two fold (relating to the cost efficiency and the statistical properties of non‑random assignment) and is embedded within the results for Questions~2 and~3.  

Due to space limitations, only a subset of the single‑centre trial results is presented in the main paper.  
The remaining single‑centre trial results, together with all results from the multi‑centre trial scenarios, are provided comprehensively in Supplementary Files~A–G.

Finally, Section~6 summarises the key findings, which are intended to guide trialists designing and conducting complex‑intervention trials.

\clearpage

\bibliographystyle{apalike}
\bibliography{Bibfile}


\title{Supplementary File D: Main Results from Medium Multi-Centre Trial}
\date{}

\maketitle

This supplementary file contains the complete set of simulation results for the medium multi-centre trial across varying levels of embedded therapist proportion and therapist availability. Throughout the tables, $N, n_B$ and $n_T^{(C)}$ denote the total sample size, number of batches and number of therapists within a centre, respectively. The proportion of embedded therapists within a centre is $a_E$. Parameter $a$ denotes that at least $100a\%$ of the total therapists available within a centre in each batch. $\hat{\delta_1}$ and $SE(\hat{\delta_1})$ denote the average estimate and standard error of the intervention effect $\delta_1$. $\hat{\sigma}_{u_1}^2$ and $\hat{\sigma}_{v_1}^2$, T-2 Error, Boundary, Fail and RMSE denote the average estimates of $\sigma_{u_1}^2$, $\sigma_{v_1}^2$, type-II error rate, boundary estimate rate, REML failure rate and root mean square error, respectively. Similarly, $\hat{\sigma}_{w_1}^2$ and $\hat{\sigma}_{z_1}^2$ denote the average estimates of $\sigma_{w_1}^2$ and $\sigma_{z_1}^2$, respectively.

Table~\ref{table.time.to.target} reports the randomisation time and target for all randomisation methods considered. Tables~\ref{table.single.1}--\ref{table.single.6} present parameter estimation results under Model~1, while Tables~\ref{table.single.7}--\ref{table.single.12} report the corresponding results for Model~2. 

Finally, Tables~\ref{table.boundary.1} and \ref{table.boundary.2} report the proportions of boundary estimates obtained under REML when estimating the variance components from Model~1 and 2, respectively. Throughout the tables, the labels I, T, B, and C denote the Intervention, Therapist, Batch, and Centre effects, respectively. In the Model~2 tables, the subscript \textit{cor} denotes the correlation between the main (intercept) and learning (slope) effects.

\begin{table}[ht]
\centering \caption{Randomisation to time and target against the fraction of therapist availability for the medium trial}
\setlength{\tabcolsep}{2pt}  
\begin{tabular}{p{2cm}|c|ccc|ccc|ccc|ccc|ccc}
\hline
&&\multicolumn{15}{c}{Randomisation to Time and Target: Number of batches actually required to meet target} \\
\cline{2-17}
 & $a_E$ & \multicolumn{3}{c|}{$0$}  
  & \multicolumn{3}{c|}{$0.25$} 
 & \multicolumn{3}{c|}{$0.5$} 
 & \multicolumn{3}{c|}{$0.75$} 
 & \multicolumn{3}{c}{$1$} \\
\cline{2-17}
&Design $\downarrow$ a $\rightarrow$ & $0.5$ & $0.75$ & $1$ & $0.5$ & $0.75$ & $1$ & $0.5$ & $0.75$ & $1$ & $0.5$ & $0.75$ & $1$&$0.5$ & $0.75$ & $1$\\
\hline
\multirow{6}{*}{
  \parbox[c]{2cm}{
    \centering
    $N=640$\\
    $n_B=5$\\  
$n_T^{(C)} \sim Poisson(8)$  }}  
& E1  & -&-&4.35&-&-&7.29&-&-&7.32&-&-&7.28&-&-&7.28\\
 & E2      &5.25&4.63&4.35&8.41&7.63&7.29&8.33&7.46&7.32&8.31&7.54&7.28&8.88&7.79&7.28\\
& E3 &4.69&4.15&3.89&4.78&4.33&4.13&5.13&4.59&4.51&5.49&5.03&4.87&6.70&5.88&5.50\\
&N1&7.68&5.32&3.89&8.35&5.72&4.13&9.09&6.25&4.51&9.76&6.76&4.87&11.24&7.66&5.50\\
 & N2     &5.06&4.62&4.43&5.11&4.80&4.61&5.29&4.95&4.84&5.56&5.22&5.09&6.72&5.91&5.58\\
 & N3 &4.69&4.15&3.89&4.79&4.33&4.13&5.24&4.60&4.51&5.78&5.10&4.87&6.85&5.93&5.50\\ 
 \hline\multirow{6}{*}{\parbox[c]{2cm}{
    $N=640$\\   
    $n_B=5$\\   
$n_T^{(C)} \sim Poisson(4)$  }}
& E1   &-&-&5.67&-&-&8.01& -&-&-& -&-&-& -&-&- \\
 & E2  &6.45&5.81&5.67&8.83&8.15&8.01& -&-&-& -&-&-& -&-&- \\
& E3&5.81&5.24&5.11&5.74&5.29&5.23& -&-&-& -&-&-& -&-&- \\ 
  & N1 &10.48&7.35&5.11&10.85&7.53&5.23& -&-&-& -&-&-& -&-&- \\
& N2&5.88&5.35&5.21&5.82&5.39 &5.32& -&-&-& -&-&-& -&-&- \\
 & N3 &5.81&5.24&5.11&5.74&5.29&5.23& -&-&-& -&-&-& -&-&- \\
 \hline 
 \multirow{6}{*}{
  \parbox[c]{2cm}{
    \centering
    $N=640$\\
    $n_B=8$\\
$n_T^{(C)} \sim Poisson(8)$ }}
& E1 &-&-&7.26& -&-&7.28&-&-&7.29&-&-&7.30&-&-&7.27\\
 & E2 &8.88&7.80&7.26&8.43&7.60&7.28&8.29&7.46&7.29&8.33&7.54&7.30&8.88&7.79&7.27\\
& E3 &8.57 &7.52&7.02&8.14&7.35&7.05&8.04&7.23&7.07 &8.09&7.33&7.10&8.67&7.60&7.10\\
  & N1 &14.49&9.83&7.02&14.51&9.83&7.05&14.33&9.90&7.07&14.58&9.92&7.10&14.64&9.94&7.10\\
& N2 &8.72&7.72&7.38&8.29&7.59&7.39&8.17&7.50&7.40&8.22&7.56&7.42&8.79&7.76&7.41\\
 & N3&8.61&7.53&7.02&8.17&7.36&7.05&8.07&7.23&7.07&8.12&7.34&7.10&8.70&7.61&7.10\\
\hline
\multirow{6}{*}{
  \parbox[c]{2cm}{
    \centering
    $N=640$\\
    $n_B=8$\\
$n_T^{(C)} \sim Poisson(4)$  }}
& E1       &-&-&7.42&12.69&12.75&12.75& -&-&-& -&-&-& -&-&- \\
 & E2    & 8.55&7.68&7.42&14.10&12.96&12.75& -&-&-& -&-&-& -&-&-\\
& E3  &8.43&7.59&7.33&8.39&7.72&7.59 & -&-&-& -&-&-& -&-&-\\
  & N1      &15.44&10.67&7.33&15.99&11.08&7.59& -&-&-& -&-&-& -&-&- \\
& N2   & 8.59&7.85&7.66&8.53&7.94&7.85& -&-&-& -&-&-& -&-&-\\
 & N3&8.43&7.59&7.33&8.39&7.72&7.59& -&-&-& -&-&-& -&-&-\\
 \hline    
\end{tabular}
\parbox{0.95\textwidth}{%
 \footnotesize
\textit{Note:} Here $N, n_B$ and $n_T^{(C)}$ denote the total sample size, number of batches and number of therapists within a centre, respectively.  The proportion of embedded therapists within a centre is $a_E$. Parameter $a$ denotes that at least $100a\%$ of the total therapists available within a centre in each batch. The expected capacities of embedded and full-time research therapists $(E_E,E_F)$ is fixed to (12,20) or (25,30) for $n_T^{(C)} \sim Poisson(8)$ or $n_T^{(C)} \sim Poisson(4)$, respectively. The standard deviation of therapist capacities is fixed to $(\sigma_E,\sigma_F)= (6,2)$. Parameter $frac$ is fixed at $0.25$.}\label{table.time.to.target}
\end{table}

\clearpage

\begin{table}[ht]
\centering\caption{Estimation results for the medium trial for model-1}

\setlength{\tabcolsep}{3pt}  
\begin{tabular}{ p{2cm}| p{2cm}| c| c  c c  c| c  c c| c c| c c| c c c}
  \hline
&& Design&$\hat{\delta_1}$& $SE(\hat{\delta_1})$ & $\hat{\sigma}_{u_1}^2$ & $\hat{\sigma}_{v_1}^2$ &T-2 Error & Boundary& Fail& RMSE\\
\hline
&True Values&&0.4& -& 0.1& 0.1& -&-&-&-\\
\hline
\multirow{36}{*}{  \parbox[c]{2cm}{
    \centering
$N=640$,\\
$n_T^{(C)} \sim Poisson(8)$
  }}
  & \multirow{6}{*}{
  \parbox[c]{2cm}{
    \centering
    $n_B=5$\\
    $a_E = 0$\\
    $a=0.5$
  }}
& E1 & - & - &- & - & - & - & -& -  \\
&& E2 & 0.403 & 0.138 & 0.1 & 0.1 & 0.33 & 0.39 & 0.14 & 0.95 \\
&& E3 & 0.402 & 0.144 & 1.09 & 0.1 & 0.38 & 0.86 & 0 & 1.7 \\
&& N1 & 0.4 & 0.125 & 0.1 & 0.1 & 0.21 & 0.33 & 0 & 0.96 \\
&& N2 & 0.402 & 0.135 & 0.1 & 0.1 & 0.3 & 0.37 & 0 & 0.95 \\
&& N3 & 0.398 & 0.139 & 0.1 & 0.1 & 0.35 & 0.39 & 0 & 0.95 \\
\hline
\end{tabular}
\parbox{0.95\textwidth}{%
 \footnotesize
\textit{Note:} The value of the expected capacities $(E_P,E_F)$ is fixed to (12,20). The standard deviation of therapist capacities is fixed to $(\sigma_e,\sigma_f)= (6,2)$. Parameter $frac$ is fixed at $0.25$.}
\label{table.single.1}
\end{table}

\clearpage

\begin{table}[ht]
\centering\caption{Estimation results for the medium trial for model-1}
\setlength{\tabcolsep}{3pt}  
    \begin{tabular}{ p{2cm}| p{2cm}| c| c  c c  c| c  c c| c c| c c| c c c}
  \hline
&& Design&$\hat{\delta_1}$& $SE(\hat{\delta_1})$ & $\hat{\sigma}_{u_1}^2$ & $\hat{\sigma}_{v_1}^2$ &T-2 Error & Boundary& Fail&RMSE\\
\hline
&True Values&&0.4& -& 0.1& 0.1& -&-&-&-\\
\hline
\multirow{36}{*}{  \parbox[c]{2cm}{
    \centering
$N=640$,\\
$n_T^{(C)} \sim Poisson(8)$
  }}
   & \multirow{6}{*}{
  \parbox[c]{2cm}{
    \centering
    $n_B=5$\\
    $a_E = 0.5$\\
    $a=1$
  }}& E1 & - & - & - & - & - & - & 1 & - \\
&& E2 & - & - & - & - & - & - & 1 & - \\
&& E3 & 0.4 & 0.145 & 1.09 & 0.1 & 0.4 & 0.86 & 0 & 1.7 \\
&& N1 & 0.398 & 0.14 & 0.1 & 0.1 & 0.34 & 0.4 & 0 & 0.95 \\
&& N2 & 0.397 & 0.137 & 0.1 & 0.1 & 0.33 & 0.39 & 0 & 0.95 \\
&& N3 & 0.403 & 0.139 & 0.1 & 0.1 & 0.33 & 0.39 & 0 & 0.95 \\
\hline
\end{tabular}
\parbox{0.95\textwidth}{%
 \footnotesize
\textit{Note:} The value of the expected capacities $(E_P,E_F)$ is fixed to (12,20). The standard deviation of therapist capacities is fixed to $(\sigma_e,\sigma_f)= (6,2)$. Parameter $frac$ is fixed at $0.25$.}
\end{table}

\clearpage

\begin{table}[ht]
\centering\caption{Estimation results for the medium trial for model-1}
   \setlength{\tabcolsep}{3pt}  
 \begin{tabular}{ p{2cm}| p{2cm}| c| c  c c  c| c  c c| c c| c c| c c c}
  \hline
&& Design&$\hat{\delta_1}$& $SE(\hat{\delta_1})$ & $\hat{\sigma}_{u_1}^2$ & $\hat{\sigma}_{v_1}^2$ &T-2 Error & Boundary& Fail&RMSE\\
\hline
&True Values&&0.4& -& 0.1& 0.1& -&-&-&-\\
\hline
\multirow{36}{*}{  \parbox[c]{2cm}{
    \centering
$N=640$,\\
$n_T^{(C)} \sim Poisson(4)$
  }}& \multirow{6}{*}{
  \parbox[c]{2cm}{
    \centering
    $n_B=5$\\
    $a_E = 0$\\
    $a=0.5$
  }}
& E1 & - & - &- & - & - & - & - & - \\
&& E2 & 0.401 & 0.135 & 0.1 & 0.1 & 0.28 & 0.41 & 0 & 0.96 \\
&& E3 & 0.399 & 0.143 & 1.05 & 0.1 & 0.37 & 0.86 & 0 & 1.69 \\
&& N1 & 0.402 & 0.124 & 0.1 & 0.1 & 0.2 & 0.44 & 0 & 0.96 \\
&& N2 & 0.398 & 0.137 & 0.1 & 0.1 & 0.31 & 0.43 & 0 & 0.95 \\
&& N3 & 0.399 & 0.138 & 0.1 & 0.1 & 0.31 & 0.43 & 0 & 0.96 \\
\hline
\end{tabular}
\parbox{0.95\textwidth}{%
 \footnotesize
\textit{Note:} The value of the expected capacities $(E_P,E_F)$ is fixed to (25,30). The standard deviation of therapist capacities is fixed to $(\sigma_e,\sigma_f)= (6,2)$. Parameter $frac$ is fixed at $0.25$.}
\end{table}

\clearpage

\begin{table}[ht]
\centering\caption{Estimation results for the medium trial for model-1}
\setlength{\tabcolsep}{3pt}  
    \begin{tabular}{ p{2cm}| p{2cm}| c| c  c c  c| c  c c| c c| c c| c c c}
  \hline
&& Design&$\hat{\delta_1}$& $SE(\hat{\delta_1})$ & $\hat{\sigma}_{u_1}^2$ & $\hat{\sigma}_{v_1}^2$ &T-2 Error & Boundary& Fail& RMSE\\
\hline
&True Values&&0.4& -& 0.1& 0.1& -&-&-&-\\
\hline
\multirow{36}{*}{  \parbox[c]{2cm}{
    \centering
$N=640$,\\
$n_T^{(C)} \sim Poisson(8)$
  }}
  & \multirow{6}{*}{
  \parbox[c]{2cm}{
    \centering
    $n_B=8$\\
    $a_E = 0$\\
    $a=0.5$
  }}
& E1 & - & - & - & - & - & - & 1 & - \\
&& E2 & - & - & - & - & - & - & 1 & - \\
&& E3 & 0.401 & 0.129 & 1.09 & 0.1 & 0.25 & 0.86 & 0 & 1.7 \\
&& N1 & 0.401 & 0.112 & 0.1 & 0.1 & 0.13 & 0.33 & 0.15 & 0.97 \\
&& N2 & 0.399 & 0.123 & 0.1 & 0.1 & 0.19 & 0.37 & 0.15 & 0.96 \\
&& N3 & 0.4 & 0.122 & 0.1 & 0.1 & 0.19 & 0.36 & 0.15 & 0.96 \\
\hline
\end{tabular}
\parbox{0.95\textwidth}{%
 \footnotesize
\textit{Note:} The value of the expected capacities $(E_P,E_F)$ is fixed to (12,20). The standard deviation of therapist capacities is fixed to $(\sigma_e,\sigma_f)= (6,2)$. Parameter $frac$ is fixed at $0.25$.}
\end{table}

\clearpage

\begin{table}[ht]
\centering\caption{Estimation results for the medium trial for model-1}
\setlength{\tabcolsep}{3pt}  
    \begin{tabular}{ p{2cm}| p{2cm}| c| c  c c  c| c  c c| c c| c c| c c c}
  \hline
&& Design&$\hat{\delta_1}$& $SE(\hat{\delta_1})$ & $\hat{\sigma}_{u_1}^2$ & $\hat{\sigma}_{v_1}^2$ &T-2 Error & Boundary& Fail&RMSE\\
\hline
&True Values&&0.4& -& 0.1& 0.1& -&-&-&-\\
\hline
\multirow{36}{*}{  \parbox[c]{2cm}{
    \centering
$N=640$,\\
$n_T^{(C)} \sim Poisson(8)$
  }}
   & \multirow{6}{*}{
  \parbox[c]{2cm}{
    \centering
    $n_B=8$\\
    $a_E = 0.5$\\
    $a=1$
  }}& E1 & - & - & - & - & - & - & 1 & - \\
&& E2 & - & - & - & - & - & - & 1 & - \\
&& E3 & 0.4 & 0.133 & 1.09 & 0.1 & 0.28 & 0.86 & 0 & 1.7 \\
&& N1 & 0.4 & 0.127 & 0.1 & 0.1 & 0.22 & 0.38 & 0.19 & 0.96 \\
&& N2 & 0.399 & 0.126 & 0.1 & 0.1 & 0.21 & 0.37 & 0.21 & 0.96 \\
&& N3 & 0.4 & 0.127 & 0.1 & 0.1 & 0.22 & 0.38 & 0.19 & 0.96 \\
\hline
\end{tabular}
\parbox{0.95\textwidth}{%
 \footnotesize
\textit{Note:} The value of the expected capacities $(E_P,E_F)$ is fixed to (12,20). The standard deviation of therapist capacities is fixed to $(\sigma_e,\sigma_f)= (6,2)$. Parameter $frac$ is fixed at $0.25$.}
\end{table}

\clearpage

\begin{table}[ht]
\centering\caption{Estimation results for the medium trial for model-1}
   \setlength{\tabcolsep}{3pt}  
 \begin{tabular}{ p{2cm}| p{2cm}| c| c  c c  c| c  c c| c c| c c| c c c}
  \hline
&& Design&$\hat{\delta_1}$& $SE(\hat{\delta_1})$ & $\hat{\sigma}_{u_1}^2$ & $\hat{\sigma}_{v_1}^2$ &T-2 Error & Boundary& Fail&RMSE\\
\hline
&True Values&&0.4& -& 0.1& 0.1& -&-&-&-\\
\hline
\multirow{36}{*}{  \parbox[c]{2cm}{
    \centering
$N=640$,\\
$n_T^{(C)} \sim Poisson(4)$
  }}& \multirow{6}{*}{
  \parbox[c]{2cm}{
    \centering
    $n_B=8$\\
    $a_E = 0$\\
    $a=0.5$
  }}
& E1 & - & - &- & - & - & - & - & - \\
&& E2 & 0.402 & 0.128 & 0.1 & 0.1 & 0.23 & 0.37 & 0.01 & 0.96 \\
&& E3 & 0.398 & 0.135 & 1.05 & 0.1 & 0.3 & 0.86 & 0 & 1.69 \\
&& N1 & 0.401 & 0.117 & 0.1 & 0.1 & 0.17 & 0.38 & 0 & 0.97 \\
&& N2 & 0.398 & 0.127 & 0.1 & 0.1 & 0.23 & 0.37 & 0 & 0.96 \\
&& N3 & 0.403 & 0.128 & 0.1 & 0.1 & 0.22 & 0.37 & 0 & 0.96 \\
\hline
\end{tabular}
\parbox{0.95\textwidth}{%
 \footnotesize
\textit{Note:} The value of the expected capacities $(E_P,E_F)$ is fixed to (25,30). The standard deviation of therapist capacities is fixed to $(\sigma_e,\sigma_f)= (6,2)$. Parameter $frac$ is fixed at $0.25$.}\label{table.single.6}
\end{table}

\clearpage

\begin{table}[ht]
\centering\caption{Estimation results for the medium trial for model-2}
 \setlength{\tabcolsep}{3pt}  
   \begin{tabular}{ p{2cm}| p{2cm}| c| c  c c  c| c  c c| c c| c c| c c c}
  \hline
&& Design&$\hat{\delta_1}$& $SE(\hat{\delta_1})$ & $\hat{\sigma}_{w_1}^2$ & $\hat{\sigma}_{z_1}^2$ &T-2 Error & Boundary& Fail&RMSE\\
\hline
&True Values&&0.4& -& 0.1& 0.1& -&-&-&-\\
\hline
\multirow{36}{*}{  \parbox[c]{2cm}{
    \centering
$N=640$,\\
$n_T^{(C)} \sim Poisson(8)$
  }}& \multirow{6}{*}{
  \parbox[c]{2cm}{
    \centering
    $n_B=5$\\
    $a_E = 0$\\
    $a=0.5$
  }}
& E1 & - & - & - & - & - & - &-&- \\
&& E2 & 0.4 & 0.091 & 0.1 & 0.1 & 0.07 & 0.43 & 0.02 & 2.16 \\
&& E3 & 0.401 & 0.101 & 1.08 & 0.1 & 0.13 & 0.62 & 0.02 & 2.61 \\
&& N1 & 0.4 & 0.092 & 0.1 & 0.1 & 0.08 & 0.45 & 0.02 & 2.49 \\
&& N2 & 0.399 & 0.091 & 0.1 & 0.1 & 0.07 & 0.43 & 0.02 & 2.32 \\
&& N3 & 0.401 & 0.091 & 0.1 & 0.1 & 0.07 & 0.43 & 0.02 & 2.19 \\
\hline
\end{tabular}
\parbox{0.95\textwidth}{%
 \footnotesize
\textit{Note:} The value of the expected capacities $(E_P,E_F)$ is fixed to (12,20). The standard deviation of therapist capacities is fixed to $(\sigma_e,\sigma_f)= (6,2)$. Parameter $frac$ is fixed at $0.25$.}\label{table.single.7}
\end{table}

\clearpage

\begin{table}[ht]
\centering\caption{Estimation results for the medium trial for model-2}
\setlength{\tabcolsep}{3pt}  
    \begin{tabular}{ p{2cm}| p{2cm}| c| c  c c  c| c  c c| c c| c c| c c c}
  \hline
&& Design&$\hat{\delta_1}$& $SE(\hat{\delta_1})$ & $\hat{\sigma}_{w_1}^2$ & $\hat{\sigma}_{z_1}^2$ &T-2 Error & Boundary& Fail&RMSE\\
\hline
&True Values&&0.4& -& 0.1& 0.1& -&-&-&-\\
\hline
\multirow{36}{*}{  \parbox[c]{2cm}{
    \centering
$N=640$,\\
$n_T^{(C)} \sim Poisson(8)$
  }}
   & \multirow{6}{*}{
  \parbox[c]{2cm}{
    \centering
    $n_B=5$\\
    $a_E = 0.5$\\
    $a=1$
  }}& E1 & 0.4 & 0.09 & 0.1 & 0.1 & 0.07 & 0.43 & 0.02 & 1.92 \\
&& E2 & 0.4 & 0.09 & 0.1 & 0.1 & 0.07 & 0.43 & 0.02 & 1.92 \\
&& E3 & 0.4 & 0.102 & 1.08 & 0.1 & 0.14 & 0.64 & 0.02 & 2.71 \\
&& N1 & 0.4 & 0.091 & 0.1 & 0.1 & 0.08 & 0.44 & 0.02 & 2.3 \\
&& N2 & 0.401 & 0.092 & 0.1 & 0.1 & 0.07 & 0.45 & 0.02 & 2.39 \\
&& N3 & 0.399 & 0.091 & 0.1 & 0.1 & 0.08 & 0.43 & 0.02 & 2.31 \\
\hline
\end{tabular}
\parbox{0.95\textwidth}{%
 \footnotesize
\textit{Note:} The value of the expected capacities $(E_P,E_F)$ is fixed to (12,20). The standard deviation of therapist capacities is fixed to $(\sigma_e,\sigma_f)= (6,2)$. Parameter $frac$ is fixed at $0.25$.}
\end{table}

\clearpage

\begin{table}[ht]
\centering\caption{Estimation results for the medium trial for model-2}
   \setlength{\tabcolsep}{3pt}  
 \begin{tabular}{ p{2cm}| p{2cm}| c| c  c c  c| c  c c| c c| c c| c c c}
  \hline
&& Design&$\hat{\delta_1}$& $SE(\hat{\delta_1})$ & $\hat{\sigma}_{w_1}^2$ & $\hat{\sigma}_{z_1}^2$ &T-2 Error & Boundary& Fail&RMSE\\
\hline
&True Values&&0.4& -& 0.1& 0.1& -&-&-&-\\
\hline
\multirow{36}{*}{  \parbox[c]{2cm}{
    \centering
$N=640$,\\
$n_T^{(C)} \sim Poisson(4)$
  }}& \multirow{6}{*}{
  \parbox[c]{2cm}{
    \centering
    $n_B=5$\\
    $a_E = 0$\\
    $a=0.5$
  }}
& E1 & - & - &- & - & - & - & - & - \\
&& E2 & 0.398 & 0.095 & 0.1 & 0.1 & 0.1 & 0.57 & 0.43 & 3.67 \\
&& E3 & 0.401 & 0.105 & 1.05 & 0.1 & 0.16 & 0.71 & 0.43 & 4 \\
&& N1 & 0.402 & 0.097 & 0.1 & 0.1 & 0.1 & 0.58 & 0.43 & 4.37 \\
&& N2 & 0.401 & 0.095 & 0.1 & 0.1 & 0.09 & 0.57 & 0.44 & 3.77 \\
&& N3 & 0.398 & 0.095 & 0.1 & 0.1 & 0.1 & 0.56 & 0.43 & 3.73 \\
\hline
\end{tabular}
\parbox{0.95\textwidth}{%
 \footnotesize
\textit{Note:} The value of the expected capacities $(E_P,E_F)$ is fixed to (25,30). The standard deviation of therapist capacities is fixed to $(\sigma_e,\sigma_f)= (6,2)$. Parameter $frac$ is fixed at $0.25$.}
\end{table}

\clearpage

\begin{table}[ht]
\centering\caption{Estimation results for the medium trial for model-2}
 \setlength{\tabcolsep}{3pt}  
   \begin{tabular}{ p{2cm}| p{2cm}| c| c  c c  c| c  c c| c c| c c| c c c}
  \hline
&& Design&$\hat{\delta_1}$& $SE(\hat{\delta_1})$ & $\hat{\sigma}_{w_1}^2$ & $\hat{\sigma}_{z_1}^2$ &T-2 Error & Boundary& Fail&RMSE\\
\hline
&True Values&&0.4& -& 0.1& 0.1& -&-&-&-\\
\hline
\multirow{36}{*}{  \parbox[c]{2cm}{
    \centering
$N=640$,\\
$n_T^{(C)} \sim Poisson(8)$
  }}& \multirow{6}{*}{
  \parbox[c]{2cm}{
    \centering
    $n_B=8$\\
    $a_E = 0$\\
    $a=0.5$
  }}
& E1 & - & - & - & - & - & - &-&- \\
&& E2 & 0.398 & 0.09 & 0.1 & 0.1 & 0.07 & 0.44 & 0.02 & 2.01 \\
&& E3 & 0.398 & 0.102 & 1.08 & 0.1 & 0.14 & 0.64 & 0.02 & 2.65 \\
&& N1 & 0.401 & 0.092 & 0.1 & 0.1 & 0.08 & 0.44 & 0.02 & 2.43 \\
&& N2 & 0.399 & 0.091 & 0.1 & 0.1 & 0.07 & 0.44 & 0.02 & 2.1 \\
&& N3 & 0.4 & 0.091 & 0.1 & 0.1 & 0.07 & 0.43 & 0.02 & 2.12 \\
\hline
\end{tabular}
\parbox{0.95\textwidth}{%
 \footnotesize
\textit{Note:} The value of the expected capacities $(E_P,E_F)$ is fixed to (12,20). The standard deviation of therapist capacities is fixed to $(\sigma_e,\sigma_f)= (6,2)$. Parameter $frac$ is fixed at $0.25$.}
\end{table}

\clearpage

\begin{table}[ht]
\centering\caption{Estimation results for the medium trial for model-2}
\setlength{\tabcolsep}{3pt}  
    \begin{tabular}{ p{2cm}| p{2cm}| c| c  c c  c| c  c c| c c| c c| c c c}
  \hline
&& Design&$\hat{\delta_1}$& $SE(\hat{\delta_1})$ & $\hat{\sigma}_{w_1}^2$ & $\hat{\sigma}_{z_1}^2$ &T-2 Error & Boundary& Fail&RMSE\\
\hline
&True Values&&0.4& -& 0.1& 0.1& -&-&-&-\\
\hline
\multirow{36}{*}{  \parbox[c]{2cm}{
    \centering
$N=640$,\\
$n_T^{(C)} \sim Poisson(8)$
  }}
   & \multirow{6}{*}{
  \parbox[c]{2cm}{
    \centering
    $n_B=8$\\
    $a_E = 0.5$\\
    $a=1$
  }}& E1 & 0.401 & 0.09 & 0.1 & 0.1 & 0.07 & 0.43 & 0.02 & 1.92 \\
&& E2 & 0.398 & 0.091 & 0.1 & 0.1 & 0.08 & 0.42 & 0.02 & 1.94 \\
&& E3 & 0.4 & 0.101 & 1.08 & 0.1 & 0.13 & 0.64 & 0.02 & 2.57 \\
&& N1 & 0.401 & 0.091 & 0.1 & 0.1 & 0.07 & 0.44 & 0.02 & 2.28 \\
&& N2 & 0.399 & 0.091 & 0.1 & 0.1 & 0.07 & 0.44 & 0.02 & 2.09 \\
&& N3 & 0.402 & 0.091 & 0.1 & 0.1 & 0.07 & 0.43 & 0.02 & 2.12 \\
\hline
\end{tabular}
\parbox{0.95\textwidth}{%
 \footnotesize
\textit{Note:} The value of the expected capacities $(E_P,E_F)$ is fixed to (12,20). The standard deviation of therapist capacities is fixed to $(\sigma_e,\sigma_f)= (6,2)$. Parameter $frac$ is fixed at $0.25$.}
\end{table}

\clearpage

\begin{table}[ht]
\centering\caption{Estimation results for the medium trial for model-2}
   \setlength{\tabcolsep}{3pt}  
 \begin{tabular}{ p{2cm}| p{2cm}| c| c  c c  c| c  c c| c c| c c| c c c}
  \hline
&& Design&$\hat{\delta_1}$& $SE(\hat{\delta_1})$ & $\hat{\sigma}_{w_1}^2$ & $\hat{\sigma}_{z_1}^2$ &T-2 Error & Boundary& Fail&RMSE\\
\hline
&True Values&&0.4& -& 0.1& 0.1& -&-&-&-\\
\hline
\multirow{36}{*}{  \parbox[c]{2cm}{
    \centering
$N=640$,\\
$n_T^{(C)} \sim Poisson(4)$
  }}& \multirow{6}{*}{
  \parbox[c]{2cm}{
    \centering
    $n_B=8$\\
    $a_E = 0$\\
    $a=0.5$
  }}
& E1 & - & - &- & - & - & - & - & - \\
&& E2 & 0.399 & 0.095 & 0.1 & 0.1 & 0.09 & 0.56 & 0.44 & 3.25 \\
&& E3 & 0.4 & 0.109 & 1.04 & 0.1 & 0.19 & 0.73 & 0.44 & 4.29 \\
&& N1 & 0.401 & 0.097 & 0.1 & 0.1 & 0.11 & 0.56 & 0.44 & 4.04 \\
&& N2 & 0.399 & 0.096 & 0.1 & 0.1 & 0.11 & 0.58 & 0.44 & 3.75 \\
&& N3 & 0.4 & 0.096 & 0.1 & 0.1 & 0.1 & 0.58 & 0.44 & 4.05 \\
\hline
\end{tabular}
\parbox{0.95\textwidth}{%
 \footnotesize
\textit{Note:} The value of the expected capacities $(E_P,E_F)$ is fixed to (25,30). The standard deviation of therapist capacities is fixed to $(\sigma_e,\sigma_f)= (6,2)$. Parameter $frac$ is fixed at $0.25$.}\label{table.single.12}
\end{table}

\clearpage
\begin{table}[ht]
\setlength{\tabcolsep}{4pt}  
\centering\caption{Proportions of boundary estimates for Model 1}
\renewcommand{\arraystretch}{0.9} 
\begin{tabular}{c| c| c| c  c c  c c  c c c c c }
\hline
& & Design & $\hat{\sigma}^2_{u_1}$ & $\hat{\sigma}^2_{u_2}$ & $\hat{\sigma}^2_{u_4}$ & $\hat{\sigma}^2_{v_1}$ & $\hat{\sigma}^2_{v_2}$ & $\hat{\sigma}^2_{v_4}$ & $\hat{\sigma}^2_{u_3}$ & $\hat{\sigma}^2_{u_5}$&$\hat{\sigma}^2_{v_3}$&$\hat{\sigma}^2_{v_5}$  \\
\hline
\multirow{36}{*}{
  \parbox[c]{2cm}{\centering
  $N=640$\\$n_T\sim Poisson(8)$}}
   & \multirow{6}{*}{
  \parbox[c]{2cm}{\centering
  $n_B=5$\\$a_E=0$\\$a=0.5$}}
& E1 & - & - & - & - & - & - & - & - & - & - \\
&& E2& 0.01 & 0.22 & 0.21 & 0 & 0.03 & 0.06 & 0 & 0.06 & 0.01 & 0.01 \\
&& E3 & 0 & 0.23 & 0.38 & 0.06 & 0.09 & 0.12 & 0.28 & 0.18 & 0.34 & 0.2 \\

&& N1 & 0.01 & 0.14 & 0.22 & 0 & 0.01 & 0.05 & 0.01 & 0.05 & 0.02 & 0.01 \\
&& N2 & 0.01 & 0.19 & 0.21 & 0 & 0.03 & 0.05 & 0 & 0.06 & 0.01 & 0.01 \\
&& N3& 0.01 & 0.2 & 0.22 & 0 & 0.03 & 0.06 & 0.01 & 0.06 & 0.01 & 0.01 \\
\cline{2-13}
& \multirow{6}{*}{
  \parbox[c]{2cm}{\centering
  $n_B=5$\\$a_E=0$\\$a=0.75$}}
& E1 & - & - & - & - & - & - & - & - & - & - \\
\hline
\end{tabular}
\parbox{0.95\textwidth}{%
 \footnotesize
\textit{Note:} The value of the expected capacities $(E_P,E_F)$ is fixed at (12,20) is fixed to (12,20) or (25,30) for $n_T \sim Poisson(8)$ or $n_T \sim Poisson(4)$, respectively. The standard deviation of therapist capacities is fixed to $(\sigma_e,\sigma_f)= (6,2)$. Parameter $frac$ is fixed at $0.25$.}\label{table.boundary.1}
\end{table}

\clearpage

\begin{table}[ht]
\setlength{\tabcolsep}{1.1pt}  
\centering\caption{Proportions of boundary estimates for Model 2}
\renewcommand{\arraystretch}{0.9} 
\begin{tabular}{c| c| c| c  c c  c c  c c c c c c c}
\hline
& & Design & $\hat{\sigma}^2_{w_1}$ &$\hat{\sigma}^2_{w_3}$ & $\hat{\sigma}^2_{z_1}$ & $\hat{\sigma}^2_{z_3}$ & $\hat{\sigma}^2_{w_2}$ & $\hat{\sigma}^2_{w_4}$&$\hat{\sigma}^2_{z_2}$&$\hat{\sigma}^2_{z_4}$ &$Cor_{\hat{\sigma}^2_{w_1},\hat{\sigma}^2_{w_2}}$&$Cor_{\hat{\sigma}^2_{w_3},\hat{\sigma}^2_{w_4}}$ &$Cor_{\hat{\sigma}^2_{z_1},\hat{\sigma}^2_{z_2}}$&$Cor_{\hat{\sigma}^2_{z_3},\hat{\sigma}^2_{z_4}}$  \\
\hline
\multirow{36}{*}{
  \parbox[c]{2cm}{\centering
  $N=640$\\$n_T\sim Poisson(8)$}}
  & \multirow{6}{*}{
  \parbox[c]{2cm}{\centering
  $n_B=5$\\$a_E=0$\\$a=0.5$}}
& E1 & - & - & - & - & - & - & - & - & - & - & - & - \\
&& E2& 0.02 & 0.06 & 0 & 0.02 & 0 & 0.03 & 0 & 0 & 0.01 & 0.48 & 0 & 0.12 \\
&& E3 & 0 & 0.17 & 0.04 & 0.06 & 0 & 0.05 & 0 & 0 & 0 & 0.58 & 0.05 & 0.23 \\
&& N1 & 0.03 & 0.06 & 0 & 0.03 & 0 & 0.03 & 0 & 0 & 0.01 & 0.47 & 0 & 0.14 \\
&& N2 & 0.02 & 0.06 & 0 & 0.02 & 0 & 0.03 & 0 & 0 & 0.01 & 0.47 & 0 & 0.13 \\
&& N3& 0.02 & 0.06 & 0 & 0.02 & 0 & 0.03 & 0 & 0 & 0.01 & 0.47 & 0 & 0.12 \\
\cline{2-15}
& \multirow{6}{*}{
  \parbox[c]{2cm}{\centering
  $n_B=5$\\$a_E=0$\\$a=0.75$}}
& E1 & - & - & - & - & - & - & - & - & - & - & - & - \\
&& E2& 0.02 & 0.07 & 0 & 0.02 & 0 & 0.03 & 0 & 0 & 0.01 & 0.47 & 0 & 0.12 \\
&& E3 & 0 & 0.17 & 0.04 & 0.06 & 0 & 0.05 & 0 & 0 & 0 & 0.58 & 0.04 & 0.23 \\
&& N1 & 0.02 & 0.06 & 0 & 0.02 & 0 & 0.03 & 0 & 0 & 0.01 & 0.47 & 0 & 0.13 \\
&& N2 & 0.02 & 0.06 & 0 & 0.02 & 0 & 0.03 & 0 & 0 & 0.01 & 0.49 & 0 & 0.14 \\
&& N3& 0.02 & 0.06 & 0 & 0.02 & 0 & 0.03 & 0 & 0 & 0.01 & 0.47 & 0 & 0.12 \\
\cline{2-15}
& \multirow{6}{*}{
  \parbox[c]{2cm}{\centering
  $n_B=5$\\$a_E=0$\\$a=1$}}
& E1 & 0.02 & 0.06 & 0 & 0.02 & 0 & 0.03 & 0 & 0 & 0.01 & 0.46 & 0 & 0.12 \\
&& E2& 0.02 & 0.06 & 0 & 0.02 & 0 & 0.03 & 0 & 0 & 0.01 & 0.47 & 0 & 0.11 \\
&& E3 & 0 & 0.16 & 0.04 & 0.06 & 0 & 0.05 & 0 & 0.01 & 0 & 0.58 & 0.04 & 0.22 \\
&& N1 & 0.03 & 0.06 & 0 & 0.02 & 0 & 0.03 & 0 & 0 & 0.01 & 0.47 & 0 & 0.12 \\
&& N2 & 0.03 & 0.06 & 0 & 0.02 & 0 & 0.03 & 0 & 0 & 0.01 & 0.49 & 0 & 0.14 \\
&& N3& 0.02 & 0.07 & 0 & 0.02 & 0 & 0.03 & 0 & 0 & 0.01 & 0.47 & 0 & 0.12 \\
\cline{2-15}
 & \multirow{6}{*}{
  \parbox[c]{2cm}{\centering
  $n_B=5$\\$a_E=0.5$\\$a=0.5$}}
& E1 & - & - & - & - & - & - & - & - & - & - & - & - \\
&& E2& 0.02 & 0.06 & 0 & 0.02 & 0 & 0.03 & 0 & 0 & 0.01 & 0.47 & 0 & 0.11 \\
&& E3 & 0 & 0.17 & 0.05 & 0.07 & 0 & 0.05 & 0 & 0.01 & 0 & 0.59 & 0.06 & 0.25 \\
&& N1 & 0.03 & 0.06 & 0 & 0.02 & 0 & 0.03 & 0 & 0 & 0.02 & 0.48 & 0 & 0.15 \\
&& N2 & 0.02 & 0.06 & 0 & 0.02 & 0 & 0.03 & 0 & 0.01 & 0.01 & 0.47 & 0 & 0.13 \\
&& N3& 0.02 & 0.06 & 0 & 0.02 & 0 & 0.03 & 0 & 0 & 0.01 & 0.47 & 0 & 0.13 \\
\cline{2-15}
& \multirow{6}{*}{
  \parbox[c]{2cm}{\centering
  $n_B=5$\\$a_E=0.5$\\$a=0.75$}}
& E1 & - & - & - & - & - & - & - & - & - & - & - & - \\
&& E2& 0.02 & 0.06 & 0 & 0.02 & 0 & 0.03 & 0 & 0 & 0.01 & 0.47 & 0 & 0.12 \\
&& E3 & 0 & 0.17 & 0.05 & 0.07 & 0 & 0.05 & 0 & 0.01 & 0 & 0.58 & 0.05 & 0.25 \\
&& N1 & 0.03 & 0.07 & 0 & 0.02 & 0 & 0.03 & 0 & 0 & 0.02 & 0.48 & 0 & 0.14 \\
&& N2 & 0.03 & 0.07 & 0 & 0.02 & 0 & 0.03 & 0 & 0 & 0.01 & 0.48 & 0 & 0.13 \\
&& N3& 0.02 & 0.06 & 0 & 0.02 & 0 & 0.03 & 0 & 0 & 0.01 & 0.46 & 0 & 0.13 \\
\cline{2-15}
& \multirow{6}{*}{
  \parbox[c]{2cm}{\centering
  $n_B=5$\\$a_E=0.5$\\$a=1$}}
& E1 & 0.02 & 0.07 & 0 & 0.02 & 0 & 0.03 & 0 & 0 & 0.01 & 0.48 & 0 & 0.12 \\
&& E2& 0.02 & 0.06 & 0 & 0.02 & 0 & 0.03 & 0 & 0 & 0.01 & 0.47 & 0 & 0.12 \\
&& E3 & 0 & 0.17 & 0.05 & 0.07 & 0 & 0.05 & 0 & 0.01 & 0 & 0.58 & 0.05 & 0.25 \\
&& N1 & 0.03 & 0.07 & 0 & 0.02 & 0 & 0.03 & 0 & 0 & 0.02 & 0.48 & 0 & 0.14 \\
&& N2 & 0.03 & 0.07 & 0 & 0.02 & 0 & 0.03 & 0 & 0 & 0.01 & 0.48 & 0 & 0.13 \\
&& N3& 0.02 & 0.06 & 0 & 0.02 & 0 & 0.03 & 0 & 0 & 0.01 & 0.46 & 0 & 0.13 \\
\hline
\multirow{18}{*}{
  \parbox[c]{2cm}{\centering
  $N=640$\\$n_T\sim Poisson(4)$}}
 & \multirow{6}{*}{
  \parbox[c]{2cm}{\centering
  $n_B=5$\\$a_E=0$\\$a=0.5$}}
& E1 & - & - & - & - & - & - & - & - & - & - & - & - \\
&& E2& 0.04 & 0.07 & 0 & 0.04 & 0 & 0.04 & 0 & 0.01 & 0.07 & 0.54 & 0 & 0.26 \\
&& E3 & 0 & 0.2 & 0.03 & 0.08 & 0 & 0.07 & 0 & 0.01 & 0.02 & 0.64 & 0.05 & 0.32 \\
&& N1 & 0.04 & 0.07 & 0 & 0.04 & 0 & 0.04 & 0 & 0.01 & 0.09 & 0.54 & 0 & 0.29 \\
&& N2 & 0.04 & 0.07 & 0 & 0.04 & 0 & 0.04 & 0 & 0.01 & 0.07 & 0.55 & 0 & 0.26 \\
&& N3& 0.04 & 0.08 & 0 & 0.04 & 0 & 0.04 & 0 & 0.01 & 0.07 & 0.55 & 0 & 0.24 \\
\cline{2-15}
& \multirow{6}{*}{
  \parbox[c]{2cm}{\centering
  $n_B=5$\\$a_E=0$\\$a=0.75$}}
& E1 & - & - & - & - & - & - & - & - & - & - & - & - \\
&& E2& 0.04 & 0.07 & 0 & 0.03 & 0 & 0.04 & 0 & 0.01 & 0.06 & 0.52 & 0 & 0.25 \\
&& E3 & 0 & 0.2 & 0.02 & 0.08 & 0 & 0.08 & 0 & 0.01 & 0.02 & 0.65 & 0.03 & 0.32 \\
&& N1 & 0.04 & 0.07 & 0 & 0.04 & 0 & 0.04 & 0 & 0.01 & 0.07 & 0.56 & 0 & 0.26 \\
&& N2 & 0.04 & 0.08 & 0 & 0.04 & 0 & 0.04 & 0 & 0.01 & 0.07 & 0.52 & 0 & 0.25 \\
&& N3& 0.04 & 0.08 & 0 & 0.03 & 0 & 0.04 & 0 & 0.01 & 0.07 & 0.53 & 0 & 0.24 \\
\cline{2-15}
& \multirow{6}{*}{
  \parbox[c]{2cm}{\centering
  $n_B=5$\\$a_E=0$\\$a=1$}}
& E1 & 0.04 & 0.07 & 0 & 0.04 & 0 & 0.04 & 0 & 0.01 & 0.07 & 0.54 & 0 & 0.24 \\
&& E2& 0.04 & 0.08 & 0 & 0.04 & 0 & 0.04 & 0 & 0.01 & 0.07 & 0.54 & 0 & 0.24 \\
&& E3 & 0 & 0.2 & 0.02 & 0.08 & 0 & 0.06 & 0 & 0.01 & 0.02 & 0.65 & 0.04 & 0.33 \\
&& N1 & 0.04 & 0.07 & 0 & 0.03 & 0 & 0.04 & 0 & 0.01 & 0.06 & 0.54 & 0 & 0.25 \\
&& N2 & 0.04 & 0.08 & 0 & 0.04 & 0 & 0.04 & 0 & 0.01 & 0.07 & 0.54 & 0 & 0.25 \\
&& N3& 0.04 & 0.07 & 0 & 0.04 & 0 & 0.04 & 0 & 0.01 & 0.06 & 0.54 & 0 & 0.25 \\
\hline
\end{tabular}
\parbox{0.95\textwidth}{%
 \footnotesize
\textit{Note:} The value of the expected capacities $(E_P,E_F)$ is fixed at (12,20) is fixed to (12,20) or (25,30) for $n_T \sim Poisson(8)$ or $n_T \sim Poisson(4)$, respectively. The standard deviation of therapist capacities is fixed to $(\sigma_e,\sigma_f)= (6,2)$. Parameter $frac$ is fixed at $0.25$.}\label{table.boundary.2}
\end{table}


\title{Supplementary File F: Extra Results from the Single-Centre Trial}
\date{}

\maketitle
This supplementary file contains estimates of the remaining parameters from Model~1 and Model~2 that are not reported in Supplementary File~C, for the single‑centre trial across varying levels of embedded therapist proportion and therapist availability. Throughout the tables, $N, n_B$ and $n_T$ denote the total sample size, number of batches and number of therapists, respectively. The proportion of embedded therapists is $a_E$. Parameter $a$ denotes that at least $100a\%$ of the total therapists available in each batch. In all the tables the value of the expected capacities $(E_P,E_F)$ is fixed to (18,24) and (35,40) for $n_T=12$ and $n_T=6$, respectively. The standard deviation of therapist capacities is fixed to $(\sigma_e,\sigma_f)= (8,2)$. Parameter $frac$ is fixed at $0.25$.

Tables~\ref{table.single.1}--\ref{table.single.6} present parameter estimation results under Model~1, while Tables~\ref{table.single.7}--\ref{table.single.12} report the corresponding results for Model~2.
Tables~\ref{table.single.12.1}--\ref{table.single.12.2} summarise results obtained by generating data under Model~1 and fitting Model~2.  
Conversely, Tables~\ref{table.single.21.1}--\ref{table.single.21.2} report results from data generated under Model~2 and analysed using Model~1. 

\begin{table}[ht]
\setlength{\tabcolsep}{4pt}  
\centering\caption{Estimation results for the single centre trial for model-1}
\renewcommand{\arraystretch}{0.9} 
\begin{tabular}{c| c| c| c  c c  c c  c c c  }
\hline
& & Design & $\hat{\sigma}_{u_2}^2$ & $\hat{\sigma}_{u_4}^2$ & $\hat{\sigma}_{u_3}^2$ & $\hat{\sigma}_{v_2}^2$ & $\hat{\sigma}_{u_5}^2$ & $\hat{\sigma}_{v_4}^2$ & $\hat{\sigma}_{v_3}^2$ & $\hat{\sigma}_{v_5}^2$  \\
\hline
& True Values & - & 0.1 & - & 0.1 & 0.1 & - & - & 0.1 & -  \\
\hline
\multirow{24}{*}{
  \parbox[c]{2cm}{\centering
  $N=240$\\$n_T=12$}}
 & \multirow{6}{*}{
  \parbox[c]{2cm}{\centering
  $n_B=5$\\$a_E=0$\\$a=0.5$}}
& E1 & - & - & - & - & - & - & - & -  \\
&& E2 & 0.11 & - & 0.1 & 0.09 & - & - & 0.1 & -  \\
&& E3 & 0.11 & - & 0.11 & 0.09 & - & - & 0.11 & -  \\
&& N1 & 0.1 & - & 0.1 & 0.1 & - & - & 0.1 & -  \\
&& N2 & 0.11 & - & 0.1 & 0.09 & - & - & 0.1 & -  \\
&& N3 & 0.11 & - & 0.1 & 0.09 & - & - & 0.1 & -  \\
\hline 
\end{tabular}
\label{table.single.1}
\end{table}

\clearpage

\begin{table}[ht]
\setlength{\tabcolsep}{4pt}  
\centering\caption{Estimation results for the single centre trial for model-1}
\renewcommand{\arraystretch}{0.9} 
\begin{tabular}{c| c| c| c  c c  c c  c c c  }
\hline
& & Design & $\hat{\sigma}_{u_2}^2$ & $\hat{\sigma}_{u_4}^2$ & $\hat{\sigma}_{u_3}^2$ & $\hat{\sigma}_{v_2}^2$ & $\hat{\sigma}_{u_5}^2$ & $\hat{\sigma}_{v_4}^2$ & $\hat{\sigma}_{v_3}^2$ & $\hat{\sigma}_{v_5}^2$ \\
\hline
& True Values & - & 0.1 & - & 0.1 & 0.1 & - & - & 0.1 & -  \\
\hline
& \multirow{6}{*}{
  \parbox[c]{2cm}{\centering
  $n_B=5$\\$a_E=0.25$\\$a=1$}}
& E1 & 0.11 & - & 0.1 & 0.09 & - & - & 0.1 & -  \\
&& E2 & 0.11 & - & 0.1 & 0.09 & - & - & 0.1 & -  \\
&& E3 & 0.11 & - & 0.1 & 0.09 & - & - & 0.1 & -  \\
&& N1 & 0.11 & - & 0.1 & 0.09 & - & - & 0.1 & -  \\
&& N2 & 0.11 & - & 0.1 & 0.09 & - & - & 0.1 & -  \\
&& N3 & 0.11 & - & 0.1 & 0.09 & - & - & 0.1 & -  \\
\hline
\end{tabular}
\end{table}

\clearpage

\begin{table}[ht]
\setlength{\tabcolsep}{4pt}  
\centering\caption{Estimation results for the single centre trial for model-1}
\renewcommand{\arraystretch}{0.9} 
\begin{tabular}{c| c| c| c  c c  c c  c c c  }
\hline
& & Design & $\hat{\sigma}_{u_2}^2$ & $\hat{\sigma}_{u_4}^2$ & $\hat{\sigma}_{u_3}^2$ & $\hat{\sigma}_{v_2}^2$ & $\hat{\sigma}_{u_5}^2$ & $\hat{\sigma}_{v_4}^2$ & $\hat{\sigma}_{v_3}^2$ & $\hat{\sigma}_{v_5}^2$ \\
\hline
& True Values & - & 0.1 & - & 0.1 & 0.1 & - & - & 0.1 & -  \\
\hline
\multirow{24}{*}{
  \parbox[c]{2cm}{\centering
  $N=240$\\$n_T=6$}}
& \multirow{6}{*}{
  \parbox[c]{2cm}{\centering
  $n_B=5$\\$a_E=0$\\$a=0.5$}}
& E1 & - & - & - & - & - & - & - & -  \\
&& E2 & 0.11 & - & 0.1 & 0.09 & - & - & 0.1 & -  \\
&& E3 & 0.11 & - & 0.1 & 0.09 & - & - & 0.1 & - \\
&& N1 & 0.11 & - & 0.1 & 0.1 & - & - & 0.1 & -  \\
&& N2 & 0.11 & - & 0.1 & 0.09 & - & - & 0.1 & - \\
&& N3 & 0.11 & - & 0.1 & 0.09 & - & - & 0.1 & -  \\
\hline
\end{tabular}

\end{table}

\clearpage

\begin{table}[ht]
\setlength{\tabcolsep}{4pt}  
\centering\caption{Estimation results for the single centre trial for model-1}
\renewcommand{\arraystretch}{0.9} 
\begin{tabular}{c| c| c| c  c c  c c  c c c  }
\hline
& & Design & $\hat{\sigma}_{u_2}^2$ & $\hat{\sigma}_{u_4}^2$ & $\hat{\sigma}_{u_3}^2$ & $\hat{\sigma}_{v_2}^2$ & $\hat{\sigma}_{u_5}^2$ & $\hat{\sigma}_{v_4}^2$ & $\hat{\sigma}_{v_3}^2$ & $\hat{\sigma}_{v_5}^2$ \\
\hline
& True Values & - & 0.1 & - & 0.1 & 0.1 & - & - & 0.1 & -  \\
\hline
\multirow{24}{*}{
  \parbox[c]{2cm}{\centering
  $N=240$\\$n_T=12$}}
 & \multirow{6}{*}{
  \parbox[c]{2cm}{\centering
  $n_B=10$\\$a_E=0$\\$a=0.5$}}
& E1 & - & - & - & - & - & - & - & - \\
&& E2 & - & - & - & - & - & - & - & - \\
&& E3 & 0.11 & - & 0.11 & 0.1 & - & - & 0.13 & - \\
&& N1 & 0.1 & - & 0.1 & 0.1 & - & - & 0.12 & - \\
&& N2 & 0.1 & - & 0.09 & 0.1 & - & - & 0.11 & - \\
&& N3 & 0.1 & - & 0.1 & 0.09 & - & - & 0.12 & - \\
\hline
\end{tabular}

\end{table}

\clearpage

\begin{table}[ht]
\setlength{\tabcolsep}{4pt}  
\centering\caption{Estimation results for the single centre trial for model-1}
\renewcommand{\arraystretch}{0.9} 
\begin{tabular}{c| c| c| c  c c  c c  c c c  }
\hline
& & Design & $\hat{\sigma}_{u_2}^2$ & $\hat{\sigma}_{u_4}^2$ & $\hat{\sigma}_{u_3}^2$ & $\hat{\sigma}_{v_2}^2$ & $\hat{\sigma}_{u_5}^2$ & $\hat{\sigma}_{v_4}^2$ & $\hat{\sigma}_{v_3}^2$ & $\hat{\sigma}_{v_5}^2$  \\
\hline
& True Values & - & 0.1 & - & 0.1 & 0.1 & - & - & 0.1 & - \\
\hline
\multirow{24}{*}{
  \parbox[c]{2cm}{\centering
  $N=240$\\$n_T=12$}}
& \multirow{6}{*}{
  \parbox[c]{2cm}{\centering
  $n_B=10$\\$a_E=0.75$\\$a=0.5$}}
& E1 & - & - & - & - & - & - & - & -  \\
&& E2 & - & - & - & - & - & - & - & - \\
&& E3 & 0.1 & - & 0.1 & 0.1 & - & - & 0.11 & -  \\
&& N1 & 0.1 & - & 0.1 & 0.1 & - & - & 0.11 & - \\
&& N2 & 0.1 & - & 0.1 & 0.1 & - & - & 0.11 & - \\
&& N3 & 0.1 & - & 0.1 & 0.1 & - & - & 0.11 & -  \\
\hline
\end{tabular}

\end{table}

\clearpage

\begin{table}[ht]
\setlength{\tabcolsep}{4pt}  
\centering\caption{Estimation results for the single centre trial for model-1}
\renewcommand{\arraystretch}{0.9} 
\begin{tabular}{c| c| c| c  c c  c c  c c c  }
\hline
& & Design & $\hat{\sigma}_{u_2}^2$ & $\hat{\sigma}_{u_4}^2$ & $\hat{\sigma}_{u_3}^2$ & $\hat{\sigma}_{v_2}^2$ & $\hat{\sigma}_{u_5}^2$ & $\hat{\sigma}_{v_4}^2$ & $\hat{\sigma}_{v_3}^2$ & $\hat{\sigma}_{v_5}^2$  \\
\hline
& True Values & - & 0.1 & - & 0.1 & 0.1 & - & - & 0.1 & - \\
\hline
\multirow{24}{*}{
  \parbox[c]{2cm}{\centering
  $N=240$\\$n_T=6$}}
& \multirow{6}{*}{
  \parbox[c]{2cm}{\centering
  $n_B=10$\\$a_E=0$\\$a=0.5$}}
& E1 & - & - & - & - & - & - & - & -  \\
&& E2 & 0.1 & - & 0.1 & 0.1 & - & - & 0.1 & -  \\
&& E3 & 0.1 & - & 0.1 & 0.1 & - & - & 0.1 & -  \\
&& N1 & 0.1 & - & 0.1 & 0.1 & - & - & 0.1 & -  \\
&& N2 & 0.1 & - & 0.1 & 0.1 & - & - & 0.1 & -\\
&& N3 & 0.1 & - & 0.1 & 0.1 & - & - & 0.1 & -  \\
\hline
\end{tabular}
\label{table.single.6}
\end{table}

\clearpage

\begin{table}[ht]
\setlength{\tabcolsep}{4pt}  
\centering\caption{Estimation results for the single centre trial for model-2}
\renewcommand{\arraystretch}{0.9} 
\begin{tabular}{c| c| c| c  c c  c c  c c c  }
\hline
& & Design & $\delta_2$ & $\hat{\sigma}_{w_3}^2$ & $\hat{\sigma}_{w_2}^2$ & $\delta_3$ & $\hat{\sigma}_{w_4}^2$ & $\hat{\sigma}_{z_3}^2$ & $\hat{\sigma}_{z_2}^2$ & $\hat{\sigma}_{z_4}^2$  \\
\hline
& True Values & - & 0.1 & - & 0.1 & 0.1 & - & - & 0.1 & -  \\
\hline
\multirow{24}{*}{
  \parbox[c]{2cm}{\centering
  $N=240$\\$n_T=12$}}
  & \multirow{6}{*}{
  \parbox[c]{2cm}{\centering
  $n_B=5$\\$a_E=0$\\$a=0.5$}}
& E1 & - & - & - & - & - & - & - & -  \\
&& E2 & 0.1 & - & 0.1 & 0.1 & - & - & 0.1 & - \\
&& E3 & 0.1 & - & 0.1 & 0.1 & - & - & 0.1 & - \\
&& N1 & 0.1 & - & 0.1 & 0.1 & - & - & 0.1 & - \\
&& N2 & 0.1 & - & 0.1 & 0.1 & - & - & 0.1 & - \\
&& N3 & 0.1 & - & 0.1 & 0.1 & - & - & 0.1 & - \\
\hline
\end{tabular}
\label{table.single.7}
\end{table}

\clearpage

\begin{table}[ht]
\setlength{\tabcolsep}{4pt}  
\centering\caption{Estimation results for the single centre trial for model-2}
\renewcommand{\arraystretch}{0.9} 
\begin{tabular}{c| c| c| c  c c  c c  c c c  }
\hline
& & Design & $\delta_2$ & $\hat{\sigma}_{w_3}^2$ & $\hat{\sigma}_{w_2}^2$ & $\delta_3$ & $\hat{\sigma}_{w_4}^2$ & $\hat{\sigma}_{z_3}^2$ & $\hat{\sigma}_{z_2}^2$ & $\hat{\sigma}_{z_4}^2$  \\
\hline
& True Values & - & 0.1 & - & 0.1 & 0.1 & - & - & 0.1 & -  \\
\hline
\multirow{24}{*}{
  \parbox[c]{2cm}{\centering
  $N=240$\\$n_T=12$}}
& \multirow{6}{*}{
  \parbox[c]{2cm}{\centering
  $n_B=5$\\$a_E=0.75$\\$a=0.5$}}
& E1 & - & - & - & - & - & - & - & -  \\
&& E2 & 0.1 & - & 0.1 & 0.1 & - & - & 0.1 & - \\
&& E3 & 0.1 & - & 0.1 & 0.1 & - & - & 0.1 & - \\
&& N1 & 0.1 & - & 0.1 & 0.1 & - & - & 0.1 & - \\
&& N2 & 0.1 & - & 0.1 & 0.1 & - & - & 0.1 & - \\
&& N3 & 0.1 & - & 0.1 & 0.1 & - & - & 0.1 & - \\
\hline
\end{tabular}

\end{table}

\clearpage

\begin{table}[ht]
\setlength{\tabcolsep}{4pt}  
\centering\caption{Estimation results for the single centre trial for model-2}
\renewcommand{\arraystretch}{0.9} 
\begin{tabular}{c| c| c| c  c c  c c  c c c  }
\hline
& & Design & $\delta_2$ & $\hat{\sigma}_{w_3}^2$ & $\hat{\sigma}_{w_2}^2$ & $\delta_3$ & $\hat{\sigma}_{w_4}^2$ & $\hat{\sigma}_{z_3}^2$ & $\hat{\sigma}_{z_2}^2$ & $\hat{\sigma}_{z_4}^2$  \\
\hline
& True Values & - & 0.1 & - & 0.1 & 0.1 & - & - & 0.1 & -  \\
\hline
\multirow{24}{*}{
  \parbox[c]{2cm}{\centering
  $N=240$\\$n_T=6$}}
& \multirow{6}{*}{
  \parbox[c]{2cm}{\centering
  $n_B=5$\\$a_E=0$\\$a=0.5$}}
& E1 & - & - & - & - & - & - & - & -  \\
&& E2 & 0.1 & - & 0.11 & 0.1 & - & - & 0.09 & - \\
&& E3 & 0.1 & - & 0.11 & 0.1 & - & - & 0.09 & - \\
&& N1 & 0.1 & - & 0.11 & 0.1 & - & - & 0.09 & - \\
&& N2 & 0.1 & - & 0.11 & 0.1 & - & - & 0.09 & - \\
&& N3 & 0.1 & - & 0.11 & 0.1 & - & - & 0.09 & - \\
\hline
\end{tabular}

\end{table}

\clearpage

\begin{table}[ht]
\setlength{\tabcolsep}{4pt}  
\centering\caption{Estimation results for the single centre trial for model-2}
\renewcommand{\arraystretch}{0.9} 
\begin{tabular}{c| c| c| c  c c  c c  c c c  }
\hline
& & Design & $\delta_2$ & $\hat{\sigma}_{w_3}^2$ & $\hat{\sigma}_{w_2}^2$ & $\delta_3$ & $\hat{\sigma}_{w_4}^2$ & $\hat{\sigma}_{z_3}^2$ & $\hat{\sigma}_{z_2}^2$ & $\hat{\sigma}_{z_4}^2$  \\
\hline
& True Values & - & 0.1 & - & 0.1 & 0.1 & - & - & 0.1 & -  \\
\hline
\multirow{24}{*}{
  \parbox[c]{2cm}{\centering
  $N=240$\\$n_T=12$}}
  & \multirow{6}{*}{
  \parbox[c]{2cm}{\centering
  $n_B=10$\\$a_E=0$\\$a=0.5$}}
& E1 & - & - & - & - & - & - & - & -  \\
&& E2 & 0.1 & - & 0.1 & 0.1 & - & - & 0.1 & - \\
&& E3 & 0.1 & - & 0.1 & 0.1 & - & - & 0.1 & - \\
&& N1 & 0.1 & - & 0.1 & 0.1 & - & - & 0.1 & - \\
&& N2 & 0.1 & - & 0.1 & 0.1 & - & - & 0.1 & - \\
&& N3 & 0.1 & - & 0.1 & 0.1 & - & - & 0.1 & - \\
\hline
\end{tabular}

\end{table}

\clearpage

\begin{table}[ht]
\setlength{\tabcolsep}{4pt}  
\centering\caption{Estimation results for the single centre trial for model-2}
\renewcommand{\arraystretch}{0.9} 
\begin{tabular}{c| c| c| c  c c  c c  c c c  }
\hline
& & Design & $\delta_2$ & $\hat{\sigma}_{w_3}^2$ & $\hat{\sigma}_{w_2}^2$ & $\delta_3$ & $\hat{\sigma}_{w_4}^2$ & $\hat{\sigma}_{z_3}^2$ & $\hat{\sigma}_{z_2}^2$ & $\hat{\sigma}_{z_4}^2$  \\
\hline
& True Values & - & 0.1 & - & 0.1 & 0.1 & - & - & 0.1 & -  \\
\hline
\multirow{24}{*}{
  \parbox[c]{2cm}{\centering
  $N=240$\\$n_T=12$}}
  & \multirow{6}{*}{
  \parbox[c]{2cm}{\centering
  $n_B=10$\\$a_E=0.75$\\$a=0.5$}}
& E1 & - & - & - & - & - & - & - & -  \\
&& E2 & 0.1 & - & 0.1 & 0.1 & - & - & 0.1 & - \\
&& E3 & 0.1 & - & 0.1 & 0.1 & - & - & 0.1 & - \\
&& N1 & 0.1 & - & 0.1 & 0.1 & - & - & 0.1 & - \\
&& N2 & 0.1 & - & 0.1 & 0.1 & - & - & 0.1 & - \\
&& N3 & 0.1 & - & 0.1 & 0.1 & - & - & 0.1 & - \\
\hline
\end{tabular}

\end{table}

\begin{table}[ht]
\setlength{\tabcolsep}{4pt}  
\centering\caption{Estimation results for the single centre trial for model-2}
\renewcommand{\arraystretch}{0.9} 
\begin{tabular}{c| c| c| c  c c  c c  c c c  }
\hline
& & Design & $\delta_2$ & $\hat{\sigma}_{w_3}^2$ & $\hat{\sigma}_{w_2}^2$ & $\delta_3$ & $\hat{\sigma}_{w_4}^2$ & $\hat{\sigma}_{z_3}^2$ & $\hat{\sigma}_{z_2}^2$ & $\hat{\sigma}_{z_4}^2$  \\
\hline
& True Values & - & 0.1 & - & 0.1 & 0.1 & - & - & 0.1 & -  \\
\hline
\multirow{24}{*}{
  \parbox[c]{2cm}{\centering
  $N=240$\\$n_T=6$}}
& \multirow{6}{*}{
  \parbox[c]{2cm}{\centering
  $n_B=10$\\$a_E=0$\\$a=0.5$}}
& E1 & - & - & - & - & - & - & - & -  \\
&& E2 & 0.1 & - & 0.11 & 0.1 & - & - & 0.09 & - \\
&& E3 & 0.1 & - & 0.11 & 0.1 & - & - & 0.09 & - \\
&& N1 & 0.1 & - & 0.11 & 0.1 & - & - & 0.09 & - \\
&& N2 & 0.1 & - & 0.11 & 0.1 & - & - & 0.09 & - \\
&& N3 & 0.1 & - & 0.11 & 0.1 & - & - & 0.09 & - \\
\hline
\end{tabular}
\label{table.single.12}
\end{table}

\clearpage

\begin{table}[ht]
\setlength{\tabcolsep}{4pt}  
\centering\caption{Estimation results for the data generated from Model 1 and fitted using Model 2.}
\renewcommand{\arraystretch}{0.9} 
\begin{tabular}{c| c| c| c  c c  c c  c c c  }
\hline
& & Design & $\delta_2$ & $\hat{\sigma}_{w_3}^2$ & $\hat{\sigma}_{w_2}^2$ & $\delta_3$ & $\hat{\sigma}_{w_4}^2$ & $\hat{\sigma}_{z_3}^2$ & $\hat{\sigma}_{z_2}^2$ & $\hat{\sigma}_{z_4}^2$  \\
\hline
& True Values & - & - & - & - & - & - & - & - & -  \\
\hline
\multirow{18}{*}{
  \parbox[c]{2cm}{\centering
  $N=240$\\$n_T=12$}}
 & \multirow{6}{*}{
  \parbox[c]{2cm}{\centering
  $n_B=5$\\$a_E=0$\\$a=0.5$}}
& E1 & - & - & - & - & - & - & - & -  \\
&& E2 & 0 & - & 0 & 0.13 & - & - & 0 & - \\
&& E3 & 0 & - & 0 & 1.12 & - & - & 0 & - \\
&& N1 & 0 & - & 0 & 0.15 & - & - & 0 & - \\
&& N2 & 0 & - & 0 & 0.13 & - & - & 0 & - \\
&& N3 & 0 & - & 0 & 0.13 & - & - & 0 & - \\
\hline
\end{tabular}
\label{table.single.12.1}
\end{table}

\clearpage

\begin{table}[ht]
\setlength{\tabcolsep}{4pt}  
\centering\caption{Estimation results for the data generated from Model 1 and fitted using Model 2.}
\renewcommand{\arraystretch}{0.9} 
\begin{tabular}{c| c| c| c  c c  c c  c c c  }
\hline
& & Design & $\delta_2$ & $\hat{\sigma}_{w_3}^2$ & $\hat{\sigma}_{w_2}^2$ & $\delta_3$ & $\hat{\sigma}_{w_4}^2$ & $\hat{\sigma}_{z_3}^2$ & $\hat{\sigma}_{z_2}^2$ & $\hat{\sigma}_{z_4}^2$  \\
\hline
& True Values & - & - & - & - & - & - & - & - & -  \\
\hline
\multirow{18}{*}{
  \parbox[c]{2cm}{\centering
  $N=240$\\$n_T=12$}}
 & \multirow{6}{*}{
  \parbox[c]{2cm}{\centering
  $n_B=5$\\$a_E=1$\\$a=0.5$}}
& E1 & - & - & - & - & - & - & - & -  \\
&& E2 & 0 & - & 0 & 0.11 & - & - & 0 & - \\
&& E3 & 0 & - & 0 & 1.12 & - & - & 0 & - \\
&& N1 & 0 & - & 0 & 0.15 & - & - & 0 & - \\
&& N2 & 0 & - & 0 & 0.13 & - & - & 0 & - \\
&& N3 & 0 & - & 0 & 0.13 & - & - & 0 & - \\
\hline
\end{tabular}
\label{table.single.12.2}
\end{table}

\clearpage

\begin{table}[ht]
\setlength{\tabcolsep}{4pt}  
\centering\caption{Estimation results for the data generated from Model 2 and fitted using Model 1.}
\renewcommand{\arraystretch}{0.9} 
\begin{tabular}{c| c| c| c  c c  c c  c c c  }
\hline
& & Design & $\hat{\sigma}_{u_2}^2$ & $\hat{\sigma}_{u_4}^2$ & $\hat{\sigma}_{u_3}^2$ & $\hat{\sigma}_{v_2}^2$ & $\hat{\sigma}_{u_5}^2$ & $\hat{\sigma}_{v_4}^2$ & $\hat{\sigma}_{v_3}^2$ & $\hat{\sigma}_{v_5}^2$ \\
\hline
& True Values & - & - & - & - & - & - & - & - & -  \\
\hline
\multirow{18}{*}{
  \parbox[c]{2cm}{\centering
  $N=240$\\$n_T=12$}}
 & \multirow{6}{*}{
  \parbox[c]{2cm}{\centering
  $n_B=5$\\$a_E=0$\\$a=0.5$}}
& E1 & - & - & - & - & - & - & - & -  \\
&& E2 & 0.02 & - & 4.32 & 1.2 & - & - & 4.26 & - \\
&& E3 & 0.89 & - & 4.66 & 3.86 & - & - & 5.14 & - \\
&& N1 & 0.51 & - & 5.11 & 5.96 & - & - & 5.1 & - \\
&& N2 & 0.06 & - & 4.62 & 2.15 & - & - & 4.82 & - \\
&& N3 & 0.07 & - & 4.62 & 2.73 & - & - & 4.72 & - \\
\hline
\end{tabular}
\label{table.single.21.1}
\end{table}

\clearpage

\begin{table}[ht]
\setlength{\tabcolsep}{4pt}  
\centering\caption{Estimation results for the data generated from Model 2 and fitted using Model 1.}
\renewcommand{\arraystretch}{0.9} 
\begin{tabular}{c| c| c| c  c c  c c  c c c  }
\hline
& & Design & $\hat{\sigma}_{u_2}^2$ & $\hat{\sigma}_{u_4}^2$ & $\hat{\sigma}_{u_3}^2$ & $\hat{\sigma}_{v_2}^2$ & $\hat{\sigma}_{u_5}^2$ & $\hat{\sigma}_{v_4}^2$ & $\hat{\sigma}_{v_3}^2$ & $\hat{\sigma}_{v_5}^2$ \\
\hline
& True Values & - & - & - & - & - & - & - & - & -  \\
\hline
\multirow{18}{*}{
  \parbox[c]{2cm}{\centering
  $N=240$\\$n_T=12$}}
 & \multirow{6}{*}{
  \parbox[c]{2cm}{\centering
  $n_B=5$\\$a_E=1$\\$a=0.5$}}
& E1 & - & - & - & - & - & - & - & -  \\
&& E2 & 0 & - & 3.78 & 0.97 & - & - & 3.9 & - \\
&& E3 & 1.8 & - & 5.68 & 6.81 & - & - & 5.88 & - \\
&& N1 & 1.37 & - & 5.82 & 9.55 & - & - & 6.01 & - \\
&& N2 & 0.13 & - & 4.69 & 2.93 & - & - & 4.91 & - \\
&& N3 & 0.25 & - & 4.91 & 4.05 & - & - & 4.99 & - \\
\hline
\end{tabular}
\label{table.single.21.2}
\end{table}



\title{Figures of randomisation strategies}
\date{}

\begin{figure}[ht]
\centering
\includegraphics[width=12cm]{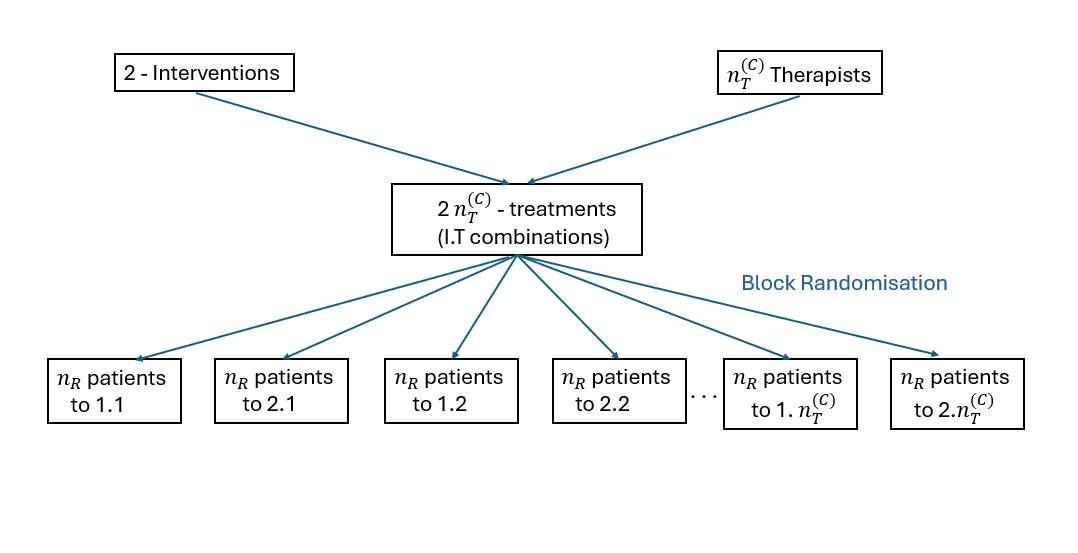}
\caption{Flowchart of randomisation strategy E1}
\label{fig.E1.rand} 
\end{figure}

\begin{figure}[ht]
\centering
\includegraphics[width=12cm]{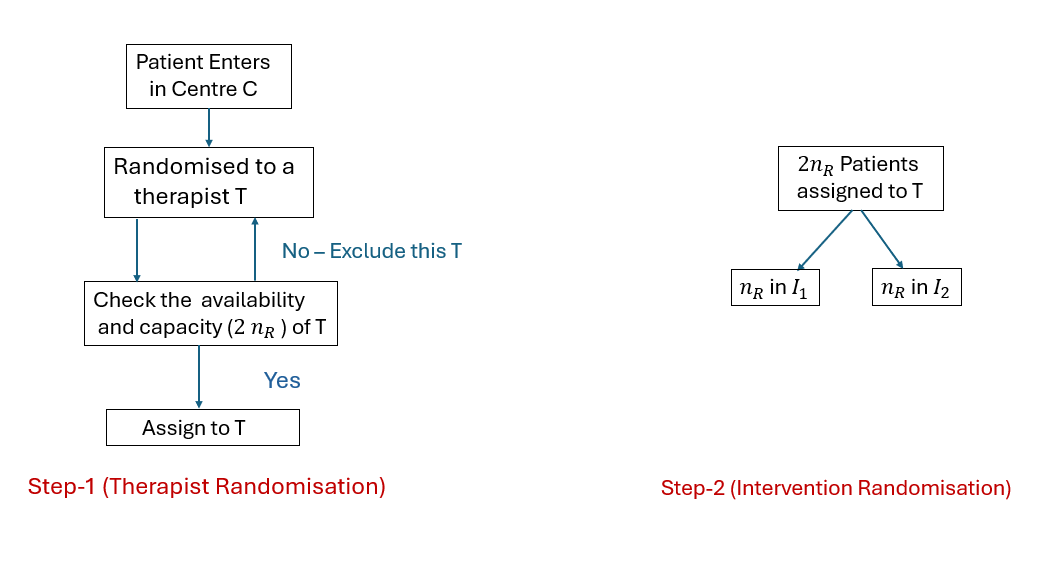}
\caption{Flowchart of randomisation strategy E2}
\label{fig.shapiro.rand} 
\end{figure}

\begin{figure}[ht]
\centering
\includegraphics[width=12cm]{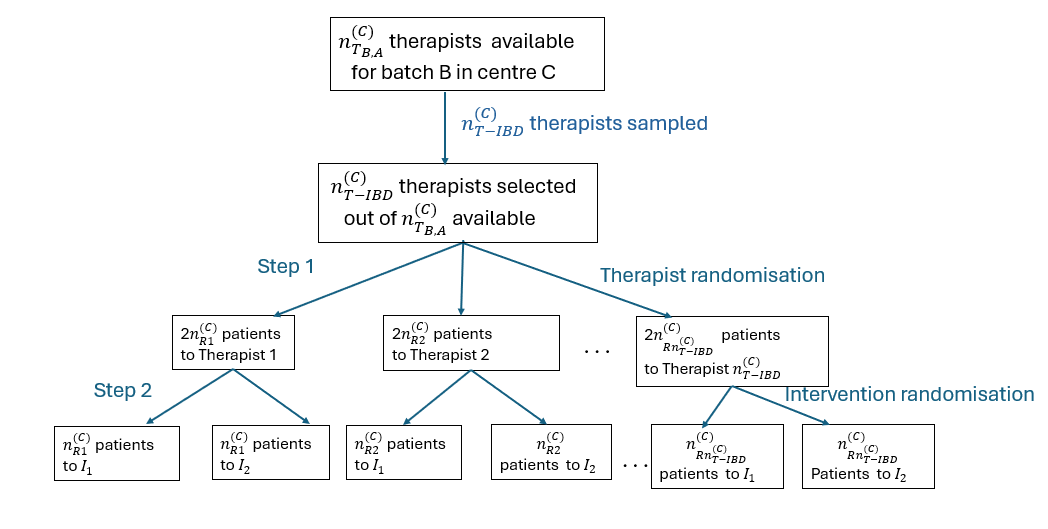}
\caption{Flowchart of randomisation strategy N1}
\label{fig.N1.rand} 
\end{figure}

\begin{figure}[ht]
\centering
\includegraphics[width=12cm]{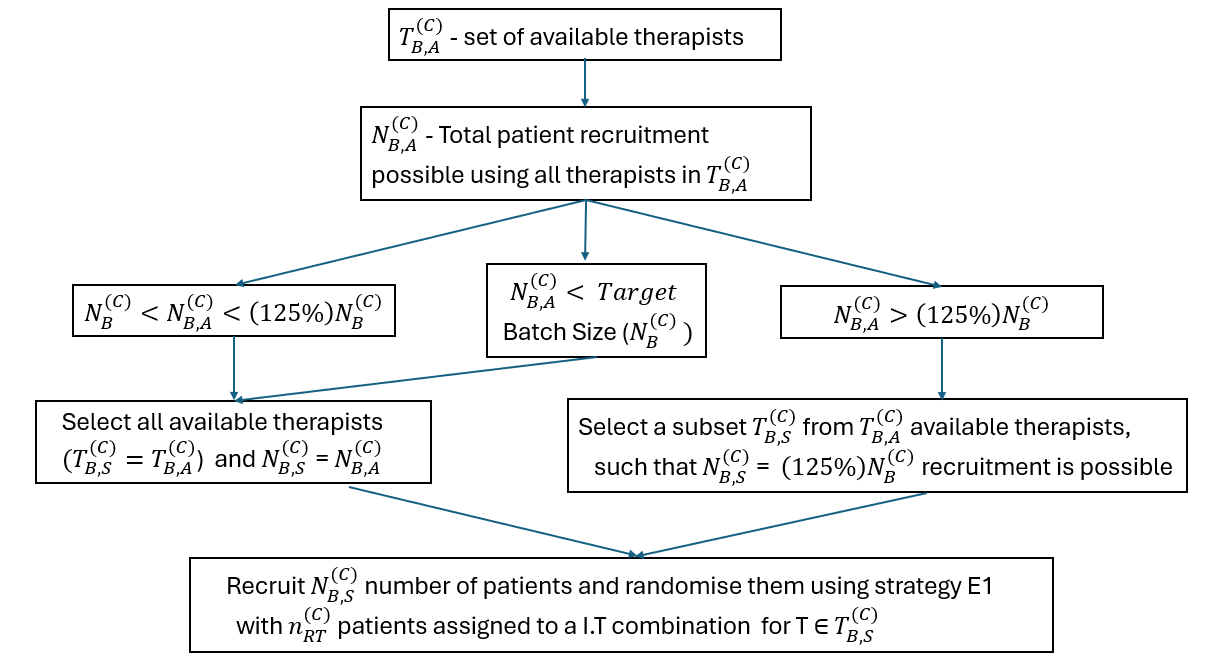}
\caption{Flowchart of randomisation strategy N2}
\label{fig.flex.rand} 
\end{figure}

\maketitle
Here we present flowcharts for several randomisation strategies to aid interpretation.  
Figure~\ref{fig.E1.rand} illustrates the E1 randomisation procedure, while Figures~\ref{fig.shapiro.rand}, \ref{fig.N1.rand}, and~\ref{fig.flex.rand} correspond to the E2, N1, and N2 randomisation strategies, respectively.

